\documentclass[twocolumn,pra,letterpaper,aps,superscriptaddress,showpacs]{revtex4-1}

\usepackage{amsmath,upgreek,amssymb,graphicx,subfigure,dsfont,color,bbm}
\usepackage{soul}
\usepackage{enumitem}
\setitemize{leftmargin=*}
\usepackage{comment}
\usepackage{mathtools}
\usepackage[hidelinks]{hyperref}
\usepackage{xcolor}
\hypersetup{
    colorlinks,
    linkcolor={red!50!black},
    citecolor={blue!50!black},
    urlcolor={blue!80!black}
}

\newcommand*\ourbar[1]{%
  \hbox{%
    \vbox{%
      \hrule height 0.4pt 
      \kern0.3ex
      \hbox{%
        \kern0em
        \ensuremath{#1}%
        \kern0em
      }%
    }%
  }%
} 
\newcommand*\ourbarr[1]{%
  \hbox{%
    \vbox{%
      \hrule height 0.4pt 
      \kern0.2ex
      \hbox{%
        \kern0em
        \ensuremath{#1}%
        \kern0em
      }%
    }%
  }%
} 

\newcommand{\overbarm}[1]{\mkern 2mu\ourbar{\mkern-2mu#1\mkern-2mu}\mkern 2mu}
\newcommand{\overbarmm}[1]{\mkern 2mu\ourbarr{\mkern-2mu#1\mkern-2mu}\mkern 2mu}
\newcommand{\overbarms}[1]{\scriptsize{\mkern 2mu\ourbar{\mkern-2mu#1\mkern-2mu}\mkern 2mu}}
\newcommand{\overbarmms}[1]{\scriptsize{\mkern 2mu\ourbarr{\mkern-2mu#1\mkern-2mu}\mkern 2mu}}
\newcommand{\overbar}[1]{\mkern 1.7mu\ourbar{\mkern-1.7mu#1\mkern-1.7mu}\mkern 1.7mu}
\newcommand{\overbarr}[1]{\mkern 1.7mu\ourbarr{\mkern-1.7mu#1\mkern-1.7mu}\mkern 1.7mu}

\usepackage{pgfplots}
 
\pgfplotsset{compat = newest}

\usepackage{lipsum}

\usepackage{fancyhdr}
\usepackage{tikz}
\usepackage{3dplot} 
\usepackage{tkz-euclide}
\usetikzlibrary{calc,through,angles,quotes,3d}

\usepackage[T1]{fontenc}
\usepackage{lmodern}

\usepackage{multirow}
\usepackage[english]{babel}
\usepackage{fullpage}
\usepackage{verbatim}
\usepackage{float}
\usepackage{ bbold }
\usepackage{physics}

\usepackage{scalerel}

\usepackage{natbib}
\usepackage{graphicx}
\usepackage{siunitx}
\usepackage{amssymb}
\usepackage{xspace}
\usepackage[utf8]{inputenc}

\graphicspath{{/home/victor/Escritorio/primerarticulo/Imagenes/}}

\DeclareMathAlphabet\mathbfcal{OMS}{cmsy}{b}{n}

\newcommand{\bbar}[1]{\overbarr{\overbar{#1}}}
\newcommand{\bbarr}[1]{\overbarmm{\overbarm{#1}}}

\newcommand{\bbarrs}[1]{\overbarmms{\overbarms{#1}}}

\newcommand{\larmornon}{\omega_0}

\newcommand{\wrffirstnon}{\omega_{1}}

\newcommand{\wrfsecondnon}{\omega_{2}}

\newcommand{\rotfirst}{\theta_1}
\newcommand{\rotsecond}{\theta_2}
\newcommand{\barm}{\overbar{m}}
\newcommand{\barM}{\overbar{M}}

\newcommand{\barms}{\overbarms{m}}
\newcommand{\barMs}{\overbarms{M}}

\newcommand{\bbarm}{\bbarr{m}}
\newcommand{\bbarM}{\bbarr{M}}

\newcommand{\bbarms}{\bbarrs{m}}
\newcommand{\bbarMs}{\bbarrs{M}}

\newcommand{\rf}{{\mathrm{rf}}}

\newcommand{\e}{\mathrm{e}}
\newcommand{\im}{\mathrm{i}}

\newcommand{\U}[2]{U_{\bold{#1}}(#2)}
\newcommand{\R}[2]{\mathcal{R}_{\bold{#1}}(#2)}

\renewcommand{\ss}{\mathrm{s}}
\newcommand{\ds}{\mathrm{d}}
\newcommand{\sd}{\mathrm{sd}}

\newcommand{\omegasecond}{\bbar{\omega}_{0}}
\newcommand{\omegafirst}{\overbar{\omega}_{0}}

\newcommand{\ca}{$^{40}\mathrm{Ca}^+$\xspace}
\newcommand{\slevel}{$^2\mathrm{S}_{1/2}$\xspace}
\newcommand{\dlevel}{$^2\mathrm{D}_{5/2}$\xspace}
\newcommand{\clocktrans}{$^2\mathrm{S}_{1/2} \leftrightarrow\ ^2\mathrm{D}_{5/2}$}
\newcommand{\groundstate}{$^2\mathrm{S}_{1/2}, \mathrm{m} = -1/2$}
\newcommand{\DDsplus}{ $\barm=\frac{1}{2}$}
\newcommand{\DDsminus}{$\barm=-\frac{1}{2}$}
\newcommand{\ts}{\textsuperscript}

\usepackage{color}

\usepackage{titlesec}

\titleformat*{\subsubsection}{\bfseries}

\usepackage{natbib}
\begin{document}

\title{Quadrupole transitions and quantum gates protected by continuous dynamic decoupling}

\author{V. J. Mart{\'{\i}}nez-Lahuerta}
\affiliation{Institute for Theoretical Physics, Leibniz University Hannover, Appelstrasse 2, 30167 Hannover, Germany}

\author{L. Pelzer}
\affiliation{Physikalisch-Technische Bundesanstalt, Bundesallee 100, 38116 Braunschweig, Germany}

\author{K. Dietze}
\affiliation{Physikalisch-Technische Bundesanstalt, Bundesallee 100, 38116 Braunschweig, Germany}

\author{L. Krinner}
\affiliation{Physikalisch-Technische Bundesanstalt, Bundesallee 100, 38116 Braunschweig, Germany}
\affiliation{Institute for Quantum Optics, Leibniz University Hannover, Welfengarten 1, 30167 Hannover, Germany.}

\author{P. O. Schmidt}
\affiliation{Physikalisch-Technische Bundesanstalt, Bundesallee 100, 38116 Braunschweig, Germany}
\affiliation{Institute for Quantum Optics, Leibniz University Hannover, Welfengarten 1, 30167 Hannover, Germany.}

\author{K. Hammerer}
\affiliation{Institute for Theoretical Physics, Leibniz University Hannover, Appelstrasse 2, 30167 Hannover, Germany}

\begin{abstract}
Dynamical decoupling techniques are a versatile tool for engineering quantum states with tailored properties. In trapped ions, nested layers of continuous dynamical decoupling by means of radio-frequency field dressing can cancel dominant magnetic and electric shifts and therefore provide highly prolonged coherence times of electronic states. Exploiting this enhancement for frequency metrology, quantum simulation or quantum computation, poses the challenge to combine the decoupling with laser-ion interactions for the quantum control of electronic and motional states of trapped ions. Ultimately, this will require running quantum gates on qubits from dressed decoupled states. We provide here a compact representation of nested continuous dynamical decoupling in trapped ions, and apply it to electronic $S$ and $D$ states and optical quadrupole transitions. Our treatment provides all effective transition frequencies and Rabi rates, as well as the effective selection rules of these transitions. On this basis, we discuss the possibility of combining continuous dynamical decoupling and M{\o}lmer-S{\o}rensen gates.  
\end{abstract}

\date\today

\maketitle


\section{Introduction}

Since the early work of Hahn on spin echoes in nuclear magnetic resonance (NMR)~\cite{Hahn_1950}, techniques for dynamically decoupling a quantum system from its environment to increase its coherence times have become indispensable tools of quantum technology~\cite{Souza2012}, with applications in quantum simulations, computation, and metrology. Robust dynamic decoupling methods by applying external pulses have been intensively developed both in theory~\cite{Viola_1998, Viola_1999,Zanardi_1999,Byrd2002,Facchi2005, Khodjasteh2008, Khodjasteh_2009, Khodjasteh_2009a, Khodjasteh2010, LidarReview2012, aharon_Robust_2019, green_arbitrary_2013, west_high_2010, uhrig_keeping_2007, haeberlen_coherent_1968} and in experiment~\cite{Biercuk2009, Du2009, Damodarakurup2009, lange_universal_2010, Souza2011, Naydenov2011, Sar_2012, shaniv_quadrupole_2019, wang_single-qubit_2017, qiu_suppressing_2021, zhou_quantum_2020, manovitz_fast_2017, shaniv_atomic_2016, piltz_protecting_2013, lange_universal_2010}. In recent years, continuous dynamical decoupling (CDD), where control pulses are applied in the form of continuous time periodic fields in the spirit of Floquet engineering~\cite{Kuwahara_2016}, have been proposed and demonstrated~\cite{Fonseca_2005, Chen_2006,yalcinkaya_Continuous_2019, Clausen_2010,xu_coherence-protected_2012, Fanchini_2007, Fanchini_prot_2007, Fanchini_2015, Rabl_2009, Chaudhry_2012, Cai_2012, Laraoui_2011, shaniv_quadrupole_2019,Bermudez_2011, Bermudez_2012,Timoney_2011,Doherty_2013, Albrecht_2014, Golter_2014, finkelstein_continuous_2021, trypogeorgos_synthetic_2018, anderson_continuously_2018, laucht_dressed_2017, sarkany_controlling_2014, webster_simple_2013, tan_demonstration_2013, aharon_general_2013, zanon-willette_magic_2012}. 

The design of long-lived quantum states using CDD has promising perspectives, especially for trapped ion frequency metrology as proposed and studied in~\cite{Cai_2012, aharon_Robust_2019,shaniv_quadrupole_2019}. The statistical uncertainty for a given clock species can be improved by extending the probe time, which will ultimately be limited by the lifetime of the excited states~\cite{kessler_heisenberg-limited_2014}. Nevertheless, in practice, it is usually limited by the coherence time of the clock laser~\cite{Peik_2005,Leroux_2017}. We can also improve the statistical uncertainty by interrogating many atoms simultaneously~\cite{Keller_ProbingTime_2019,arnold_prospects_2015,herschbach_Linear_2012,Champenois_2010}. But increasing the number of ions stored in a Paul trap entails further obstacles to overcome. Depending on the ion species chosen,  inhomogeneous or time-dependent frequency shifts, such as the Zeeman shift, the Quadrupole shift, or the radio frequency (rf) electric field-induced tensor ac Stark shift~\cite{Itano2000, berkeland_Minimization_1998, arnold_prospects_2015}, pose a limitation. These effects can contribute to the decoherence of the state or broaden the joint linewidth of the ions, thus limiting the usable probe time. Several approaches exist to constrain the tensor-like electric field shifts even without exact knowledge of the electric field gradient. One approach consists in averaging over different transitions or directions to exploit the different scaling of the shift with the angular momentum component~\cite{Itano2000, Schneider2005, dube_electric_2005}, or by chosing a magnetic field direction along which the tensor shifts have a zero crossing~\cite{tan_suppressing_2019}. Another method dynamically changes the static offset B-field direction within the clock interrogation~\cite{lange_Coherent_2020} to mimic the magic angle spinning technique of nuclear magnetic resonance spectroscopy~\cite{andrew_Nuclear_1958}. Elimination of these shifts can also be achieved by suitable hyperfine or Zeeman averaging using DD~\cite{kaewuam_hyperfine_2020, shaniv_quadrupole_2019}. Achieving robust optical clock transitions protected by CDD has been explored by Aharon \textit{et al.}~\cite{aharon_Robust_2019}.

In order to exploit these tailored states for quantum metrology, possibly involving entangled states of many ions, dynamical decoupling has to be combined with laser-ion interaction on optical quadrupole transitions, which will be the focus of the current work. Following the work of Aharon et. al.~\cite{aharon_Robust_2019} we reformulate the CDD description to easily treat the laser ion interaction. We begin by recapitulating the dynamical decoupling principle for a particular spin manifold, which is subject to a Zeeman splitting controlled by a static dc magnetic field, showing the effective Hamiltonian in the so-called doubly-dressed basis. Here, modulated external rf magnetic fields are employed to mitigate the amplitude-induced line shifts~\cite{aharon_Robust_2019}. Then, with appropriate CDD parameters, we achieve suppression of Zeeman and quadrupole shifts in this basis. Next, we consider optical quadrupole transitions between two of these spin manifolds and characterize the laser-ion interaction needed to drive the above transitions. We will show that there is no selection rule for transitions in the doubly-dressed basis. The only necessary condition will be the proper detuning of the laser. The suppression of Zeeman and quadrupole shifts will come at the cost of a reduction of the effective Rabi frequency for these transitions, and therefore, the characterization of these transitions will allow us to choose an appropriate candidate for a clock transition. We compare our analytical treatment with measurements of CDD states of a single \ca ion. Measurements of the energy spectrum between different spin manifolds as well as their relative optical coupling are in good agreement with the predictions. We will finish by discussing the application of a M\o lmer-S\o rensen gate in the doubly dressed basis, discussing its challenges and calculating a theoretical prediction for the gate time.

The article is organized as follows: In Section~\ref{DynamicalDecoupling} we reformulate the CDD description showing the suppression of Zeeman and quadrupole shifts for the appropiate parameters. The characterization of the optical transitions among two doubly-dressed manifolds through laser interaction, as well as the application to a trapped \ca ion is discussed in Section~\ref{QuadrupoletransitionsDDS}. In Section~\ref{Experiment} the experiment is described along with a comparison of the predicted and measured first stage CDD spectrum. Finally in Section~\ref{MolmerSorensen} we motivate the application of a M\o lmer-S\o rensen gate and study the time gate for the case of a trapped \ca ion.

\section{Dynamical Decoupling}\label{DynamicalDecoupling}

In this section, we recapitulate the principle of dynamical decoupling for the suppression of Zeeman and quadrupole shifts by applying radiofrequency magnetic fields~\cite{Cai_2012, aharon_Robust_2019,shaniv_quadrupole_2019}. Although we will eventually consider quadrupole transitions from a spin-$S^\mathrm{s}$ manifold of ground states to a spin-$S^\mathrm{d}$ manifold of excited states, it will be useful to first examine how the dressing fields affect a single spin-$S$ manifold. This will facilitate the discussion of the physical principle of dynamical decoupling. Moreover, this will separate the effects associated with the problem of a single manifold from those associated with the cross-coupling of spin manifolds, which we will consider later.

\subsection{Doubly-dressed basis}\label{DoublyDressedBasis}

We will consider a manifold of total spin $S$ with $\mathbf{S}=(S_x,S_y,S_z)$, basis states $\ket*{M}$, $|M|\leq S$, and quantization axis along $z$. If a static magnetic field $B$ along the $z$-axis is present, the internal states $\ket*{M}$ will be shifted by a value proportional to their spin, due to the linear Zeeman effect. Therefore, the Hamiltonian will have the expression
\begin{align}
    H_\mathrm{dc}= g\mu_B BS_z=\larmornon S_z,
\end{align}
where $g$ is the gyromagnetic factor, the corresponding Larmor frequency is $\larmornon=g\mu_BB$, with $\mu_B$ being the Bohr magnetron, and we set $\hbar=1$. The eigenstates of this Hamiltonian will be referred to as bare states. A radio-frequency field $B_\rf(t)$ is applied with a polarisation in the $x-y$ plane, which for the sake of generality we consider enclosing an angle $\alpha$ with the $x$-axis. The $\rf$ field $B_\rf(t)$ is assumed to comprise frequency components at a fundamental frequency $\omega_1$ and sideband frequencies $\omega_1\pm\omega_2$, where $\omega_2<\omega_1$, such that the Hamiltonian for the rf fields is
\begin{align}\label{eq:rffields}
    H_{\rf}&= g \qty\big(\Omega_{1} \cos(\wrffirstnon t)-\Omega_{2} \sin(\wrffirstnon t)\cos(\wrfsecondnon t))\nonumber\\
    &\quad\times(S_x \cos\alpha+S_y \sin\alpha),
\end{align}
where $\Omega_1$ and $\Omega_2$ are set by the amplitudes of the fundamental and sideband components of the  rf-magnetic field, respectively. Therefore, the total Hamiltonian for the spin $S$ in the laboratory frame (LF) is
\begin{align}\label{eq:H_QA0_LF}
    H^\mathrm{LF}=H_\mathrm{dc}+H_{\rf}.
\end{align}

To help characterize the rf or dressing fields, we are going to introduce a series of transformations into several frames. In this sequence of transformations we will denote a unitary rotation around an axis $\mathbf{n}$ about an angle $\theta$ by
\begin{align}
    \U{n}{\theta}=\exp\qty(\im\theta\mathbf{n}\mathbf{S}),
\end{align}
and use the notation  
\begin{align}
    \R{n}{\theta}A\coloneqq U^{\phantom\dagger}_{\bold{n}}(\theta)AU^\dagger_{\bold{n}}(\theta),
\end{align}
for the superoperator corresponding to the conjugation of an operator $A$ with $\U{n}{\theta}$. Bold symbols denote three-vectors. To determine the Hamiltonian operator $H$ in a new reference system, we consider the transformation of the operator $H-\im\dv{t}$ in each case so that the time dependence of the transformation is properly accounted for. This will be useful when dealing with sequences of transformations.

First, we go into a frame rotating around the $z$-axis at the rf frequency $\omega_1$
\begin{align}\label{eq:Ham0_RF}
&\R{z}{\omega_1 t}\qty[H^\mathrm{LF}-\im\textstyle{\dv{t}}]\nonumber\\
&=\qty(\R{z}{\omega_1 t}H^\mathrm{LF})-\omega_1 S_z-\im\dv{t}\nonumber\\
&=\Delta_{1}S_z+\frac{g\Omega_{1}}{2}(S_x \cos\alpha+S_y\sin\alpha)
    \\
&\quad+\frac{g\Omega_{2}}{2}\cos(\wrfsecondnon t)(S_y \cos\alpha-S_x \sin\alpha)-\im\dv{t}.\nonumber
\end{align}
Here, we have defined the detuning of the rf-field with respect to the Larmor frequency $\Delta_{1}=\larmornon-\wrffirstnon $. We have also used a rotating wave approximation (RWA) and dropped terms oscillating at $2\omega_1$, assuming $2\omega_1\gg g\Omega_1/2$. The effective contribution of these counter rotating terms on the bare states is addressed in Appendix~\ref{appendixC}. 

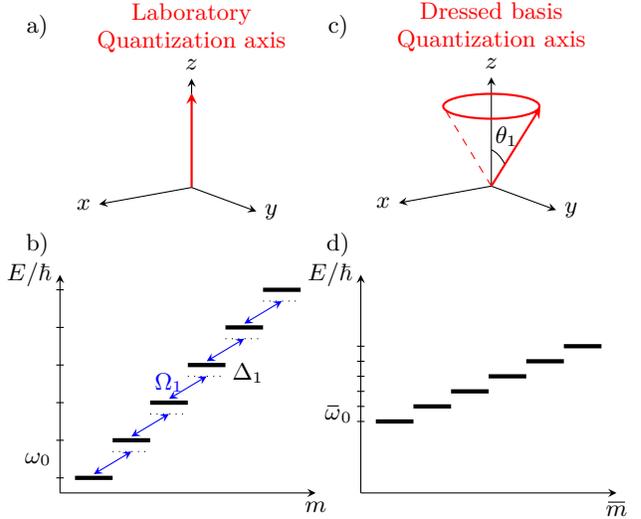
\begin{figure}[h]

\usetikzlibrary{calc,through,shapes.callouts, decorations.pathreplacing,angles,quotes,3d}%
\tdplotsetmaincoords{75}{145} 
\tikzoption{canvas is plane}[]{\@setOxy#1}%
\def\@setOxy O(#1,#2,#3)x(#4,#5,#6)y(#7,#8,#9)%
  {\def\tikz@plane@origin{\pgfpointxyz{#1}{#2}{#3}}%
   \def\tikz@plane@x{\pgfpointxyz{#4}{#5}{#6}}%
   \def\tikz@plane@y{\pgfpointxyz{#7}{#8}{#9}}%
   \tikz@canvas@is@plane%
  }%
\tikzset{%
  level/.style   = { ultra thick, black },%
  connect/.style = { ultra thin, gray },%
  line/.style = { dashed , gray },%
  notice/.style  = { draw, rectangle callout, callout relative pointer={#1} },%
  label/.style   = { text width=2cm }%
}%
\hspace*{-1.5cm}
\begin{tikzpicture}[scale=1,tdplot_main_coords,>=stealth,dot/.style={draw,fill,circle,inner sep=1pt}]%
\coordinate (O) at ($(-0.8191520442889919,0.5735764363510459,0)$);
\draw[->] (O) -- ($(O)+(1.5,0,0)$) node[anchor=east]{$x$};
\draw[->] (O) -- ($(O)+(0,1.5,0)$) node[anchor=west]{$y$};
\draw[->] (O) -- ($(O)+(0,0,1.5)$) node[anchor=south]{$z$};
\draw[->, thick, color=red] (O) -- ($(O)+(0,0,1.3)$) ;
\draw ($(O)+(0,0,2.2)$) node[align=center, color = red]{Laboratory\\ Quantization axis};
\coordinate (DIS) at ($0.5*({-8*cos(30)},{8*sin(30)},-0.15)$);

\coordinate (P) at ($(O)+(DIS)$);
\draw[->] (P) -- ($(1.5,0,0)+(P)$) node[anchor=east]{$x$};
\draw[->] (P) -- ($(0,1.5,0)+(P)$) node[anchor=west]{$y$};
\draw[->] (P) -- ($(0,0,1.5)+(P)$) node[anchor=south]{$z$};
\draw[->, thick, color=red] (P) -- ($(-0.825,-0.04,0.95)+(P)$);
\draw ($(P)+(0,0,2.2)$) node[align=center, color = red]{Dressed basis\\ Quantization axis};

\draw[dashed, color=red] (P) -- ($(0.56,-0.3,1.1)+(P)$);

\draw[thick, color=red] ($(0,0,1.1)+(P)$) circle (.64 and .64);

\tdplotsetthetaplanecoords{88}
	
	\tdplotdrawarc[tdplot_rotated_coords]{(P)}{0.5}{0}{40}{above}{  $\ \ \theta_1$} 
	

\begin{scope}[canvas is plane={O(0,0,-4)x(-0.8191520442889919,0.5735764363510459,-4)y(0,0,-4+1.0362694300518136)}]
\begin{scope}[xshift=-0.75cm]
\hspace*{0.7cm}
\coordinate (DIS2) at (4,0);
    \draw (-1,6) node {a)};
    \draw (-1,3.1) node{b)};
    \draw ($(-1,6)+(DIS2)$) node {c)};
    \draw ($(-1,3.1)+(DIS2)$) node{d)};

  \coordinate (A) at (-0.5,0);
  \coordinate (B) at (0,0.5);
  \coordinate (C) at (0.5,1);
  \coordinate (D) at (1,1.5);
  \coordinate (E) at (1.5,2);
  \coordinate (F) at (2,2.5);
  
  \coordinate (P1) at (-0.75, 0);
  \coordinate (P2) at (-0.65, 0);
  \coordinate (P3) at ($(P1)+(DIS2)$);
  \coordinate (P4) at ($(P2)+(DIS2)$);
  \coordinate (G) at (5.5,1);

  \foreach \x in {A, B, C, D, E, F}
    \draw[level] (\x) -- ($(\x)+(0.5,0)$);
  \foreach \x in {B, C, D, E, F}
    \draw[dotted] ($(\x)+(0,-0.15)$) -- ($(\x)+(0.5,-0.15)$);

  \foreach \x/\y in {A/B, B/C, C/D, D/E, E/F}
    \draw [<->,blue] ($(\x)+(0.25,0.05)$) to ($(\y)+(0.25,-0.15)$);

\draw ($(D)+(0.8,-0.1)$) node[] {$\Delta_1$};
\draw ($(C)+(0.25,0.27)$) node[blue] {$\Omega_1$};

\foreach \x in {A, B, C, D, E, F}
    \draw (P1 |- \x) -- (P2 |- \x);

  \draw (-0.7,0.25) node[anchor=east] {$\omega_0$};
  
  \draw[->] (-0.7,-0.2) -- (-0.7,2.7) node[anchor=east]{$E/\hbar$};
  \draw[->] (-0.7,-0.2) -- (2.7,-0.2) node[anchor=north]{$m$};


  \coordinate (H) at (3.5,0.75);
  \coordinate (I) at (4,0.95);
  \coordinate (J) at (4.5,1.15);
  \coordinate (K) at (5,1.35);
  \coordinate (L) at (5.5,1.55);
  \coordinate (M) at (6,1.75);
  
  \foreach \x in {H, I, J, K, L, M}
    \draw[level] (\x) -- ($(\x)+(0.5,0)$);
    
  \foreach \x in {H, I, J, K, L, M}
    \draw (P3 |- \x) -- (P4 |- \x);

  \draw (3.3,0.85) node[anchor=east] {$\omegafirst$};

  \draw[->] ($(-0.7,-0.2)+(DIS2)$) -- ($(-0.7,2.7)+(DIS2)$) node[anchor=east]{$E/\hbar$};
  \draw[->] ($(-0.7,-0.2)+(DIS2)$) -- ($(2.7,-0.2)+(DIS2)$) node[anchor=north]{$\barm$};

 \end{scope}
\end{scope}

\end{tikzpicture}
\caption{\label{fig:LevelScheme}Sketch of dynamical decoupling effect on a spin manifold (here $S=3/2$).  (a) illustrates the quantization axis and (b) the energy splitting $\omega_1$ of bare basis states $\ket{m}$ and the rf drive at Rabi frequency $\Omega_1$ and detuning $\Delta_1$. (c) shows the quantization with one layer of dressing, and (d) the effective level scheme of the dressed levels $\ket{\barm}$ with splitting $\omegafirst$. (b) and (d) are not to scale as $\omegafirst \ll \omega_0$.}
\end{figure}

In the next step, the Hamiltonian is rewritten in the dressed state basis corresponding to the eigenstates of the time-independent part of the Hamiltonian on the right-hand side of Eq.~\eqref{eq:Ham0_RF}, which correspond to the first line in the right-hand side. We achieve this by a rotation around an axis $\bold{n}_1=(-\sin\alpha,\cos\alpha,0)$ and an angle $\rotfirst \in \left[0,\pi\right]$ defined by $\cos\rotfirst=\Delta_{1}/\omegafirst$, where
\begin{align}
    \omegafirst=\qty(\Delta_{1}^2+g^2\Omega^2_{1}/4)^{1/2}.
\end{align}
The Hamiltonian in this first dressed basis is
\begin{align}\label{eq:FirstLayer}
&\R{n_\text{1}}{\rotfirst}\R{z}{\omega_1 t}\qty[H^\mathrm{LF}-\im\textstyle{\dv{t}}]\nonumber\\
&=\omegafirst S_z
+\frac{g\Omega_{2}}{2}\cos(\wrfsecondnon t)(S_y \cos\alpha-S_x \sin\alpha)-\im\dv{t}.
\end{align}
This Hamiltonian refers to a new time-dependent quantization axis enclosing an angle $\rotfirst$ with the $z$-axis. The relation between the bare basis and the dressed basis and their respective quantization axis and energy splittings $\hbar\larmornon$ and $\hbar\omegafirst$ are shown in Fig.~\ref{fig:LevelScheme}. In the regime considered here, these frequencies satisfy the hierarchy $\larmornon\gg\omegafirst$.

\makeatletter
\makeatother

The next dressing layer consists of the same two types of transformations as the first one. First, the system is transformed into the rotating frame with frequency $\omega_2$ around the new quantization axis, where fast oscillating terms $2\omega_2\gg g\Omega_2/4$ are neglected. Then, a transformation is applied in a new basis that diagonalizes the Hamiltonian, now independent of time. The transformation that achieves this corresponds to a rotation by an axis $\bold{n}_2=(-\cos\alpha,-\sin\alpha,0)$ and the angle $\rotsecond$ where $\cos\rotsecond=\Delta_{2}/\omegasecond$, and
\begin{align}
    \omegasecond=\qty(\Delta_{2}^2+g^2\Omega^2_{2}/16)^{1/2}.
\end{align}
The detuning at the second dressing layer is $\Delta_{2}=\omegafirst-\wrfsecondnon $. This results in the final, doubly-dressed Hamiltonian
\begin{align}\label{eq:HamT2}
&\R{n_\text{2}}{\rotsecond}\R{z}{\omega_2 t}
\R{n_\text{1}}{\rotfirst}\R{z}{\omega_1 t}\qty[H^\mathrm{LF}-\im\textstyle{\dv{t}}]\nonumber\\
&=\bbar{H}-\im\dv{t}
\end{align}
where the Hamiltonian in the doubly-dressed basis is
\begin{align}
\bbar{H}&=\omegasecond S_z\label{eq:Ham2}.
\end{align}
The quantization axis of the Hamiltonian in Eq.~\eqref{eq:HamT2} is now again rotated at an angle $\rotsecond$ with respect to the previous one. In principle, further dressing layers can be added, which will correspond to a similar sequence of transformations. Applications of $n$ layers of dressing have been discussed by Cai \textit{et al.}~\cite{Cai_2012}. We note that we will use symbols with single and double overbars (such as $\omegafirst$ and $\omegasecond$) to denote quantities in the singly or doubly dressed frame, respectively. 

We emphasize that the dressing procedure involves two rotating wave approximations, which are implicit in Eq.~\eqref{eq:HamT2}, and are based on $2\omega_i\gg g\Omega_i/2^i$ for $i=1,2$. Thus, we have the hierarchy  $\omegasecond\ll\omega_2\ll\omega_1$. Nevertheless, the terms neglected during the RWA will be accounted for perturbatively using the Magnus expansion in appendix~\ref{appendixC}. We note that, instead of the perturbative treatment given here, it is also possible to determine the exact quasi-energy eigenstates of the time-periodic Hamiltonian in the laboratory frame in the framework of Floquet theory. However, since this analysis provides mainly numerical insight, we focus on the analytical perturbative treatment in this presentation. We checked numerically that this treatment is in excellent agreement with the dc component of the Floquet states when counter-rotating terms are accounted for in a Magnus expansion~\cite{Martinez_thesis}.

\subsection{Suppression of Zeeman and quadrupole shifts}\label{SuppresionZeemanQuadrupole}

In this section we briefly discuss how the two layers of dressing help to suppress linear Zeeman and electric quadrupole shifts. We refer to the original work of Aharon et. al.~\cite{aharon_Robust_2019} for a detailed discussion. Both effects can be modeled by adding a suitable perturbation  $V^\mathrm{LF}(t)$ to the Hamiltonian in the laboratory frame in Eq.~\eqref{eq:H_QA0_LF}. This term may be time-dependent, but is assumed to fluctuate slowly on the time scale of the dressed states energy splitting $\omegasecond^{-1}$. In the doubly-dressed basis (DB) and in an interaction picture with respect to $\bbar{H}$, Eq.~\eqref{eq:Ham2}, such an additional term will be effectively described by 
\begin{align}
V^\mathrm{DB}&=\R{z}{\omegasecond t}\R{n_\text{2}}{\rotsecond}\R{z}{\omega_2 t}
\R{n_\text{1}}{\rotfirst}\R{z}{\omega_1 t}V^\mathrm{LF}\nonumber\\
&\eqqcolon\mathcal{D}(\omega_i,g\Omega_i,t)\qty[V^\mathrm{LF}].\label{eq:trafo}
\end{align}
The last (leftmost) rotation around $z$ at frequency $\omegasecond$ accounts for the interaction picture. The complete sequence of transformations corresponding to the dynamic decoupling and the change to the interaction picture will be abbreviated by the superoperator $\mathcal{D}(\omega_i,g\Omega_i,t)$. The goal of dynamic decoupling is to reduce $V^\mathrm{DB}$ by an appropriate choice of the driving parameters, which are the rf frequencies $\omega_i$ and Rabi frequencies $g\Omega_i$ with $i=1,2$. This general reasoning can now be applied to linear-magnetic and electric-quadrupole shifts.

Let us first study the shift of the bare states created through magnetic field fluctuations. This shift can be described by
\begin{align}
    V^\mathrm{LF}_{\delta B}=  g \mu_B \delta \bold{B}(t) \bold{S},\label{eq:Bfieldfluctuation}
\end{align}
where $\delta \bold{B}(t)$ is the time dependent part of the magnetic field, being the total magnetic field $\bold{B}(t)= (0,0,B)+\delta \bold{B}(t)$. Transforming this shift into the doubly-dressed basis according to Eq.~\eqref{eq:trafo} gives rise to
\begin{align}
V^\mathrm{DB}_{\delta B}
&=\mathcal{D}(\omega_i,g\Omega_i,t)\qty[V^\mathrm{LF}_{\delta B}]\nonumber\\
&= \cos\rotfirst \cos\rotsecond\, g\mu_B \delta B_z(t) S_z.\label{eq:Bfieldsuppression}
\end{align}
The derivation of this expression is shown in Appendix~\ref{appendixA}. Under the assumption that $\delta \bold{B}(t)$ fluctuates slowly on all relevant time scales, only the component along $z$, the direction of the dc field, matters. The terms in the $x$ and $y$ components of $\delta \bold{B}(t)$ can be neglected in a rotating wave approximation after the first rotation around $z$ with frequency $\omega_1$. Eq.~\eqref{eq:Bfieldsuppression} shows that magnetic field fluctuations can be suppressed and even nulled by choosing the angle in the first and/or second stage dressing to be $\theta_{1(2)}=\pi/2$, which is fulfilled by a set of resonant parameters $\Delta_{1(2)}=0$.

A similar cancelation can be achieved for electric-quadrupole shifts, as has been shown in \cite{aharon_Robust_2019, shaniv_quadrupole_2019} for a single layer of dressing. We generalize this treatment here for two layers of dressing. The quadrupole shift is described by the Hamiltonian
\begin{align}
    V^\mathrm{LF}_{Q}=\Tr\qty{QF(t)},
\end{align}
where
    $
    Q_{ij}=\frac{3}{2}\left(S_iS_j+S_jS_i\right)-S\left(S+1\right)\mathds{1}
    $
    with $S\left(S+1\right)=\bold{S}^2$,
    $
    F_{ij}=\frac{\partial E_j}{\partial x_i}
    $
and the components of the electric field $E_j$. The change to the doubly-dressed basis and the interaction picture following Eq.~\eqref{eq:trafo} gives
\begin{align}
    V_Q^{\mathrm{DB}}
    &=\mathcal{D}(\omega_i,g\Omega_i,t)\qty[V^\mathrm{LF}_{Q}]\nonumber\\
    &=\frac{1}{4}\qty(1-3\cos^2\theta_1)\qty(1-3\cos^2\theta_2)
    \nonumber\\
    &\quad\times
    \frac{3F_{zz}(t)}{2}\qty[S(S+1)-3S_z^2].\label{eq:HQ}
\end{align}
Details of the derivation of this expression are given in Appendix~\ref{appendixA}. The first line on the right-hand side of Eq.~\eqref{eq:HQ}, whose magnitude is at most one, gives the reduction of the quadrupole shift due to dynamic decoupling. The last line is just the standard expression for the quadrupole shift of the non-degenerate levels in the rotating wave approximation. With the so-called magic angle, $\cos^2\theta_{1(2)}=1/3$, the quadrupole shift can be eliminated in either the first or the second dressing layer.

In general, with two layers of dressing, it is possible to eliminate both Zeeman and quadrupole shifts by choosing $\cos\theta_{1(2)}=0$ and
$\cos^2\theta_{2(1)}=1/3$. When determining which effect to cancel in the first layer and which in the second, it is important to consider time scales and shift magnitudes. The first dressing layer involves a coarse grain of time over a scale of $\omega_1^{-1}$ with a protective energy gap proportional to $\Omega_1$, while the second one averages over $\omega_2^{-1}>\omega_1^{-1}$ at a correspondingly smaller energy gap proportional to $\Omega_2$. Therefore, it will be advantageous to cancel the faster fluctuations with larger magnitude first. For example, in the case of \ca discussed in the next section, it is advantageous to suppress magnetic field fluctuations using the first drive and the quadrupole and other small quasi-static tensor shifts using the second drive.

\section{Laser ion interaction}
\label{Sec:laserioninteraction}
Now, we will apply this formalism to the description to two Zeeman manifolds, and study the   electric-quadrupole transitions between them. We will start by characterizing the laser-ion interaction and finding the conditions that drive each transition. After that we will apply this formalism to the particular case of \ca in order to visualize how this transitions will be spread in the frequency spectrum.

\makeatletter
\begin{figure}
\usetikzlibrary{shapes.callouts}
\tikzset{%
  level/.style   = { ultra thick, black },%
  connect/.style = { ultra thin, gray },%
  line/.style = { dashed , gray },%
  notice/.style  = { draw, rectangle callout, callout relative pointer={#1} },%
  label/.style   = { text width=2cm }%
}%

\begin{tikzpicture}%
\hspace{0.6cm}

\coordinate (DIS2) at (-3.8,0);
    \draw (-0.5,3.8) node {b)};
    \draw ($(-0.2,3.8)+(DIS2)$) node {a)};

  \coordinate (A) at (0,3.5);
  \coordinate (B) at (0.3,3.5);
  \coordinate (C) at (0.6,3.5);
  \coordinate (D) at (0.9,3.5);
  \coordinate (E) at (1.2,3.5);
  \coordinate (F) at (1.5,3.5);
  
  \coordinate (G) at (0.6,2);
  \coordinate (H) at (0.9,2);

  \foreach \x in {A, B, C, D, E, F, G, H}
    \draw[level] (\x) -- ($(\x)+(0.25,0)$);
  \foreach \x in {A, B, C, D, E, F, G, H}
    \draw[level] ($(\x)+(DIS2)$) -- ($(\x)+(0.25,0)+(DIS2)$);

  \draw [<->,blue] ($(G)+(0.1,0.05)+(DIS2)$) to ($(B)+(0.1,-0.05)+(DIS2)$);
  \draw [<->,blue] ($(H)+(0.1,0.05)+(DIS2)$) to ($(C)+(0.15,-0.05)+(DIS2)$);
  \foreach \x/\y in {G/C, H/D}
    \draw [<->,gray] ($(\x)+(0.12,0.05)+(DIS2)$) to ($(\y)+(0.12,-0.05)+(DIS2)$);
  \draw [<->,red] ($(G)+(0.15,0.05)+(DIS2)$) to ($(D)+(0.1,-0.05)+(DIS2)$);
  \draw [<->,red] ($(H)+(0.15,0.05)+(DIS2)$) to ($(E)+(0.15,-0.05)+(DIS2)$);
  \foreach \x/\y in {G/A, H/B}
    \draw [<->,green] ($(\x)+(0.075,0.05)+(DIS2)$) to ($(\y)+(0.15,-0.05)+(DIS2)$);
  \foreach \x/\y in {G/E, H/F}
    \draw [<->,green] ($(\x)+(0.2,0.05)+(DIS2)$) to ($(\y)+(0.1,-0.05)+(DIS2)$);
  \draw[<->] ($(G)+(0.12,0.05)$) to ($(C)+(0.12,-0.05)$);

  \draw (1.5,2) node {$\barm$};
  \draw (2,3.5) node {$\barM$};
  \draw ($(1.5,2)+(DIS2)$) node {$m$};
  \draw ($(2,3.5)+(DIS2)$) node {$M$};

  \filldraw ($(G)+(0.125,0)$) circle (2pt);

\end{tikzpicture}
\begin{tikzpicture}%
\hspace{-3.5cm}
\node[below right, draw, align = left,font=\fontsize{6}{6.5}\selectfont] at (0,3.5){\baselineskip=5pt
     {$\quad \ \ \Delta M$}\\
     {$\qquad +1$}\\
     {$\qquad +0$} \\
     {$\qquad -1$}\\
     {$\qquad \pm 2$}
};

\draw[thick, red] (0.2,3.08) --  (0.5,3.08);
\draw[thick] (0.2,2.86) --  (0.5,2.86);
\draw[thick, blue] (0.2,2.64) --  (0.5,2.64);
\draw[thick, green] (0.2,2.42) --  (0.5,2.42);
\useasboundingbox (0.2,2);
\end{tikzpicture}
\begin{tikzpicture}

\begin{axis}[tick label style={/pgf/number format/fixed},
    title={$\Delta \barM = \barM-\barm= 0,\quad \barm=-1/2$},
    xmin = -15, xmax = 15,
    ymin=0, ymax= 0.55,
    xtick distance = 5,
    ytick distance = 0.1,
    minor tick num = 1,
    width = 0.48\textwidth,
    height = 0.2966563146\textwidth,
    xlabel = {Laser detuning$/2\pi$(MHz)},
    ylabel = {$\abs{\bar{\Omega}^{mM}_{\barms\barMs}/\Omega_{mM}}$},
    ytick pos=left,
    xtick pos= bottom,clip=false]

\draw (-17.2,0.65) node {c)};
\coordinate (DIS) at (0,0);

\addplot [ycomb, thick] file[skip first] {1dressed_transitions_with_m=-0.5_M=-0.5_0.dat};

\addplot [ycomb, color=red, thick] file[skip first] {1dressed_transitions_with_m=-0.5_M=-0.5_1.dat};


\addplot [ycomb, color=blue, thick] file[skip first] {1dressed_transitions_with_m=-0.5_M=-0.5_-1.dat};


\addplot [ycomb, color = green, thick] file[skip first] {1dressed_transitions_with_m=-0.5_M=-0.5_2.dat};

\end{axis}

\end{tikzpicture}
\caption{\label{fig:Combinatorics} Dressed atomic levels and couplings for a singly-dressed system with $S^\ss=1/2\rightarrow S^\ds=5/2$. (a) illustrates the quadrupole selection rules among the bare basis states $\ket*{m}$ and $\ket*{M}$. For the specific case considered here, there are 10 possible transitions. In (b) we consider a particular transition in the dressed basis, $\barM=-1/2 \leftrightarrow \barm=-1/2$.
Since the dressed states are a time dependent superposition of the bare basis states, cf. Fig.~\ref{fig:LevelScheme}, this transition can be driven with any one of the 10 underlying transitions in the bare basis. This is illustrated in (c) which shows the effective Rabi frequencies $\overbar{\Omega}^{mM}_{\barms\barMs}$, scaled to the Rabi frequency $\Omega_{mM}$ for the bare states, and the effective transition frequency. Colors correspond to those of (a).}
\end{figure}
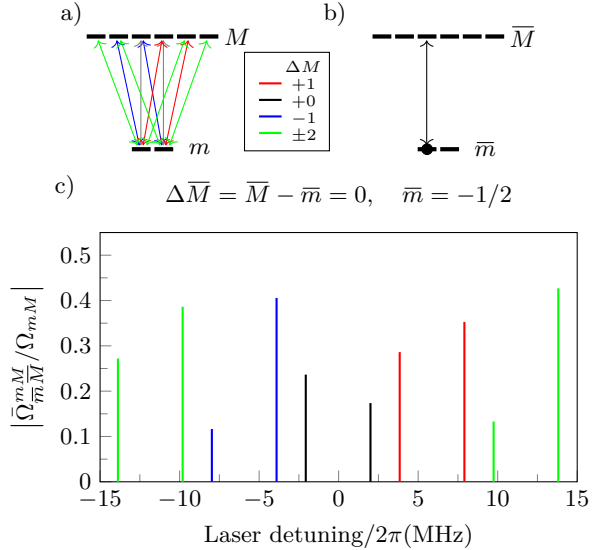
\makeatother

\label{QuadrupoletransitionsDDS}

\subsection{Quadrupole transitions in doubly-dressed basis}

We consider an ion with a manifold of ground states ($\ss$) and a manifold of excited states ($\ds$) that exhibit an electric-quadrupole allowed, optical transition at frequency $\omega_{\sd}$. The spin in the manifolds is $S^\kappa$ ($\kappa=\ss,\,\ds$) and the angular momentum operators are denoted by $\bold{S}^\kappa$, such that $\qty(\bold{S}^\kappa)^2=S^\kappa(S^\kappa+1)$. The Zeeman states in the two manifolds will be expressed with lower case letters for the ground states, $\ket{m}$, $|m|\leq S^\ss$, and upper case letters for the excited states, $\ket{M}$, $|M|\leq S^\ds$. A schematic for this transition between the two manifolds can be seen in Fig.~\ref{fig:Combinatorics} $a)$ for the case of \ca.

The dc magnetic field along the laboratory axis $z$ splits the Zeeman states by frequencies $\omega_0^\kappa=g_\kappa \mu_BB$, where $g_\kappa$ is the gyromagnetic factor of spin manifold $S^\kappa$. Both manifolds are subject to the respective dynamical decoupling rf-dressing fields with angles $\alpha_\kappa$, rf frequencies $\omega_i^\kappa$, and Rabi frequencies $g_\kappa\Omega_i^\kappa$, for $i=1,2$, as explained in Sec.~\ref{DoublyDressedBasis}. Therefore, the Hamiltonian in the laboratory frame is
\begin{align}
    H^{\mathrm{LF}}&=H^\ss_\mathrm{dc}+H^\ss_{\rf}+H^\ds_\mathrm{dc}+H^\ds_{\rf},
\end{align}
generalizing Eq.~\eqref{eq:H_QA0_LF} to the case of two spin manifolds. 
We note that this neglects an unavoidable cross-coupling through off-resonant driving of the $\ss$ manifold by the rf dressing fields of the $\ds$ manifold, and vice versa. This effect will be neglected in the following, and is treated in Appendix~\ref{appendixD}. In the doubly-dressed basis, this Hamiltonian becomes 
\begin{align}\label{eq:Hamsd}
    \bbar{H}=\omegasecond^\ss S_z^\ss+\omegasecond^\ds S_z^\ds,
\end{align}
generalizing Eq.~\eqref{eq:Ham2}. From now on, we will not include the time derivative in the Hamiltonian, since we will not perform any further time-dependent transformations.

The electric-quadrupole interaction ($E2$) of the ion with a laser of frequency $\omega_L$ and vector potential $\bold{A}(\bold{R},t)=\bold{A}^+(\bold{R})\e^{-\im\omega_L t}+\mathrm{c.c.}$ is $V_{E2}= \frac{\im e\omega_\sd}{2}\left(r_ir_j\partial_i{A}_j(\bold{R},t)-h.c.\right)$, see e.g.~\cite{James1998}. In a frame rotating at the optical transition frequency $\omega_\sd$, one obtains, in optical RWA,
\begin{align}
    V^\mathrm{LF}_{E2}
    &=\im\sum_{m,M}\qty(\Omega_{mM}\ketbra{M}{m}\e^{-\im\Delta_L t}-\mathrm{h.c.} )
\end{align}
where we used an expansion in the laboratory frame bare states $\ket{m}$ and $\ket{M}$ of the $\ss$ and $\ds$ manifolds, respectively. The Rabi frequencies are $\Omega_{mM}=\mel{M}{r_ir_j}{m}\partial_i{A}^+_j(\bold{R})/\hbar$. The matrix elements $\mel{M}{r_ir_j}{m}$ imply the quadrupole selection rules $\abs{\Delta m}=\abs{M-m}\leq 2$, see e.g. Fig.~\ref{fig:Combinatorics} $a)$. The laser detuning is $\Delta_L=\omega_L-\omega_\sd$.

We are now in a position to discuss how the dynamical decoupling affects the quadrupole interaction. To do so, we need to switch to the doubly-dressed basis and an interaction picture with respect to \eqref{eq:Hamsd}, generalizing the procedure explained in the previous section to two spin manifolds. Denoting by $\mathcal{D}^\kappa=\mathcal{D}^\kappa(\omega^\kappa_i,g_\kappa\Omega^\kappa_i,t)$ the dressing procedure of the spin manifold $\kappa$, where $\mathcal{D}$ is defined in Eq.~\eqref{eq:trafo}, the laser-ion interaction becomes

\begin{align}
    V^\mathrm{DB}_{E2}&=\mathcal{D}^\ss\otimes\mathcal{D}^\ds\qty[V^\mathrm{LF}_{E2}]\nonumber\\
    &=\im\sum_{\bbarms,\bbarMs}\left(\sum_{m,M}\Omega_{mM}    \mel*{\bbarM}{\mathcal{D}^\ss\otimes\mathcal{D}^\ds\qty\big[\ketbra{M}{m}]}{\bbarm}\right. \nonumber\\
    &\quad\times\ketbra*{\bbarM}{\bbarm}\e^{-\im\Delta_L t}-\mathrm{h.c.} \Biggr)\nonumber\\
    &=\im\sum_{\bbarms,\bbarMs}\sum_{m,M}\sum_{\barms,\barMs}\overbar{\Omega}^{mM,\barms\barMs}_{\bbarms\bbarMs}\ketbra*{\bbarM}{\bbarm}\nonumber\\
    &\quad\times\exp(\im t \Delta^{mM,\barms\barMs}_{\bbarms\bbarMs})
    -\mathrm{h.c.} .\label{eq:VE2}
\end{align}
Here, we expanded the quadrupole interaction in the basis of doubly-dressed states $\ket*{\bbarm}$ and $\ket*{\bbarM}$  of the $\ss$ and $\ds$ manifolds, respectively, and introduced the effective Rabi frequency
\begin{align}\label{eq:effRabiFreq}
    &\overbar{\Omega}^{mM,\barms\barMs}_{\bbarms\bbarMs}=\Omega_{mM}\nonumber\\
    &\quad\times
    \e^{-\im\alpha_\ds(\bbarMs-M)-\im\frac{\pi}{2}(\bbarMs-\barMs)}
    d_{M\barMs}(\rotfirst^\ds) d_{\barMs\bbarMs}(\rotsecond^\ds)\nonumber\\
    &\quad\times\e^{\im\alpha_\ss(\bbarms-m)+\im\frac{\pi}{2}(\bbarms-\barms)}
    d_{\barms m}(\rotfirst^\ss) d_{\bbarms\barms}(\rotsecond^\ss),
\end{align}
with $d_{m\barms}(\theta)$ the elements of the Wigner $d$-matrix, whose explicit expression is given in Appendix~\ref{appendixB} along with more details on the last equality. We note that the angles $\alpha_{\kappa}$ determining the direction of the second dressing fields, cf. Eq.~\eqref{eq:rffields}, contribute to the Rabi frequencies only in the form of phases. We also introduced the effective detuning
\begin{align}\label{eq:effDetuning}
    \Delta^{mM,\barms\barM}_{\bbarms\bbarMs}=&-\Delta_L+\bbarM\omegasecond^\ds+\barM\omega^\ds_2+ M\omega^\ds_1\nonumber\\
    &-\bbarm\omegasecond^\ss-\barm\omega^\ss_2- m\omega^\ss_1.
\end{align}
In Eq.~\eqref{eq:VE2} no RWA is applied with respect to these detunings.

Thus, to drive a $\bbarm\leftrightarrow\bbarM$ transition in the doubly-dressed basis, the laser detuning must be chosen such that $\Delta^{mM,\barms\barMs}_{\bbarms\bbarMs}=0$, that is
\begin{align}\label{eq:effectivedetuning}
    \Delta_L=\bbarM\omegasecond^\ds+\barM\omega^\ds_2+ M\omega^\ds_1-\bbarm\omegasecond^\ss-\barm\omega^\ss_2- m\omega^\ss_1,
\end{align}
is satisfied for one set of indices $(m,M,\barm,\barM)$. These resonance frequencies can be intuitively understood within the dressed state energy level picture including the photon energy of the rf dressing fields~\cite{dalibard_dressed-atom_1985}. The magnitude of the effective Rabi frequency is $\abs{\overbar{\Omega}^{mM,\barms\barMs}_{\bbarms\bbarMs}}\leq\abs{\Omega_{mM}}$ since the Wigner $d$-matrix is unitary, and therefore, all its elements are smaller than one in magnitude. To make efficient use of the laser power, it will be advantageous to choose $(m,M,\barm,\barM)$ such that the contribution of the Wigner $d$-matrix elements is as large as possible. In doing so, $m$ and $M$ have to respect the quadrupole selection rules, but not the pairs $(\barm,\barM)$ and $(\bbarm,\bbarM)$, since the dressed states are composed of all of the bare states.  It is worthwhile noting that the polarisation and $k$-vector dependence of the coupling strength is contained in $\Omega_{mM}$, akin to the Wigner-Eckart theorem. Thus, $\Omega_{mM}$ can be maximized independent of the selected dressed-state transition.

\subsection{Illustration for  \ca}\label{TheoreticalResults}

In this section, we will apply the above expressions to the case of the $S_{1/2}$ to $D_{5/2}$ transition in a \ca\-ion and compare them to measurements on the decoupled system. Therefore, we will have the total spin of the manifolds $S^{\ss}=\frac{1}{2}$ and  $S^{\ds} =\frac{5}{2}$. The goal is to derive the frequency spectrum and the relative coupling strengths with the parameters given in Table~\ref{table:ExperimentalistValues}, for each possible transition with a set of indices $\left(m,M,\barm,\barM,\bbarm,\bbarM\right)$.

\begin{table}\centering
\begin{tabular}{ |c|c|r|}
\hline
\multicolumn{1}{|c}{Dressing} & \multicolumn{1}{|c}{Parameter} & \multicolumn{1}{|c|}{Value} \\
\hline
   & $g_s\mu_bB_z$ & $2\pi\times\SI{10}{\mega \hertz}$ \\
   & $\Omega_{1}^s$   & $2\pi\times\SI{46805}{\hertz}$    \\
   $1$\textsuperscript{st} layer & $\Omega_{1}^d$   & $2\pi\times\SI{115600}{\hertz}$  \\
   & $\omega_{1}^{s}$ & $2\pi\times\SI{10002090}{\hertz}$ \\
   & $\omega_{1}^{d}$ & $2\pi\times\SI{5994834}{\hertz}$ \\
  \hline
   & $\Omega_{2}^s$   & $2\pi\times\SI{3469}{\hertz}$   \\ 
   \multirow{2}{*}{$2$\textsuperscript{nd} layer} & $\Omega_{2}^d$   & $2\pi\times\SI{6809}{\hertz}$   \\
   & $\omega_{2}^{s}$ & $2\pi\times\SI{72050}{\hertz}$ \\
   & $\omega_{2}^{d}$ & $2\pi\times\SI{160589}{\hertz}$ \\
\hline
\end{tabular}
\caption{Case study of double dressing of a \ca ion for the $S_{1/2}$ and $D_{5/2}$  manifolds. The upper part of the table refers to the variables in the first layer of dressing and the lower part of the second layer of dressing. The gyromagnetic factors are $g_s=2.00225664$~\cite{tommaseo_factor_2003} and $g_d=1.2003340$~\cite{chwalla_absolute_2009}.} \label{table:ExperimentalistValues}
\end{table}

Before showing the results for two layers of dressing, we first want to gain some insight by explaining just one particular transition $(\barm,\barM)$ in the case of a single layer of dressing, with the parameters given in the first part of Table~\ref{table:ExperimentalistValues}. We need to translate the equations for the effective Rabi frequency~\eqref{eq:effRabiFreq} and the effective detuning~\eqref{eq:effDetuning} for the case of a single dressing. This can be achieved by fixing $\omega_2^{\ds (\ss)}=0$ and $\Omega_2^{\ds (\ss)}=0$, which implies 
\begin{align}\label{eq:FirstOpticalCoupling}
    \overbar{\Omega}^{mM}_{\barms\barMs}=\Omega_{mM}\e^{\im(\alpha_\ds M-\alpha_\ss m)+\im\frac{\pi}{2}(\barMs-\barms)}d_{M\barMs}(\rotfirst^\ds)d_{\barms m}(\rotfirst^\ss) ,
\end{align}
and
\begin{align}\label{eq:FirstSpectrum}
    \Delta^{mM}_{\barms\barMs}=&-\Delta_L+\barM\omegafirst+ M\omega^\ds_1-\barm\omegafirst- m\omega^\ss_1,
\end{align}
where we go to an interaction picture with respect to the Hamiltonian in the first dressed basis~\eqref{eq:FirstLayer}.

The results are illustrated in Fig.~\ref{fig:Combinatorics}, where Fig.~\ref{fig:Combinatorics} c) shows the different effective Rabi frequencies for the 10 ways in which a transition in the first dressed basis depicted in Fig.~\ref{fig:Combinatorics} b) with indices $(\barm,\barM)=(-1/2,-1/2)$ can be achieved through transitions in the bare basis for the appropriate laser detunings. The colors refer to the different possible selection rules shown in Fig.~\ref{fig:Combinatorics} a).

Each singly-dressed ground state is composed of two bare states from each of which five transitions lead to the bare excited states that each of the six singly-dressed excited states are composed of. Therefore, $5\times2$ transitions are possible from a fixed singly-dressed ground to a singly-dressed excited state (see Fig.~2c)) or $10\times2\times6$ overall transitions between all singly-dressed ground (two) and excited (six) states. In turn, each doubly-dressed ground state is composed of two singly-dressed ground states, each connected via $10\times6$ transitions to a single doubly-dressed excited state composed of six singly-dressed excited states), resulting in $10\times6\times2$ transitions between two selected doubly-dressed states, $10\times12\times6$ between a single doubly-dressed ground state $\ket*{\bbarm}$ and all excited states or an overall of $10\times12\times12$ transitions between all doubly-dressed states. For the transitions with an initial state $\ket*{\bbarm}=\ket*{-1/2}$, Fig.~\ref{fig:FullComb} a) depicts the effective Rabi frequencies relative to the Rabi frequencies of the transitions in the bare basis, i.e., $\abs{\bar{\Omega}^{mM,\barms\barMs}_{\bbarms\bbarMs}/\Omega_{mM}}$. This ratio is plotted against the laser detuning, that shows for which values the transitions are resonant. The shaded area corresponds to the region defined by the pair $(m,M)=(-0.5,-1.5)$, shown in more detail in Fig.~\ref{fig:FullComb} b). Similarly, Fig.~\ref{fig:FullComb} c) shows the tuple $(m,M,\barm,\barM)=(-0.5,-1.5,-0.5,-2.5)$, where we can see the transition with higher effective Rabi frequency. Here, we can also observe that there are no selection rules for $\Delta\bbarM$. Noticeably, the relative Rabi frequencies have different weights. Efficient use of laser power can be achieved by choosing a transition with high effective Rabi frequency and, ideally, a small effective Rabi frequency of the nearest neighboring transitions. As we can see, such an optimization becomes simply a matter of engineering after the characterization of the transitions.

\begin{widetext}

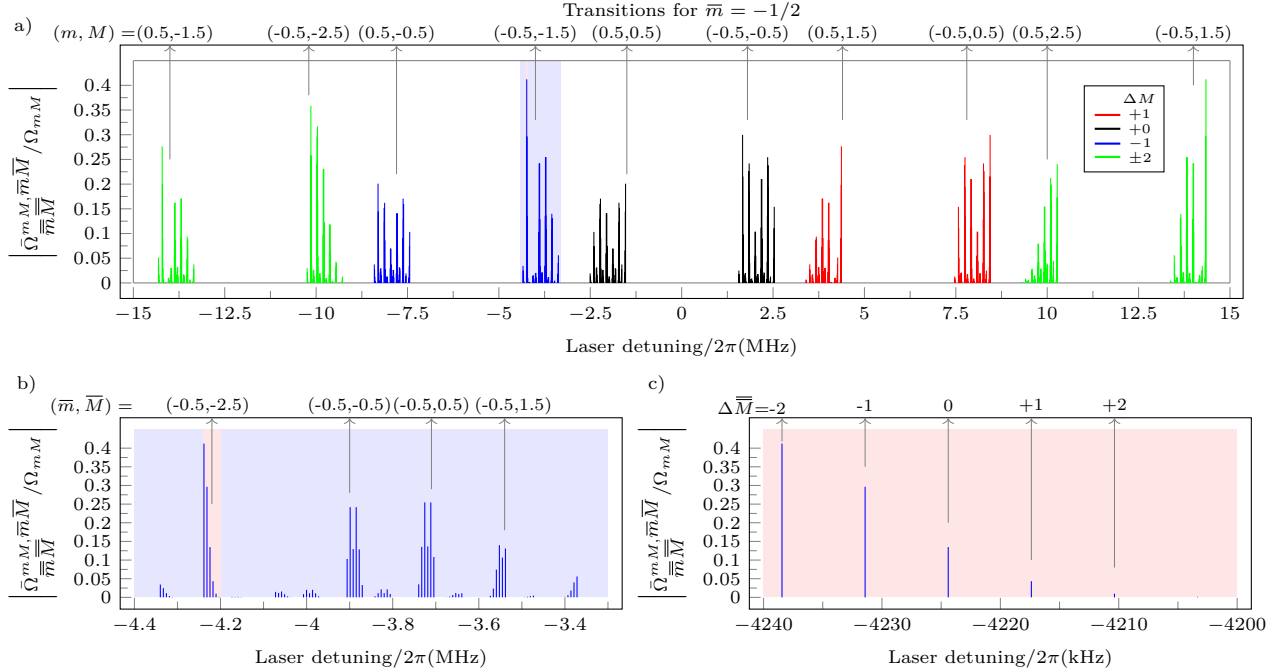
\begin{figure}[H]

\begin{tikzpicture}

\begin{axis}[tick label style={/pgf/number format/fixed},
    font=\scriptsize,
    title={Transitions for $\barm=-1/2$},
    xmin = -15, xmax = 15,
    ymin=0, ymax= 0.45,
    xtick distance = 2.5,
    ytick={0, 0.05, 0.1, 0.15, 0.2, 0.25, 0.3, 0.35, 0.4},
    ytick pos=left,
    xtick pos= bottom,
    enlarge x limits=0.012,
    enlarge y limits=0.07,
    minor tick num = 1,
    width = \textwidth,
    height = 0.3\textwidth,
    xlabel = {Laser detuning$/2\pi$(MHz)},
    ylabel = {$\abs{\bar{\Omega}^{mM,\barms\barMs}_{\bbarms\bbarMs}/\Omega_{mM}}$},clip=false]

 \path [fill=blue!10] (axis cs: -4.4,0) rectangle (-3.3,0.45);
 \path[fill=red!10] (axis cs:-4.24,0) rectangle (axis cs:-4.2,0.45);
 \draw[thin, color=gray] (axis cs:-15,0.45) -- (axis cs:15,0.45); 
 \draw[thin, color=gray] (axis cs:-15,0) -- (axis cs:-15,0.45);
 \draw[thin, color=gray] (axis cs:-15,0.) -- (axis cs:15,0.); 
 \draw[thin, color=gray] (axis cs:15,0) -- (axis cs:15,0.45);
 
\addplot [ycomb] file[skip first] {All_possible_transitions_with_mb=-0.5_Mb=-0.5.dat};
\node[font=\fontsize{6.5}{7}\selectfont] at (1.8,0.505) { (-0.5,-0.5)};
\draw[->, gray,very thin] (1.8,0.33) -- (1.8,0.48);

\addplot [ycomb] file[skip first] {All_possible_transitions_with_mb=0.5_Mb=0.5.dat};
\node[font=\fontsize{6.5}{7}\selectfont] at (-1.5,0.505) { (0.5,0.5)};
\draw[->, gray, very thin] (-1.5,0.22) -- (-1.5,0.48);

\addplot [ycomb, color=red] file[skip first] {All_possible_transitions_with_mb=-0.5_Mb=0.5.dat};
\node[font=\fontsize{6.5}{7}\selectfont] at (7.8,0.505) { (-0.5,0.5)};
\draw[->, gray, very thin] (7.8,0.33) -- (7.8,0.48);

\addplot [ycomb, color=red] file[skip first] {All_possible_transitions_with_mb=0.5_Mb=1.5.dat};
\node[font=\fontsize{6.5}{7}\selectfont] at (4.4,0.505) { (0.5,1.5)};
\draw[->, gray, very thin] (4.4,0.33) -- (4.4,0.48);

\addplot [ycomb, color=blue] file[skip first] {All_possible_transitions_with_mb=-0.5_Mb=-1.5.dat};
\node[font=\fontsize{6.5}{7}\selectfont] at (-4,0.505) { (-0.5,-1.5)};
\draw[->, gray, very thin] (-4,0.33) -- (-4,0.48);

\addplot [ycomb, color=blue] file[skip first] {All_possible_transitions_with_mb=0.5_Mb=-0.5.dat};
\node[font=\fontsize{6.5}{7}\selectfont] at (-7.8,0.505) { (0.5,-0.5)};
\draw[->, gray, very thin] (-7.8,0.22) -- (-7.8,0.48);


\addplot [ycomb, color = green] file[skip first] {All_possible_transitions_with_mb=-0.5_Mb=1.5.dat};
\node[font=\fontsize{6.5}{7}\selectfont] at (14,0.505) { (-0.5,1.5)};
\draw[->, gray, very thin] (14,0.4) -- (14,0.48);

\addplot [ycomb, color = green] file[skip first] {All_possible_transitions_with_mb=0.5_Mb=2.5.dat};
\node[font=\fontsize{6.5}{7}\selectfont] at (10,0.505) { (0.5,2.5)};
\draw[->, gray, very thin] (10,0.25) -- (10,0.48);

\addplot [ycomb, color = green] file[skip first] {All_possible_transitions_with_mb=-0.5_Mb=-2.5.dat};
\node[font=\fontsize{6.5}{7}\selectfont] at (-10.2,0.505) { (-0.5,-2.5)};
\draw[->, gray, very thin] (-10.2,0.38) -- (-10.2,0.48);

\addplot [ycomb, color = green] file[skip first] {All_possible_transitions_with_mb=0.5_Mb=-1.5.dat};
\node[font=\fontsize{6.5}{7}\selectfont] at (-15,0.505) { $(m,M)=$(0.5,-1.5)};
\draw[->, gray, very thin] (-14,0.25) -- (-14,0.48);

\draw (-18,0.52) node {a)};
\node[below right, draw, align = left,font=\fontsize{5}{5.5}\selectfont] at (axis cs:11,0.4){\baselineskip=5pt
     {$\quad \ \ \Delta M$}\\
     {$\qquad +1$}\\
     {$\qquad +0$} \\
     {$\qquad -1$}\\
     {$\qquad \pm 2$}
};

\draw[thick, red] (axis cs:11.3,0.34) --  (axis cs:12,0.34);
\draw[thick] (axis cs:11.3,0.312) --  (axis cs:12,0.312);
\draw[thick, blue] (axis cs:11.3,0.282) --  (axis cs:12,0.282);
\draw[thick, green] (axis cs:11.3,0.255) --  (axis cs:12,0.255);

\end{axis}

\end{tikzpicture}

\begin{tikzpicture}

\begin{axis}[tick label style={/pgf/number format/fixed},
    font=\scriptsize,
    xmin = -4.4, xmax = -3.3,
    ymin = 0, ymax = 0.45,
    xtick distance = 0.2,
    ytick={0, 0.05, 0.1, 0.15, 0.2, 0.25, 0.3, 0.35, 0.4},
    minor tick num = 1,
    width = 0.5\textwidth,
    height = 0.25\textwidth,
    xlabel = {Laser detuning$/2\pi$(MHz)},
    ylabel = {$\abs{\bar{\Omega}^{mM,\barms\barMs}_{\bbarms\bbarMs}/\Omega_{mM}}$},clip=false,
    enlarge x limits=0.03,
    enlarge y limits=0.07,
    ytick pos=left,
    xtick pos= bottom]
    \path[fill=blue!10] (axis cs:-4.4,0) rectangle (axis cs:-3.3,0.45);
    \path[fill=red!10] (axis cs:-4.24,0) rectangle (axis cs:-4.2,0.45);

 

\addplot [ycomb, color=blue] file[skip first] {All_possible_transitions_with_mb=-0.5_Mb=-1.5.dat};

\node[font=\fontsize{6.5}{7}\selectfont] at (-4.5,0.515) {$(\barm,\barM)=$};
\node[font=\fontsize{6.5}{7}\selectfont] at (-4.23,0.515) {(-0.5,-2.5)};
\draw[->, gray, very thin] (-4.22,0.25) -- (-4.22,0.48);

\node[font=\fontsize{6.5}{7}\selectfont] at (-3.9,0.515) {(-0.5,-0.5)};
\draw[->, gray, very thin] (-3.9,0.28) -- (-3.9,0.48);

\node[font=\fontsize{6.5}{7}\selectfont] at (-3.71,0.515) {(-0.5,0.5)};
\draw[->, gray, very thin] (-3.71,0.29) -- (-3.71,0.48);

\node[font=\fontsize{6.5}{7}\selectfont] at (-3.52,0.515) {(-0.5,1.5)};
\draw[->, gray, very thin] (-3.54,0.18) -- (-3.54,0.48);

\draw (axis cs:-4.65,0.57) node {b)};

\end{axis}

\end{tikzpicture}%
\begin{tikzpicture}
\hspace{0.1cm}

\begin{axis}[tick label style={/pgf/number format/.cd, fixed, 1000 sep={}},
    font=\scriptsize,
    xmin = -4240, xmax = -4200,
    ymin = 0, ymax = 0.45,
    xtick distance = 10,
    ytick={0, 0.05, 0.1, 0.15, 0.2, 0.25, 0.3, 0.35, 0.4},
    minor tick num = 1,
    width = 0.5\textwidth,
    height = 0.25\textwidth,
    xlabel = {Laser detuning$/2\pi$(kHz)},clip=false,
    ylabel = {$\abs{\bar{\Omega}^{mM,\barms\barMs}_{\bbarms\bbarMs}/\Omega_{mM}}$},
    enlarge x limits=0.03,
    enlarge y limits=0.07,
    ytick pos=left,
    xtick pos= bottom]

\path[fill=red!10] (axis cs:-4240,0) rectangle (axis cs:-4200,0.45);

\addplot [ycomb, color=blue] file[skip first] {Case_mb=-0.5_Mb=-1.5_ma=-0.5_Ma=-2.5.dat};

\draw (-4249,0.57) node {c)};

\node[font=\fontsize{6.5}{7}\selectfont] at (-4241,0.515) {$\Delta \bbarM$=-2};
\draw[->, gray, very thin] (-4238.443353610104,0.418) -- (-4238.443353610104,0.48);

\node[font=\fontsize{6.5}{7}\selectfont] at (-4231.422635069315,0.515) {-1};
\draw[->, gray, very thin] (-4231.422635069315,0.35) -- (-4231.422635069315,0.48);

\node[font=\fontsize{6.5}{7}\selectfont] at (-4224.401916528526,0.515) {0};
\draw[->, gray, very thin] (-4224.401916528526,0.2) -- (-4224.401916528526,0.48);

\node[font=\fontsize{6.5}{7}\selectfont] at (-4217.381197987738,0.515) {+1};
\draw[->, gray, very thin] (-4217.381197987738,0.1) -- (-4217.381197987738,0.48);

\node[font=\fontsize{6.5}{7}\selectfont] at (-4210.360479446949,0.515) {+2};
\draw[->, gray, very thin] (-4210.360479446949,0.08) -- (-4210.360479446949,0.48);

\end{axis}
\end{tikzpicture}
\caption{Normalised Rabi frequencies between dressed states. (a) shows the Rabi frequencies $\abs*{\overbar{\Omega}^{mM,\barms\barMs}_{\bbarms\bbarMs}/\Omega_{mM}}$ in Eq.~\eqref{eq:effRabiFreq} and effective transition frequencies in Eq.~\eqref{eq:effectivedetuning} for all possible transitions from the doubly-dressed ground state $\ket*{\bbarm}=-1/2$ to any one of the doubly-dressed excited state $\ket*{\bbarM}$. Each color represents a different selection rule for $\Delta M=M-m$ for a pair of bare states $\left(m, M\right)$, as shown in the inset of (a). Panels (b) and (c) are zoom-ins on the shaded regions in (a) and (b), respectively.}\label{fig:FullComb}
\end{figure}

\end{widetext}

\section{Experiment with \ca}
\label{Experiment}
\ca\ is a widely used species, e.g. in the fields of quantum information  \citep{monz_14qubit_2011,kaushal_shuttlingbased_2020, ringbauer_universal_2022, pogorelov_compact_2021, hilder_fault-tolerant_2022}, quantum simulation \citep{joshi_observing_2022, kokail_self-verifying_2019, hempel_quantum_2018} and optical ion clocks \citep{matsubara_direct_2012, huang_ca_2019, huang_liquid_2021, cao_compact_2017,chwalla_absolute_2009}. The narrow $S_{1/2}$ to $D_{5/2}$ transition in combination with a favourable level sheme for advanced laser cooling techniques \citep{li_robust_2022, morigi_Ground_2000, scharnhorst_Experimental_2018, lechner_electromagnetically-induced-transparency_2016} and efficient state readout makes it an ideal testbed for the implementation of the introduced CDD scheme. In addition, the negative static differential polarizability of the transition allows for canncellation of trap drive induced second-oder Doppler shift with the 2\ts{nd}-order Stark shift \citep{huang_ca_2019}. Especially ion clocks based on large three-dimensional ion crystals will benefit from this feature due to their unavoidable excess micromotion accross the crystal.

First, we give an overview of the used experimental setup and highlight relevant key figures for the CDD spectroscopy. Next, the hardware for generating of CDD rf-field fields is shown. Finally, the experiments for verification of the predictions are presented together with their results.

\subsection{Setup}\label{Setup}
A single \ca\ ion is trapped in a segmented Paul trap \citep{herschbach_Linear_2012, hannig_transportable_2019} with secular frequencies of ($\omega_{z}, \omega_{x}, \omega_{y}) = 2\pi \times (1.2, 1.6, 1.8)$ MHz obtained with $\Omega_{RF} =2\pi\times \SI{33}{\mega \hertz}$ trap drive frequency. All lasers needed for cooling, detection and state preparation are locked to a wave-meter~\footnote{High Finesse U10} with typical stability of $\delta \nu< \SI{1}{\mega \hertz}$ \citep{hannig_transportable_2019}. The amplified extended cavity diode laser \footnote{TA pro, Toptica} at \SI{729}{\nano \metre} addressing the \clocktrans\ transition is pre-stabilised via the Pound-Drever-Hall technique \citep{drever_Laser_1983} to an optical reference cavity. Additionally, the light is transfer-locked \citep{scharnhorst_Highbandwidth_2015} to a highly stable laser, which is locked to a cryogenic silicium cavity~\citep{matei_Lasers_2017}. Even without correction of inter-branch comb-noise~\citep{benkler_Endtoend_2019}, as well as a few metres of unstabilized fibre path length, a differential frequency stability of $\frac{\Delta \nu_L}{\nu_L} < 10^{-16}$ against the reference at a few seconds is reached.
The individual beams are switched and frequency steered by acousto-optic modulators controlled by a pulse sequencer \cite{schindler_Frequency_2008, pham_generalpurpose_2005}.  For minimizing photon scattering and light shifts during probing of the clock transition, mechanical shutters in all relevant beam paths are used.
Three pairs of orthogonal magnetic field coils generate a static magnetic field of \SI{357}{\micro \tesla} aligned with the axial trap direction resulting in a \SI{10.0000(4)}{\mega \hertz} splitting of the two \slevel\ Zeeman components. The B-field is determined by probing two Zeeman levels with resolution of $\delta \nu_L < \SI{100}{\hertz}$. The resolution limit is caused by mains line-synchronous magnetic field fluctuations.

\subsection{RF Coil Setup}
\begin{figure}[t]
\centering
\includegraphics[width=0.49\textwidth]{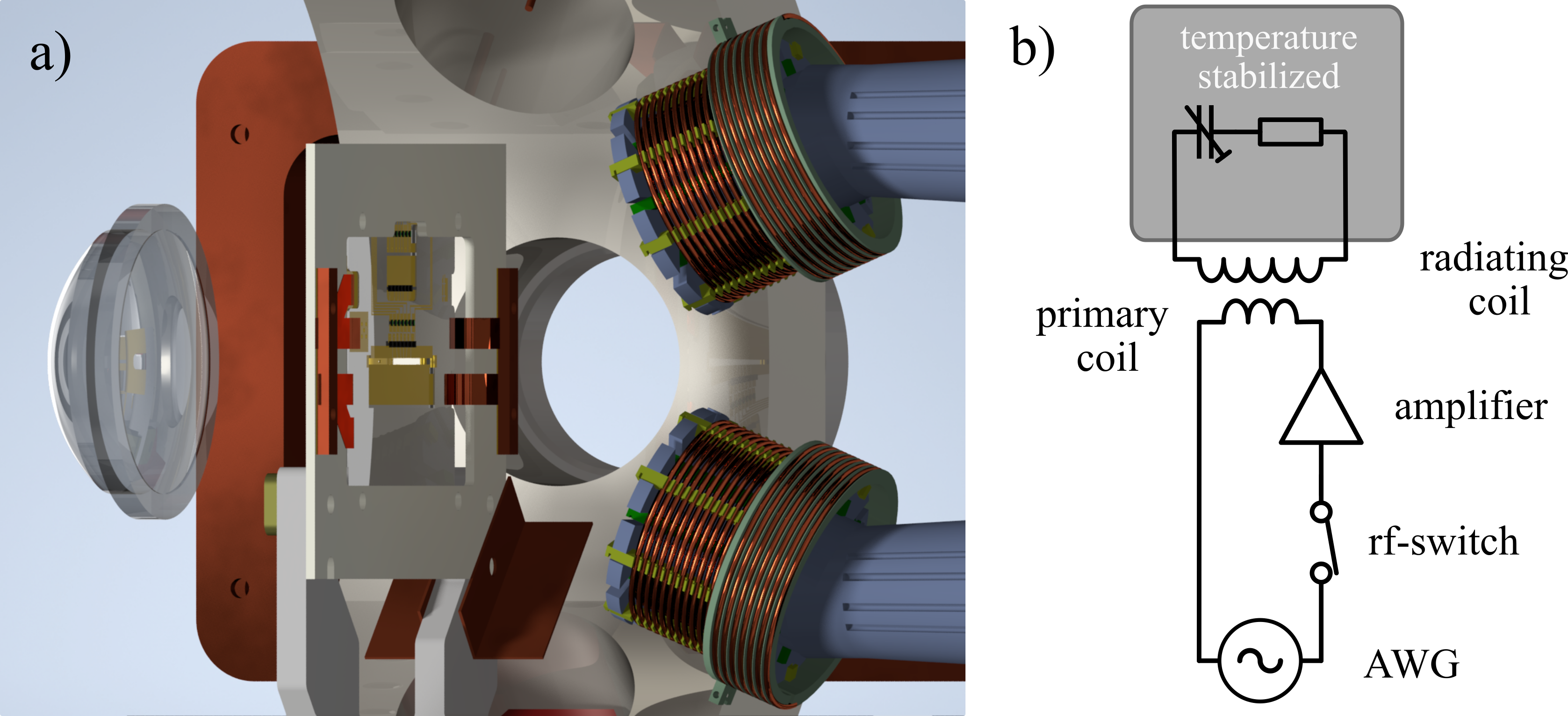}
\caption{(a) CAD image of the \textbf{CDD} coil setup. RF magnetic field coils (right) for dressing the \slevel\ and \dlevel\ are mounted at a distance of $d_c< \SI{50}{\milli \metre}$ to the Paul trap (centre). The aspheric lens (left) for imaging of the ion crystals has a distance of $d_a = \SI{36.6}{\milli \metre}$ to the trap centre. (b) Electronic schematic of the \textbf{CDD} drive.}\label{fig:coils}\label{fig:coil_CAD}
\end{figure}

Resonant tank-circuits with a radiating coil produce the rf magnet-field needed for the CDD scheme. They consist of two separate LCR-circuits with tunable capacitors to match the resonance frequency of the Zeeman manifolds (see Fig.~\ref{fig:coils}(b)). The current for each coil is supplied via an inductively-coupled, impedance-matched primary coil which is driven by an amplifier. A two-channel arbitrary voltage generator \footnote{Keysight 33622A} acts as the signal source. A pulse sequencer-controlled rf-switch ensures synchronization of the rf pulses with the remaining sequence. 

The quality factor $Q_{S(D)} = 14(30)$  of the coils is chosen as a compromise between large B-field amplitude and corresponding Rabi frequency for high Zeeman shift suppression (compare Eq.~\eqref{eq:Bfieldsuppression}) and minimal signal distortion by the coil's transfer function. The resonance frequency $\omega_0(T) =\frac{1}{\sqrt{L(T)C(T)}}$ is temperature dependent. Therefore, the coil temperature increases by up to $\SI{10}{\kelvin}$ during operation depending on the applied rf power and the duty cycle of the rf-pulses within the experimental sequence. The circuit design includes a temperature-controlled base plate for the electronic components to avoid theses temperature-induced amplitude drifts. For passive temperature stability, the inductive part of the circuit is a copper coil held by an open, mesh-like 3D printed polylactide-part. This minimizes heat build-up during longer sequences. The holders are placed on translation stages and positioned in close proximity to the ion(s) inside an inverted viewport (see Fig.~\ref{fig:coils}(a)).

\subsection{Experimental sequence}\label{experimental_sequence}
First, the \ca-ion is Doppler-cooled close to the cooling limit of $T<\SI{1}{\milli \kelvin}$. The secular modes are then cooled to a mean motional phonon number of $\overbar{n} \lesssim 0.2$ by electromagnetically-induced-transparency cooling \citep{morigi_Ground_2000, roos_experimental_2000, scharnhorst_Experimental_2018} to reduce the second-order Doppler shift. After state preparation into the \groundstate\ level by optical pumping with an axial $\sigma^-$ polarised \SI{397}{\nm} beam, the CDD sequence starts. 

A frequency and amplitude ramp is applied, realizing a rapid adiabatic passage \citep{wunderlich_Robust_2007}, to avoid populating nearby dressed states by abrupt switching of the S-drive-coils. By choosing the sweep direction, the population is transferred to the  \DDsminus\ or \DDsplus\ dressed states with success probability of $P>98\,\%$. 
After this initial switch-on sequence, the S \& D rf-drives are applied continuously together with a spectroscopy \SI{729}{\nano \metre} pulse. 

\begin{figure*}[t]
\centering
\subfigure[]{\includegraphics[width=0.49\textwidth]{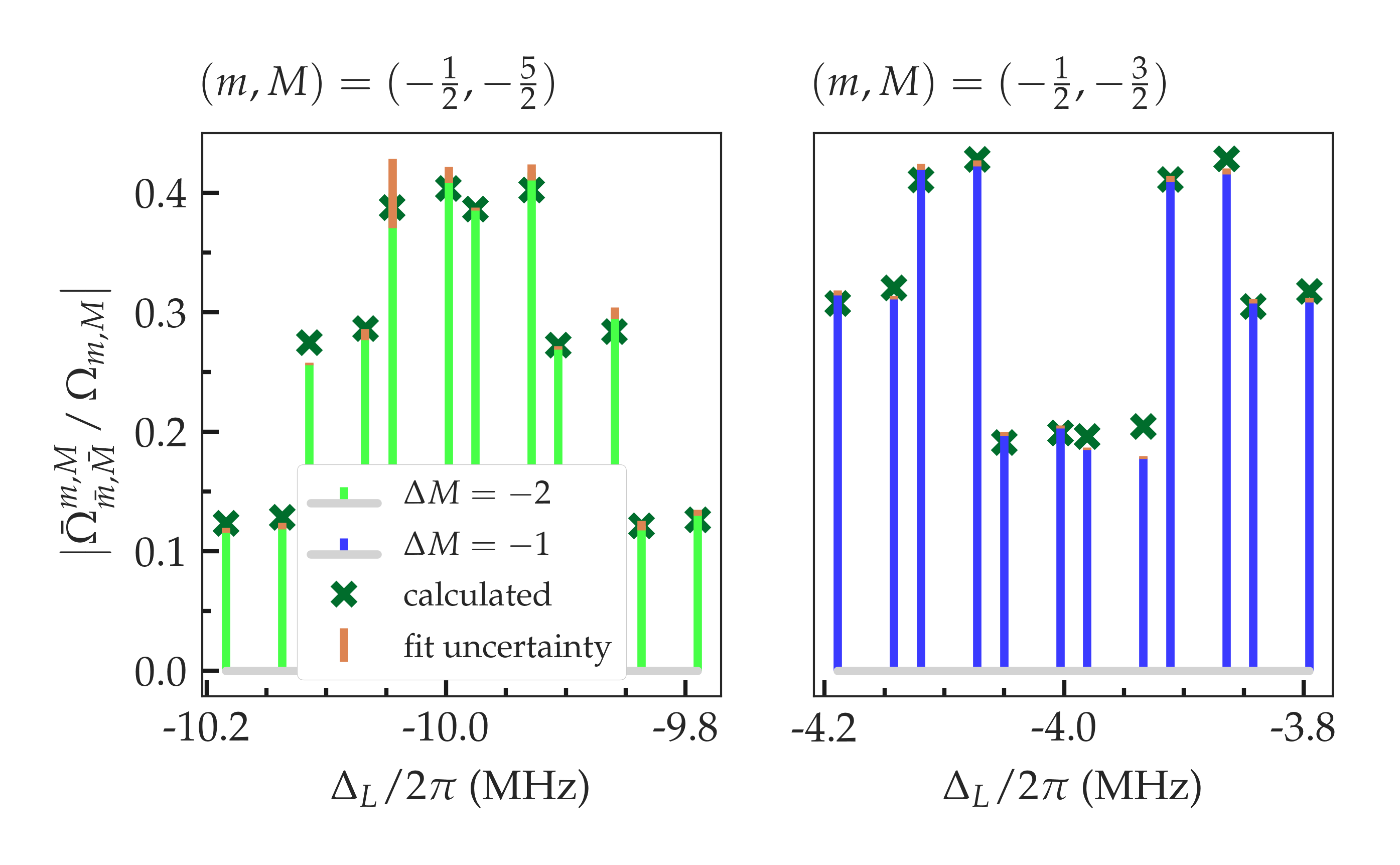}}
\subfigure[]{\includegraphics[width=0.49\textwidth]{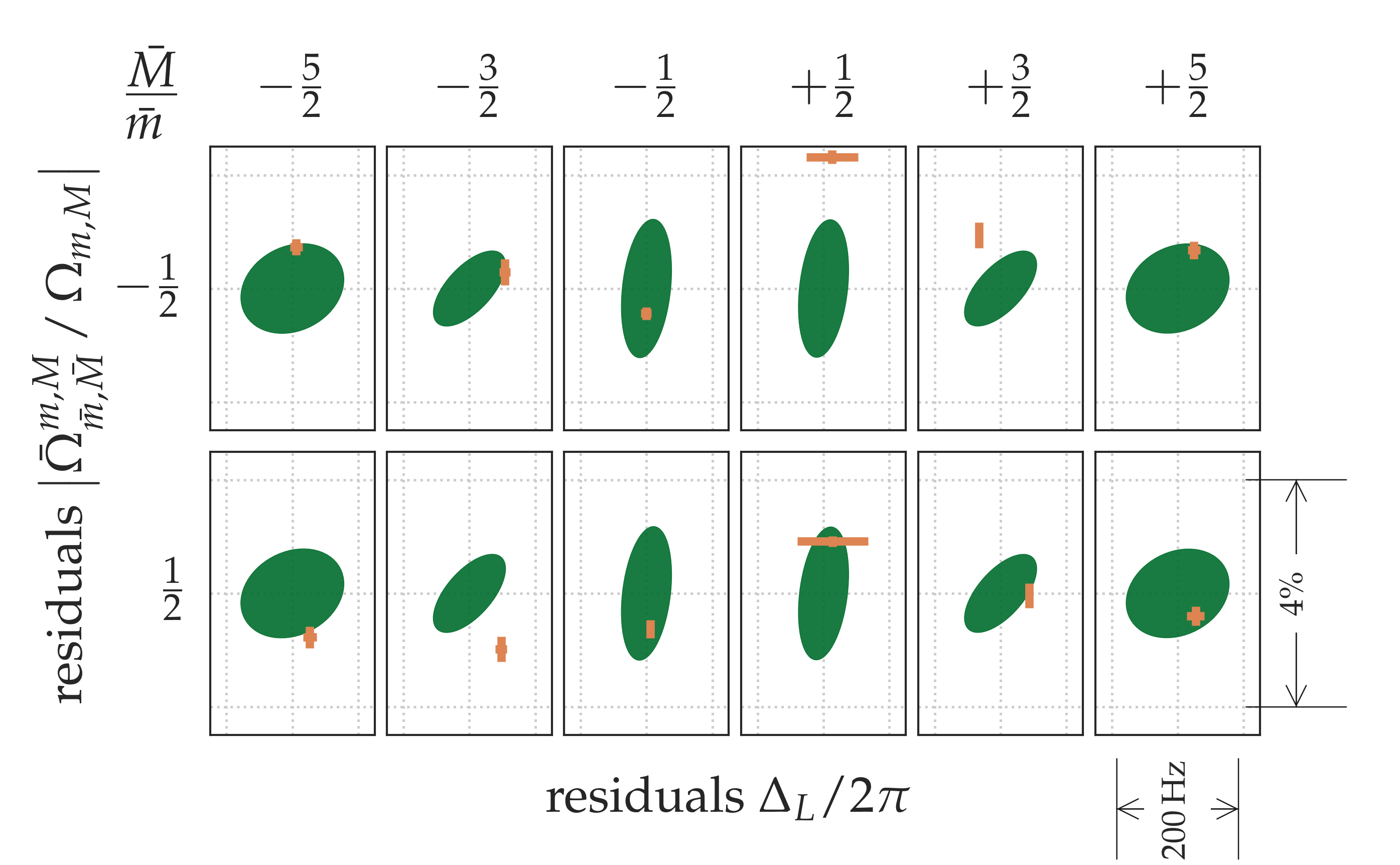}}
\caption{Comparison between experimental and theoretical coupling strengths and resonance frequencies for singly-dressed \ca. (a) Relative optical coupling strength of two 1\ts{st}-stage ensembles with $\theta_{S(D)}=\pi/2$. Pulse length spectroscopy was used to determine the optical coupling strength of each transition. The relative coupling strength of the \SI{729}{nm} beam with respect to the associated Zeeman transition is plotted against the frequency offset from the zero B-field transition frequency. (b) Residuals for the $m=-\frac{1}{2}, \, M=-\frac{3}{2}$ ensemble. The measured transitions values (orange) and the calculated (dark green) are compared. For the calculated uncertainty region, a fractional driving strength uncertainty of $\frac{\Delta \Omega_i}{\Omega_i}=4\times 10^{-4}$ and B-field uncertainty of $\Delta B_0= \SI{60}{\nano \tesla}$ is assumed. For the measured data, only the fitting uncertainty was taken into account.}
\label{fig:rel_opt_coupling}
\end{figure*}

The dressed states resonances are addressed by their frequency detuning from the field-free $S_{1/2}\rightarrow D_{5/2}$ transition by the $\SI{729}{nm}$ laser. 
If the optical coupling is much weaker than the rf-coupling ($\Omega_{m,M} \ll \overbar{\omega}^{s,d}$), the dressed system's Eigenstates are quasi-static with respect to the laser interaction. We have performed scans across the dressed state resonances to determine their frequency and on-resonance Rabi flopping to determine their coupling strength (compare appendix \ref{appendixD}).

For the prediction of the transition energies and coupling strengths of the dressed system adequate knowledge of the experimental parameters is crucial. The frequencies $\omega_i$ of the driving fields can be chosen with high precision, but the coupling strengths $\Omega_i$ must be determined experimentally via the splitting of the dressed states $\tilde{\omega}_0^i$. Therefore, resonance frequencies of four CDD transitions with opposing $\barm$ and $\barM$ are measured. With knowledge of these parameters the resonance frequencies and relative optical couplings of all 12 1\ts{st}-stage transitions per Zeeman-level can be determined (see Eq.~\eqref{eq:FirstSpectrum} and \eqref{eq:FirstOpticalCoupling}). In figure \ref{fig:rel_opt_coupling} the comparison of the measured and calculated optical coupling strengths for transitions from the $m=-\frac{1}{2}$ mainfold to the $M=-\frac{5}{2}$ and $M=-\frac{3}{2}$ manifolds are compared. The Rabi frequencies of the CDD states are normalized to the underlying bare Zeeman transition. The theoretical predictions are in good agreement with the measured transition frequencies and relative optical coupling strengths. Deviations arise from calibration imperfections and thermally induced drive strength fluctuations in combination with a drifting offset magnetic field. Equation.~\eqref{eq:FirstOpticalCoupling} predicts scaling of each CDD manifold with the underlying bare Zeeman transition.  This was qualitatively confirmed by using different beam propagation directions. Especially, strict vanishing of dressed states together with an underlying bare Zeeman transitions with vanishing optical coupling (e.g. $|\Delta M|\neq1$ for axial interrogation) was also confirmed. 

\section{M\o lmer-S\o rensen gates. }
\label{MolmerSorensen}
We proceed to discuss the feasibility of executing a quantum gate on qubits defined by dressed states. Optical clocks based on entangled particles can provide a stability gain with the ion number $N$ over the standard quantum limit $\sigma_y \propto 1/\sqrt{N} \rightarrow  1/N$, the so-called Heisenberg limit \citep{leibfried_toward_2004, kessler_heisenberg-limited_2014, nichol_elementary_2022}. Therefore, suitably entangled states pose a promising way towards fast averaging ion clocks, even with moderate ion number \citep{schulte_prospects_2020}. For performing e.g. a M{\o}lmer-S{\o}rensen (MS) gate~\cite{HAFFNER_2008}, this requires to drive sideband transitions off-resonantly in a way which is compatible with the dressing procedure explained in the previous sections.

We consider first a monochromatic driving field tuned close to one of the sideband transitions. In first order Lamb-Dicke expansion, the laser-ion interaction in the laboratory frame bare basis is \cite{James1998}
\begin{align}
    V^\mathrm{LF}_{E2}
    &=\im\sum_{m,M}\Bigl\{\Omega_{mM}\ketbra{M}{m}\e^{-\im\Delta_L t}\nonumber\\
    &\quad\times\left(1 
    +\im\bar{\eta}\qty(\hat{a}\e^{-\im\nu t}+\hat{a}^\dagger\e^{\im\nu t})\right)-\mathrm{h.c.} \Bigr\}.
\end{align}
Here $\bar{\eta}$ is the effective Lamb Dicke parameter, for which we assume $\bar{\eta}\ll 1$, and $\hat{a}$ and $\hat{a}^\dagger$ are creation/annhilation operators referring to one of the normal motional modes of the crystal. The laser detuning from the carrier transition in the bare basis is $\Delta_L$. 

As an example, we consider the case where the detuning is chosen close to the red sideband of one of the transitions in the doubly dressed basis characterized by the set of quantum numbers $(m,M,\barm,\barM,\bbarm,\bbarM)$. This means, the detuning $\Delta_L$ satisfies 
\begin{align}
    \Delta^{mM,\barms\barM}_{\bbarms\bbarMs}+\nu=\delta,
\end{align}
where $\Delta^{mM,\barms\barM}_{\bbarms\bbarMs}$ is given in Eq.~\eqref{eq:effDetuning}, and $\delta$ is the detuning from the sideband transition (aka M{\o}lmer-S{\o}rensen detuning). In a rotating wave approximation with respect to all other terms, the Hamiltonian for a red sideband (rsb) transition becomes 
\begin{align}
    V^\mathrm{DB}_{rsb}
    \approx&-\bar{\eta}\Omega\ketbra*{\bbarM}{\bbarm} \hat{a}\e^{-\im\delta t}+h.c.,
\end{align}
where $\Omega=\bar{\Omega}^{mM,\barms\barMs}_{\bbarms\bbarMs}$, as given in Eq.~\eqref{eq:effRabiFreq}. Given that $\omegasecond^d$ will be the smallest frequency scale in the comb of frequencies induced by the dressing fields, the closest neighbouring transitions will be $\Delta^{mM,\barms\barM}_{\bbarms\bbarMs\pm 1}$, which will be separated by $\omegasecond^d$. We therefore require $\delta\ll \omegasecond^d$ and $\abs*{\Omega}\ll\abs*{\omegasecond^d}$ in applying the rotating wave approximation.  For the blue sideband (bsb) one has instead
\begin{align}
    V^\mathrm{DB}_{bsb}
    \approx&-\bar{\eta}\Omega\ketbra*{\bbarM}{\bbarm} \hat{a}^\dagger\e^{-\im\delta t}+h.c.,
\end{align}
with $\Delta^{mM,\barms\barM}_{\bbarms\bbarMs}-\nu=\delta$. For driving a MS gate, we require $\abs*{\Omega}\ll\abs*{\delta}$. Thus, the MS detuning, the effective sideband Rabi frequency and the smallest frequency split in the double-dressed basis must therefore satisfy a hierarchy of coupling strengths $\Omega\ll\delta\ll\omegasecond^d$. 

For a bi-chromatic field driving the red and the blue sideband transitions at the same time on a crystal of ions, the time evolution operator can be expressed in a Magnus expansion~\cite{MagnusExpansion}
\begin{align}
    U(t)=\e^{\sum_{j}\sigma_x^{(j)}\qty(\alpha_j(t)a^{\dagger}-\alpha^{\star}(t)a)}\e^{-\im\sum_{j,n}\sigma_x^{(j)}\sigma_x^{(n)}\Phi(t)},
\end{align}
with the time-dependent displacement and the geometric phase
\begin{align}
    \alpha(t)&=\frac{\Omega}{\delta}\qty(\e^{-\im \delta t}-1), &
    \Phi(t)&=\frac{\Omega^2}{\delta}\left[t-\frac{1}{\delta}\sin\qty(\delta t)\right],
\end{align}
respectively. Here we used the Pauli operator $\sigma_x=\ketbra*{\bbarM}{\bbarm}+\ketbra*{\bbarm}{\bbarM}$ and write $\sigma^{(j)}_x$ for the operator referring to the $j$-th ion ($j=1,\ldots, N$). For simplicity, we assumed that the sideband Rabi frequency is the same for all particles. In order to decouple the mode of motion in the end of the gate at time $T$, we require $\delta T = 2 n \pi$ for $n\in\mathds{N}$. For achieving a maximally entangling gate, we need $T\Omega^2/\delta=2\pi K$ for  $K$ the number of loops executed in phase space.

Picking up the concrete example treated in the previous section, we can estimate the gate parameters. In view of $\Omega\ll \delta \ll \omegasecond$, we assume $3\Omega_s = \delta=\omegasecond/3$. Assuming $n=K=1$, we estimate a gate duration
\begin{align}
    T=2\pi\frac{\delta}{\Omega^2}=2\pi\frac{9}{\delta}=2\pi\frac{27}{\omegasecond}\approx 3.375 \mathrm{ms}.
\end{align}
While this will not be a competitive gate for quantum computing applications, it may well be sufficient for applications in ion clocks. For ion clocks the gate time has to be compared with the interrogation time which can be on the order of seconds. The extra time of the gate will add to the dark time of the interrogation scheme. We note that some of the conditions imposed on the parameters can be relaxed by exploiting the structure of the comb of frequencies induced by the dressing procedure.

\section{Conclusions}

In this article we developed a compact formalism to describe nested layers of continuous dynamical decoupling by rf dressing fields of ground and excited state Zeeman manifolds. We showed that two layers of dressing can be used to cancel linear Zeeman shifts and electric-quadrupole shifts, and established criteria for which shift to cancel at what layer of dressing. Our main result concerns the description of quadrupole laser-ion interaction in the basis of doubly-dressed states. We characterized the comb of transition frequencies induced by the dressing and expressed the effective Rabi and the transitions frequencies in terms of a set of quantum numbers, which allowed us also to identify the relevant selection rules for these transitions. We addressed the rotating wave approximations and the cross-field effect by treating them in an approximate manner using a Magnus expansion, and showed that both can be effectively interpreted as a shift of the Zeeman splitting for the Zeeman manifolds. With this correction, theoretical predictions are in excellent agreement with experimental data for the quadrupole transitions $S_{1/2}\rightarrow D_{5/2}$ in \ca.
We used our insights to estimate the feasibility of executing MS-gates on the level of the doubly-dressed basis, showing gate times on the order of milliseconds, which is in principle sufficient for use in ion clocks. Faster gates are possible with only one layer of dressing, at the expense of becoming more sensitive to either Zeeman or electric-quadrupole shifts. Gates can be further optimized by exploiting the selection rules and the specific structure of the comb of frequencies induced by the dressing.

\vspace*{0.5cm}
\begin{acknowledgments}
We thank PTB's unit-of-length working group for providing the stable silicium referenced laser source. Fruitful discussions with Nati Aharon, Alex Retzker and the group of Roee Ozeri helped the deepened understanding of CDD shemes. This joint research project was ﬁnancally supported
by the State of Lower Saxony, Hannover, Germany through Niedersächsisches Vorab and by the Deutsche Forschungsgemeinschaft (DFG, German Research Foundation) – Project-ID 274200144 – SFB 1227. This project also received funding from the European Metrology Programme for Innovation and Research (EMPIR) coﬁnanced by the Participating 5 States and from the European Union’s Horizon 2020 research and innovation programme (Project No. 20FUN01 TSCAC).
\end{acknowledgments}

\appendix

\section{Magnetic field fluctuations and Quadrupole shift in the interaction picture}\label{appendixA}
\vspace{0,5cm}

To calculate the energy shift of the bare states created through magnetic field fluctuations, Eq.~\eqref{eq:Bfieldfluctuation}, in the interaction picture, the changes of the spin vectors for the different transformations must be taken into account.

In a RWA one has $\R{z}{\omega t}\mathbf{S}=S_z\bold{e}_z$, therefore, applying the rotation and going to an interaction picture for one layer with a general direction of rotation $\bold{n}=\cos \varphi \bold{e}_x+ \sin \varphi \bold{e}_y$, we obtain
\begin{align}\label{eq:sztransform}
    &\R{z}{\omega t} \R{n}{\theta}S_z\nonumber\\
    &=\cos \theta S_z + \frac{\im}{2} \sin \theta \left(e^{\im t(\omega + \varphi)}S_+-e^{-\im t(\omega + \varphi)}S_-\right)\nonumber\\
    &\simeq \cos \theta S_z.
\end{align}
The RWA drops all the terms oscillating at frequency $\omega+\varphi$. This can be applied for the two dressing layers, recovering the result of Eq.~\eqref{eq:Bfieldsuppression}.

The quadrupole operator, defined by $Q_{ij}=\frac{3}{2}\left(S_iS_j+S_jS_i\right)-S\left(S+1\right)\mathds{1}$, becomes in a RWA
\begin{align}
\R{z}{\omega t}Q&\simeq\frac{3}{2}
    \begin{pmatrix}
    S_x^2+S_y^2 & 0 & 0\\
    0 & S_x^2+S_y^2 & 0\\
    0 & 0 & 2S_z^2
    \end{pmatrix}
    -S(S+1) \mathds{1}\nonumber\\
 &=\frac{S(S+1)-3S_z^2}{2}
    \begin{pmatrix}
    1 & 0 & 0\\
    0 & 1 & 0\\
    0 & 0 & -2
    \end{pmatrix}.
\end{align}
The latter expression is useful for evaluating the quadrupole shift. 
This is further simplified when using the Laplace equation $F_{xx}+F_{yy}+F_{zz}=0$ in the quadrupole shift Hamiltonian
\begin{align}
    \R{z}{\omega t}V_Q^\mathrm{LF}&=\Tr{\R{z}{\omega t}[Q]F}\nonumber\\
    &\simeq\frac{3F_{zz}}{2}\left(3S_z^2-S(S+1)\right).\label{eq:QuadrupoleShiftGeneralForm}
\end{align}
Thus, in the first layer of dressing one has to evaluate
\begin{align}
    &\R{z}{\omega t} \R{n}{\theta}S^2_z\nonumber\\
    &=\qty[\cos \theta S_z + \frac{\im}{2} \sin \theta \left(e^{\im t(\omega + \varphi)}S_+-e^{-\im t(\omega + \varphi)}S_-\right)]^2\nonumber\\
    &\simeq \cos^2\theta S^2_z+\frac{\sin^2\theta}{4}(S_+S_-+S_-S_+)\nonumber\\
    &=\frac{\sin^2\theta}{2}S(S+1)-\frac{1-3\cos^2\theta}{2}S_z^2
\end{align}
Iterating this expression another time yields Eq.~\eqref{eq:HQ}.

\section{Effective Rabi frequency in the doubled dressed basis}
\vspace{0,5cm}
\label{appendixB}

For evaluating the laser-ion interaction in the dressed basis the expression
\begin{align}
    \mel*{\bbarM}{\mathcal{D}^\ss\otimes\mathcal{D}^\ds\qty\big[\ketbra{M}{m}]}{\bbarm}=\mathcal{U}^\ds_{\bbarMs M}(t) \qty(\mathcal{U}^\ss_{\bbarms m}(t))^*
\end{align}
is used, with
\begin{align}
    &\mathcal{U}^\ds_{\bbarMs M}(t)=\nonumber\\ 
    &\mel*{\bbarM}{\U{z}{\omegasecond^\ds t}\U{n^\ds_\text{2}}{\rotsecond^\ds}\U{z}{\omega^\ds_2 t}\U{n^\ds_\text{1}}{\rotfirst^\ds}\U{z}{\omega^\ds_1 t}}{M}
\end{align}
and equivalently for $\mathcal{U}^\ss_{\bbarms m}(t)$ with $\ds\leftrightarrow\ss$ and $\bbarM,M\leftrightarrow\bbarm,m$. As an example we will evaluate the matrix elements for the $\dd$-states. 
\begin{align}
    &\mel*{\bbarM}{\U{z}{\omegasecond t}\U{n_\text{2}}{\rotsecond}\U{z}{\omega_2 t}\U{n_\text{1}}{\rotfirst}\U{z}{\omega_1 t}}{M}\nonumber\\
    &=\sum_{\barMs}
    \mel*{\bbarM}{\U{n_\text{2}}{\rotsecond}}{\barM}
    \mel*{\barM}{\U{n_\text{1}}{\rotfirst}}{M}
    \nonumber\\
    &\quad \times
    \e^{\im\qty(\bbarMs\omegasecond+\barMs\omega_2+ M\omega_1)t},
\end{align}
where we used the expansion of the identity $\mathds{1}= \sum_{\barMs}\ketbra*{\barM}{\barM}$. Finally, the remaining matrix elements of the unitary matrices corresponding to the rotations of the quantization axis are
\begin{align}
    \mel*{\barM}{\U{n_\text{1}}{\rotfirst}}{M}
    &=\mel*{\barM}{\e^{\im\rotfirst \qty(-\sin\alpha S_x+\cos\alpha S_y)}} {M}\nonumber\\
    &=\mel*{\barM}{\e^{-\im\alpha S_z}\e^{\im\rotfirst S_y}\e^{\im\alpha S_z}} {M}\nonumber\\
    &=\e^{-\im\alpha (\barMs-M)} d^S_{M\barMs}(\rotfirst),
\end{align}
and
\begin{align}
    \mel*{\bbarM}{\U{n_\text{2}}{\rotsecond}}{\barM}
    &=\e^{-\im\qty(\alpha-\pi/2) (\bbarMs-\barMs)} d^S_{\barMs\bbarMs}(\rotsecond).
\end{align}
Here, the Wigner d-matrix is used, which is defined in~\cite{Quantummechanics1A.GalindoP.Pascual} as
     \begin{align}
         &d^S_{\barMs M}(\theta)=\bra{S\barM}e^{-\im\theta S_y}\ket{SM}\nonumber\\
         &=\sqrt{(S+\barM )!(S-\barM )!(S+M)!(S-M)!}\nonumber\\
         &\times\sum_k\frac{(-1)^{k}\cos{\left(\frac{\theta}{2}\right)}^{2S+M-\barMs -2k}\qty[-\sin{\left(\frac{\theta}{2}\right)}]^{\barMs -M+2k}}{(S+M-k)!k!(\barM -M+k)!(S-\barM -k)!}.
     \end{align}
     The sum is over all $k$ that do not make negative any factorial in the denominator. We also use that $d^S_{\bbarMs M}(-\theta)=d^S_{M\bbarMs}(\theta)$.

\section{Counter rotating terms or Bloch-Siegart effect}\label{appendixC}

Now, the previously neglected effect of the counter rotating terms in the first rotating wave approximation~\eqref{eq:Ham0_RF} is investigated. We consider the full Hamiltonian
\begin{align}
    H_\mathrm{co}&=
    \frac{g}{4} \left(\Omega_{1} \qty(\e^{\im\qty(2\wrffirstnon t -\alpha)}S_++\e^{-\im\qty(2\wrffirstnon t -\alpha)}S_-)\right.\nonumber\\
    &\left.-\frac{\Omega_{2}}{\im}\cosine(\wrfsecondnon t)\qty(\e^{\im\qty(2\wrffirstnon t -\alpha)}S_+-\e^{-\im\qty(2\wrffirstnon t -\alpha)}S_-)\right).
\end{align}
We will treat this term as a correction to the detuning, thus in a rotating frame with respect to $H_\mathrm{det}=\Delta_1 S_z$ this is 
\begin{align}\label{Eq:HRFrwa}
H^\mathrm{RF}_\mathrm{co}&=\R{z}{\Delta_1 t}\qty[H_\mathrm{co}]=c(t)S_++c^*(t)S_-,    
\end{align}
where
\begin{equation}
    c(t)=\frac{g}{4} \left(\Omega_{1}-\frac{\Omega_2}{\im}\cosine(\wrfsecondnon t)\right)
    \e^{\im(\qty(\omega_0+\wrffirstnon) t-\alpha_\ds)}.
\end{equation}
Therefore, $H^\mathrm{RF}_\mathrm{co}$ will contain only terms oscillating fast at time scales $\omega_0+ \wrffirstnon$ and at sideband frequencies $\wrfsecondnon$ of these. The effect of these off-resonant driving terms, averaged over a time scale $T\gg\qty(\omega_0 + \wrffirstnon)^{-1}$, can be described by an effective Hamiltonian
\begin{align}
    H_\mathrm{co}^\mathrm{eff}&=-\frac{\im}{2 T}\int_0^T\dd{t_1}\int_0^{t_1}\dd{t_2}[H^\mathrm{RF}_\mathrm{co}(t_1),H^\mathrm{RF}_\mathrm{co}(t_2)]\nonumber\\
    &=-\frac{\im}{ T}\int_0^T\dd{t_1}\int_0^{t_1}\dd{t_2}\qty(c(t_1)c^*(t_2)-\mathrm{c.c.} )S_z\nonumber\\
    &\simeq \omega_0\frac{g^2}{8} \frac{\qty(\Omega_1)^2+\qty(\Omega_2)^2}{\omega_0\qty(\omega_0+\omega_1)}S_z.
\end{align}
Further corrections are of higher order in $\Omega_i/\abs{\omega_0 + \wrffirstnon}\ll 1$. The form of the effective Hamiltonian (first line) corresponds to the first non-vanishing term in the Magnus expansion of the time evolution operator corresponding to the Hamiltonian~\eqref{Eq:HRFrwa}. Therefore, the counter rotating terms can be accounted for by suitably shifted bare frequencies that absorb the contributions of $H_\mathrm{co}^\mathrm{eff}$.

\section{Cross-field effect}\label{appendixD}

The non-resonant rf dressing fields of the $\ds$ ($\ss$) spin manifold affect the $\ss$ ($\ds$) manifold. Here, only the former case is covered. The corresponding Hamiltonian on the $\ss$ manifold is
\begin{align}
    H_{\ds\rightarrow\ss}&=
    g_\ss \qty\big(\Omega^\ds_{1} \cosine(\wrffirstnon^\ds t)-\Omega^\ds_{2} \sine(\wrffirstnon^\ds t)\cosine(\wrfsecondnon^\ds t))\nonumber\\
    &\quad\times(S^\ss_x \cos\alpha_\ds+S^\ss_y \sin\alpha_\ds).
\end{align}
In a rotating frame with respect to the dc Hamiltonian $H^\ss_\mathrm{dc}=\omega_0^\ss S_z^\ss$, we obtain 
\begin{align}\label{Eq:HRFds}
H^\mathrm{RF}_{\ds\rightarrow\ss}&=\R{z}{\omega_0^\ss t}\qty[H_{\ds\rightarrow\ss}]=c(t)S_++c^*(t)S_-,    
\end{align}
where
\begin{align}
    c(t)=&\frac{g_\ss}{2} \left(\Omega^\ds_{1} \cosine(\wrffirstnon^\ds t)\right.\nonumber\\
    &\left.-\Omega^\ds_{2} \sine(\wrffirstnon^\ds t)\cosine(\wrfsecondnon^\ds t)\right) \e^{\im(\omega_0^\ss t-\alpha_\ds)}.
\end{align}
Thus, $H^\mathrm{RF}_{\ds\rightarrow\ss}$ will contain only terms oscillating fast at time scales $\omega_0^\ss\pm \wrffirstnon^\ds$ and at sideband frequencies $\wrfsecondnon^\ds$ of these. The effect of these off-resonant driving terms, averaged over a time scale $T\gg\qty(\omega_0^\ss\pm \wrffirstnon^\ds)^{-1}$, can be described by an effective Hamiltonian
\begin{align}
    H_{\ds\rightarrow\ss}^\mathrm{eff}&=-\frac{\im}{2 T}\int_0^T\dd{t_1}\int_0^{t_1}\dd{t_2}[H^\mathrm{RF}_{\ds\rightarrow\ss}(t_1),H^\mathrm{RF}_{\ds\rightarrow\ss}(t_2)]\nonumber\\
    &=-\frac{\im}{ T}\int_0^T\dd{t_1}\int_0^{t_1}\dd{t_2}\qty(c(t_1)c^*(t_2)-\mathrm{c.c.} )S_z^\ss\nonumber\\
    &\simeq \omega_0^\ss\frac{g_\ss^2}{4} \frac{\qty(\Omega_1^\ds)^2+\qty(\Omega_2^\ds)^2}{\qty(\omega_0^\ss)^2-\qty(\omega_1^\ds)^2}S^\ss_z.
\end{align}
Corrections to this are of higher order in $\Omega_i^\ds/\abs{\omega_0^\ss\pm \wrffirstnon^\ds}\ll 1$. The form of the effective Hamiltonian (first line) corresponds to the first non-vanishing term in the Magnus expansion of the time evolution operator corresponding to the Hamiltonian~\eqref{Eq:HRFds}. The same result holds for the effect on the other manifold with $\ss\leftrightarrow\ds.$ Thus, the cross-driving can be accounted for by suitably shifted bare frequencies absorbing the contributions of $H_{\ds(\ss)\rightarrow\ss(\ds)}^\mathrm{eff}$.

\section{Experimental data recording}\label{appendixE}

\begin{figure}[h]
\centering
\subfigure[]{\includegraphics[width=0.48\textwidth]{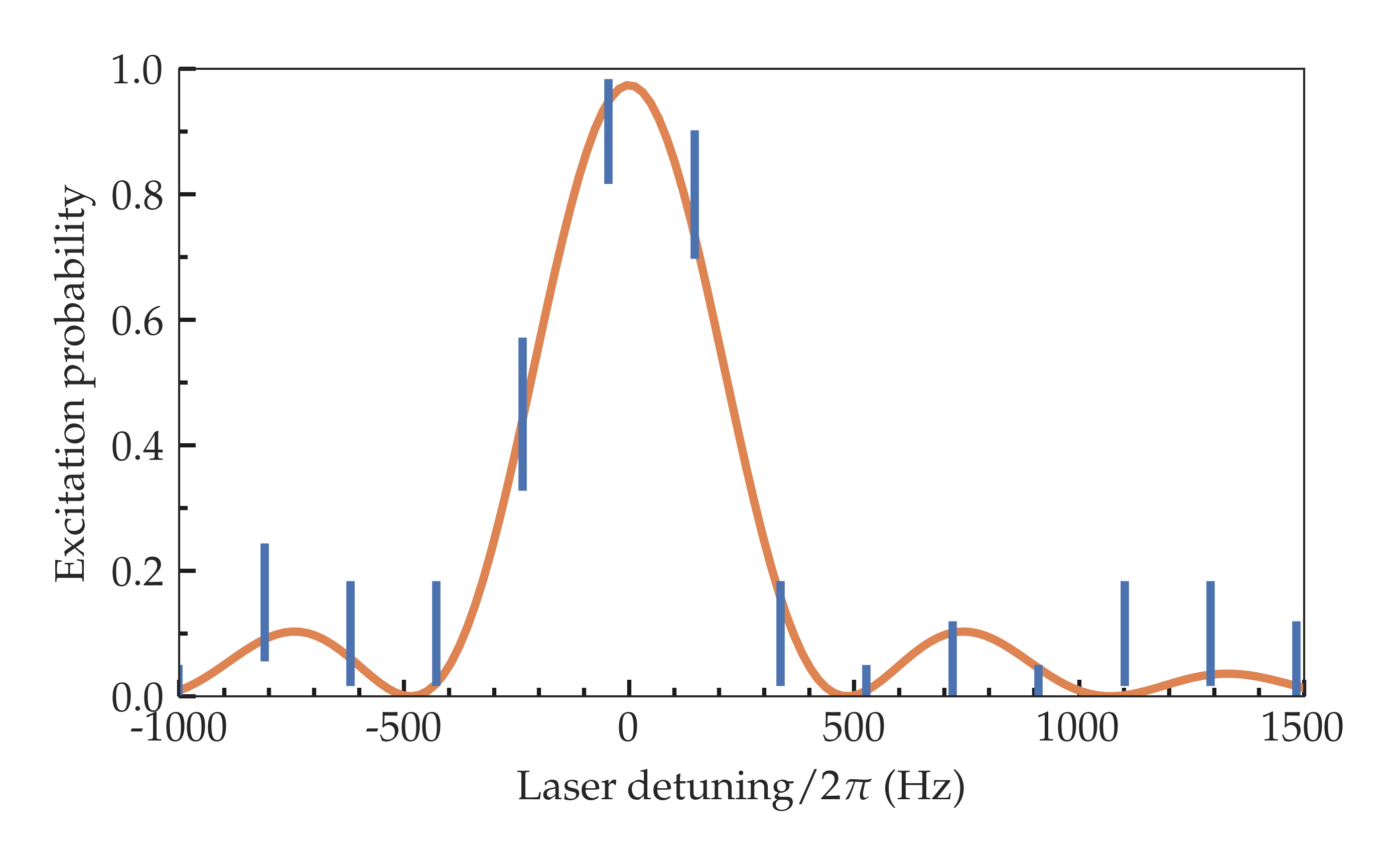}}
\subfigure[]{\includegraphics[width=0.48\textwidth]{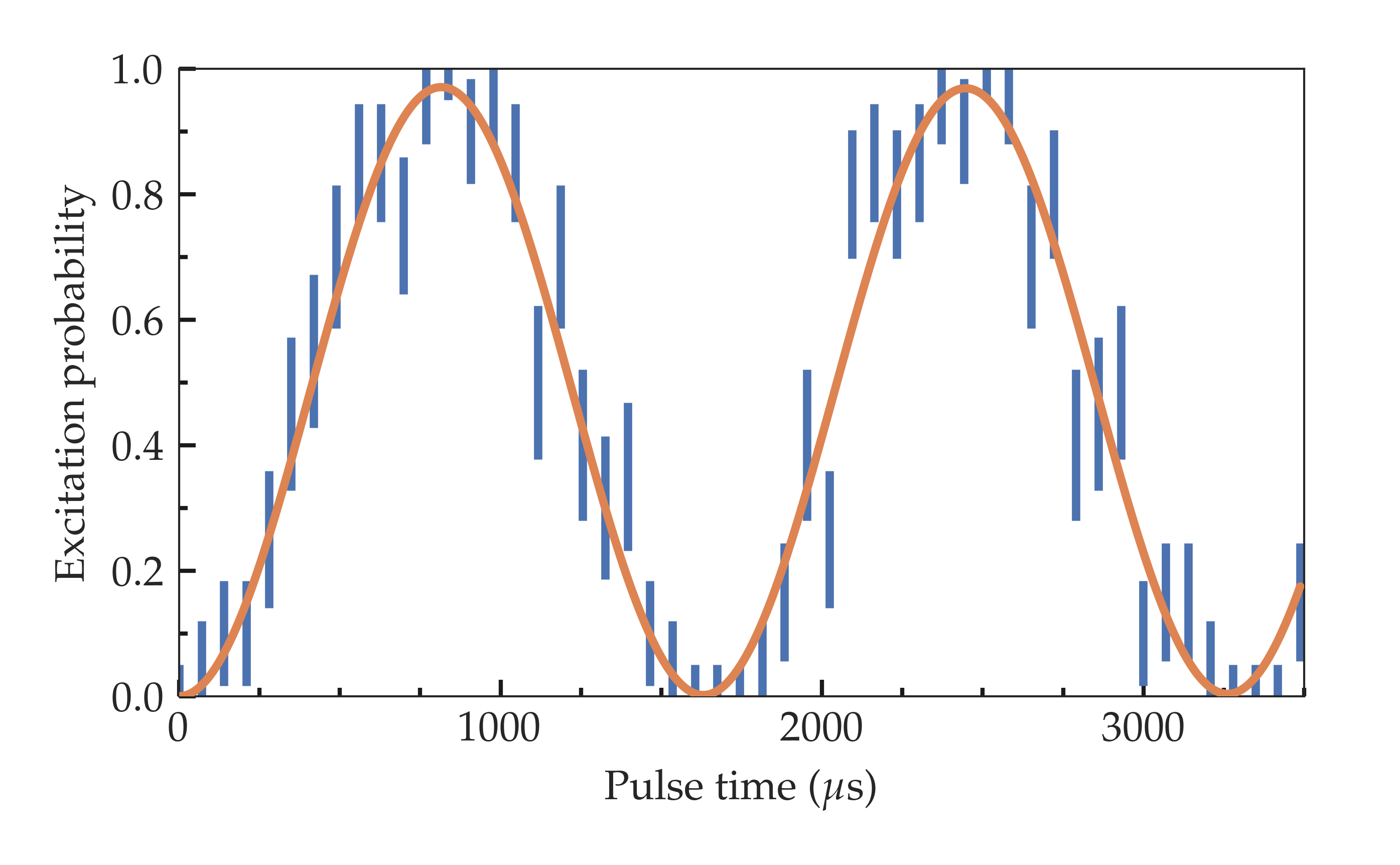}}
\caption{Example measurements for the determination of one transition frequency and coupling strength data point pair. The excitation data (blue) was fitted (orange) to extract:  (a) The center frequency  of one CDD transition using a laser detuning scan. (b) The coupling strength of the same transition using pulse time spectroscopy. For better frequency resolution, the frequency scan was taken with less optical power, thus higher resolution.}
\label{oneDatapoint}
\end{figure}

After the calibration of the rf-drive amplitudes (compare \ref{experimental_sequence}), the acquisition of the individual datapoints for figure \ref{fig:rel_opt_coupling} was performed. Therefore, two different scans were used for each datapoint (compare figure \ref{oneDatapoint}). 

For the first scan, the laser frequency was varied around the predicted CDD transition to extract the transition frequency with high resolution. For the next scan the center frequency was fixed and the pulse duration varied.

A sinosoidal fit of the Rabi flopping signal is used to extract the optical coupling strength. This procedure was repeated for all transitions. The resolution of the individual scans was chosen as a compromise between sufficient low uncertainty and data acquisition speed. The latter is important in order to minimize the uncertainties of drifting static B-field and coupling strength over the course of a complete series of measurements. The acquired data is summarized in Tab.~\ref{tab:scandata}.

\begin{table*}[t]
  \centering
\begin{tabular}{|c|c|c|c|c|c|c|c|}
\hline
   m &    M &   $\barm$ &   $\barM$ & $\Delta_{L,\text{calc}}$ (MHz) & $\Delta_{L,\text{exp}} $ (MHz) &  $\frac{\Omega^{m,M}_{ \bar{m},\bar{M}}}{\Omega_{m,M}}$ (calc) & $\Omega^{m,M}_{ \bar{m},\bar{M}}$ (kHz) \\
\toprule
-0.5 & -1.5 &  0.5 & -2.5 &                         -4.18903 &                   -4.18906 &                                              0.27493 &                        0.46923 \\
-0.5 & -1.5 & -0.5 & -2.5 &                         -4.14213 &                   -4.14214 &                                              0.28640 &                        0.45828 \\
-0.5 & -1.5 &  0.5 & -1.5 &                         -4.11965 &                   -4.11970 &                                              0.36708 &                        0.62568 \\
-0.5 & -1.5 & -0.5 & -1.5 &                         -4.07275 &                   -4.07281 &                                              0.38239 &                        0.62326 \\
-0.5 & -1.5 &  0.5 & -0.5 &                         -4.05027 &                   -4.05028 &                                              0.17087 &                        0.29412 \\
-0.5 & -1.5 & -0.5 & -0.5 &                         -4.00337 &                   -4.00337 &                                              0.17799 &                        0.29951 \\
-0.5 & -1.5 &  0.5 &  0.5 &                         -3.98089 &                   -3.98091 &                                              0.17538 &                        0.27588 \\
-0.5 & -1.5 & -0.5 &  0.5 &                         -3.93399 &                   -3.93401 &                                              0.18270 &                        0.26195 \\
-0.5 & -1.5 &  0.5 &  1.5 &                         -3.91151 &                   -3.91156 &                                              0.36743 &                        0.61065 \\
-0.5 & -1.5 & -0.5 &  1.5 &                         -3.86461 &                   -3.86458 &                                              0.38276 &                        0.61313 \\
-0.5 & -1.5 &  0.5 &  2.5 &                         -3.84213 &                   -3.84216 &                                              0.27255 &                        0.45904 \\
-0.5 & -1.5 & -0.5 &  2.5 &                         -3.79523 &                   -3.79526 &                                              0.28392 &                        0.45519 \\
\hline

-0.5 & -2.5 &  0.5 & -2.5 &                        -10.18386 &                  -10.18395 &                                              0.12331 &                        0.23861 \\
-0.5 & -2.5 & -0.5 & -2.5 &                        -10.13697 &                  -10.13702 &                                              0.12845 &                        0.22617 \\
-0.5 & -2.5 &  0.5 & -1.5 &                        -10.11448 &                  -10.11448 &                                              0.27493 &                        0.52114 \\
-0.5 & -2.5 & -0.5 & -1.5 &                        -10.06759 &                  -10.06771 &                                              0.28640 &                        0.52519 \\
-0.5 & -2.5 &  0.5 & -0.5 &                        -10.04510 &                  -10.04488 &                                              0.38769 &                        0.81061 \\
-0.5 & -2.5 & -0.5 & -0.5 &                         -9.99821 &                   -9.99805 &                                              0.40385 &                        0.77476 \\
-0.5 & -2.5 &  0.5 &  0.5 &                         -9.97572 &                   -9.97573 &                                              0.38657 &                        0.78431 \\
-0.5 & -2.5 & -0.5 &  0.5 &                         -9.92883 &                   -9.92877 &                                              0.40269 &                        0.77869 \\
-0.5 & -2.5 &  0.5 &  1.5 &                         -9.90634 &                   -9.90644 &                                              0.27255 &                        0.54841 \\
-0.5 & -2.5 & -0.5 &  1.5 &                         -9.85945 &                   -9.85939 &                                              0.28392 &                        0.55880 \\
-0.5 & -2.5 &  0.5 &  2.5 &                         -9.83696 &                   -9.83703 &                                              0.12154 &                        0.24665 \\
-0.5 & -2.5 & -0.5 &  2.5 &                         -9.79007 &                   -9.79012 &                                              0.12661 &                        0.24661 \\
\hline
\end{tabular}
\caption{Data used for Fig.~\ref{fig:rel_opt_coupling}. }
\label{tab:scandata}
\end{table*}
\vspace*{5cm}


\begin{thebibliography}{113}%
\makeatletter
\providecommand \@ifxundefined [1]{%
 \@ifx{#1\undefined}
}%
\providecommand \@ifnum [1]{%
 \ifnum #1\expandafter \@firstoftwo
 \else \expandafter \@secondoftwo
 \fi
}%
\providecommand \@ifx [1]{%
 \ifx #1\expandafter \@firstoftwo
 \else \expandafter \@secondoftwo
 \fi
}%
\providecommand \natexlab [1]{#1}%
\providecommand \enquote  [1]{``#1''}%
\providecommand \bibnamefont  [1]{#1}%
\providecommand \bibfnamefont [1]{#1}%
\providecommand \citenamefont [1]{#1}%
\providecommand \href@noop [0]{\@secondoftwo}%
\providecommand \href [0]{\begingroup \@sanitize@url \@href}%
\providecommand \@href[1]{\@@startlink{#1}\@@href}%
\providecommand \@@href[1]{\endgroup#1\@@endlink}%
\providecommand \@sanitize@url [0]{\catcode `\\12\catcode `\$12\catcode
  `\&12\catcode `\#12\catcode `\^12\catcode `\_12\catcode `\%12\relax}%
\providecommand \@@startlink[1]{}%
\providecommand \@@endlink[0]{}%
\providecommand \url  [0]{\begingroup\@sanitize@url \@url }%
\providecommand \@url [1]{\endgroup\@href {#1}{\urlprefix }}%
\providecommand \urlprefix  [0]{URL }%
\providecommand \Eprint [0]{\href }%
\providecommand \doibase [0]{http://dx.doi.org/}%
\providecommand \selectlanguage [0]{\@gobble}%
\providecommand \bibinfo  [0]{\@secondoftwo}%
\providecommand \bibfield  [0]{\@secondoftwo}%
\providecommand \translation [1]{[#1]}%
\providecommand \BibitemOpen [0]{}%
\providecommand \bibitemStop [0]{}%
\providecommand \bibitemNoStop [0]{.\EOS\space}%
\providecommand \EOS [0]{\spacefactor3000\relax}%
\providecommand \BibitemShut  [1]{\csname bibitem#1\endcsname}%
\let\auto@bib@innerbib\@empty
\bibitem [{\citenamefont {Hahn}(1950)}]{Hahn_1950}%
  \BibitemOpen
  \bibfield  {author} {\bibinfo {author} {\bibfnamefont {E.~L.}\ \bibnamefont
  {Hahn}},\ }\bibfield  {title} {\emph {\bibinfo {title} {Spin echoes},\
  }}\href {\doibase 10.1103/PhysRev.80.580} {\bibfield  {journal} {\bibinfo
  {journal} {Physical Review}\ }\textbf {\bibinfo {volume} {80}},\ \bibinfo
  {pages} {580} (\bibinfo {year} {1950})}\BibitemShut {NoStop}%
\bibitem [{\citenamefont {Souza}\ \emph {et~al.}(2012)\citenamefont {Souza},
  \citenamefont {{\'A}lvarez},\ and\ \citenamefont {Suter}}]{Souza2012}%
  \BibitemOpen
  \bibfield  {author} {\bibinfo {author} {\bibfnamefont {A.~M.}\ \bibnamefont
  {Souza}}, \bibinfo {author} {\bibfnamefont {G.~A.}\ \bibnamefont
  {{\'A}lvarez}}, \ and\ \bibinfo {author} {\bibfnamefont {D.}~\bibnamefont
  {Suter}},\ }\bibfield  {title} {\emph {\bibinfo {title} {Robust dynamical
  decoupling},\ }}\href {\doibase 10.1098/rsta.2011.0355} {\bibfield  {journal}
  {\bibinfo  {journal} {Philosophical Transactions of the Royal Society A:
  Mathematical, Physical and Engineering Sciences}\ }\textbf {\bibinfo {volume}
  {370}},\ \bibinfo {pages} {4748} (\bibinfo {year} {2012})}\BibitemShut
  {NoStop}%
\bibitem [{\citenamefont {Viola}\ and\ \citenamefont
  {Lloyd}(1998)}]{Viola_1998}%
  \BibitemOpen
  \bibfield  {author} {\bibinfo {author} {\bibfnamefont {L.}~\bibnamefont
  {Viola}}\ and\ \bibinfo {author} {\bibfnamefont {S.}~\bibnamefont {Lloyd}},\
  }\bibfield  {title} {\emph {\bibinfo {title} {Dynamical suppression of
  decoherence in two-state quantum systems},\ }}\href {\doibase
  10.1103/PhysRevA.58.2733} {\bibfield  {journal} {\bibinfo  {journal}
  {Physical Review A: Atomic, Molecular, and Optical Physics}\ }\textbf
  {\bibinfo {volume} {58}} (\bibinfo {year} {1998}),\
  10.1103/PhysRevA.58.2733}\BibitemShut {NoStop}%
\bibitem [{\citenamefont {Viola}\ \emph {et~al.}(1999)\citenamefont {Viola},
  \citenamefont {Knill},\ and\ \citenamefont {Lloyd}}]{Viola_1999}%
  \BibitemOpen
  \bibfield  {author} {\bibinfo {author} {\bibfnamefont {L.}~\bibnamefont
  {Viola}}, \bibinfo {author} {\bibfnamefont {E.}~\bibnamefont {Knill}}, \ and\
  \bibinfo {author} {\bibfnamefont {S.}~\bibnamefont {Lloyd}},\ }\bibfield
  {title} {\emph {\bibinfo {title} {Dynamical decoupling of open quantum
  systems},\ }}\href {\doibase 10.1103/PhysRevLett.82.2417} {\bibfield
  {journal} {\bibinfo  {journal} {Physical Review Letters}\ }\textbf {\bibinfo
  {volume} {82}} (\bibinfo {year} {1999}),\
  10.1103/PhysRevLett.82.2417}\BibitemShut {NoStop}%
\bibitem [{\citenamefont {Zanardi}(1999)}]{Zanardi_1999}%
  \BibitemOpen
  \bibfield  {author} {\bibinfo {author} {\bibfnamefont {P.}~\bibnamefont
  {Zanardi}},\ }\bibfield  {title} {\emph {\bibinfo {title} {Symmetrizing
  evolutions},\ }}\href {\doibase 10.1016/S0375-9601(99)00365-5} {\bibfield
  {journal} {\bibinfo  {journal} {Physics Letters A}\ }\textbf {\bibinfo
  {volume} {258}} (\bibinfo {year} {1999}),\
  10.1016/S0375-9601(99)00365-5}\BibitemShut {NoStop}%
\bibitem [{\citenamefont {Byrd}\ and\ \citenamefont {Lidar}(2002)}]{Byrd2002}%
  \BibitemOpen
  \bibfield  {author} {\bibinfo {author} {\bibfnamefont {M.~S.}\ \bibnamefont
  {Byrd}}\ and\ \bibinfo {author} {\bibfnamefont {D.~A.}\ \bibnamefont
  {Lidar}},\ }\href {\doibase 10.1023/a:1019697017584} {\bibfield  {journal}
  {\bibinfo  {journal} {Quantum Information Processing}\ }\textbf {\bibinfo
  {volume} {1}} (\bibinfo {year} {2002}),\ 10.1023/a:1019697017584}\BibitemShut
  {NoStop}%
\bibitem [{\citenamefont {Facchi}\ \emph {et~al.}(2005)\citenamefont {Facchi},
  \citenamefont {Tasaki}, \citenamefont {Pascazio}, \citenamefont {Nakazato},
  \citenamefont {Tokuse},\ and\ \citenamefont {Lidar}}]{Facchi2005}%
  \BibitemOpen
  \bibfield  {author} {\bibinfo {author} {\bibfnamefont {P.}~\bibnamefont
  {Facchi}}, \bibinfo {author} {\bibfnamefont {S.}~\bibnamefont {Tasaki}},
  \bibinfo {author} {\bibfnamefont {S.}~\bibnamefont {Pascazio}}, \bibinfo
  {author} {\bibfnamefont {H.}~\bibnamefont {Nakazato}}, \bibinfo {author}
  {\bibfnamefont {A.}~\bibnamefont {Tokuse}}, \ and\ \bibinfo {author}
  {\bibfnamefont {D.~A.}\ \bibnamefont {Lidar}},\ }\bibfield  {title} {\emph
  {\bibinfo {title} {Control of decoherence: {{Analysis}} and comparison of
  three different strategies},\ }}\href {\doibase 10.1103/physreva.71.022302}
  {\bibfield  {journal} {\bibinfo  {journal} {Physical Review A}\ }\textbf
  {\bibinfo {volume} {71}} (\bibinfo {year} {2005}),\
  10.1103/physreva.71.022302}\BibitemShut {NoStop}%
\bibitem [{\citenamefont {Khodjasteh}\ and\ \citenamefont
  {Lidar}(2008)}]{Khodjasteh2008}%
  \BibitemOpen
  \bibfield  {author} {\bibinfo {author} {\bibfnamefont {K.}~\bibnamefont
  {Khodjasteh}}\ and\ \bibinfo {author} {\bibfnamefont {D.~A.}\ \bibnamefont
  {Lidar}},\ }\bibfield  {title} {\emph {\bibinfo {title} {Rigorous bounds on
  the performance of a hybrid dynamical-decoupling quantum-computing scheme},\
  }}\href {\doibase 10.1103/physreva.78.012355} {\bibfield  {journal} {\bibinfo
   {journal} {Physical Review A}\ }\textbf {\bibinfo {volume} {78}} (\bibinfo
  {year} {2008}),\ 10.1103/physreva.78.012355}\BibitemShut {NoStop}%
\bibitem [{\citenamefont {Khodjasteh}\ and\ \citenamefont
  {Viola}(2009{\natexlab{a}})}]{Khodjasteh_2009}%
  \BibitemOpen
  \bibfield  {author} {\bibinfo {author} {\bibfnamefont {K.}~\bibnamefont
  {Khodjasteh}}\ and\ \bibinfo {author} {\bibfnamefont {L.}~\bibnamefont
  {Viola}},\ }\bibfield  {title} {\emph {\bibinfo {title} {Dynamically
  error-corrected gates for universal quantum computation},\ }}\href {\doibase
  10.1103/physrevlett.102.080501} {\bibfield  {journal} {\bibinfo  {journal}
  {Physical Review Letters}\ }\textbf {\bibinfo {volume} {102}} (\bibinfo
  {year} {2009}{\natexlab{a}}),\ 10.1103/physrevlett.102.080501}\BibitemShut
  {NoStop}%
\bibitem [{\citenamefont {Khodjasteh}\ and\ \citenamefont
  {Viola}(2009{\natexlab{b}})}]{Khodjasteh_2009a}%
  \BibitemOpen
  \bibfield  {author} {\bibinfo {author} {\bibfnamefont {K.}~\bibnamefont
  {Khodjasteh}}\ and\ \bibinfo {author} {\bibfnamefont {L.}~\bibnamefont
  {Viola}},\ }\bibfield  {title} {\emph {\bibinfo {title} {Dynamical quantum
  error correction of unitary operations with bounded controls},\ }}\href
  {\doibase 10.1103/physreva.80.032314} {\bibfield  {journal} {\bibinfo
  {journal} {Physical Review A}\ }\textbf {\bibinfo {volume} {80}} (\bibinfo
  {year} {2009}{\natexlab{b}}),\ 10.1103/physreva.80.032314}\BibitemShut
  {NoStop}%
\bibitem [{\citenamefont {Khodjasteh}\ \emph {et~al.}(2010)\citenamefont
  {Khodjasteh}, \citenamefont {Lidar},\ and\ \citenamefont
  {Viola}}]{Khodjasteh2010}%
  \BibitemOpen
  \bibfield  {author} {\bibinfo {author} {\bibfnamefont {K.}~\bibnamefont
  {Khodjasteh}}, \bibinfo {author} {\bibfnamefont {D.~A.}\ \bibnamefont
  {Lidar}}, \ and\ \bibinfo {author} {\bibfnamefont {L.}~\bibnamefont
  {Viola}},\ }\bibfield  {title} {\emph {\bibinfo {title} {Arbitrarily accurate
  dynamical control in open quantum systems},\ }}\href {\doibase
  10.1103/physrevlett.104.090501} {\bibfield  {journal} {\bibinfo  {journal}
  {Physical Review Letters}\ }\textbf {\bibinfo {volume} {104}} (\bibinfo
  {year} {2010}),\ 10.1103/physrevlett.104.090501}\BibitemShut {NoStop}%
\bibitem [{\citenamefont {D.A}(2012)}]{LidarReview2012}%
  \BibitemOpen
  \bibfield  {author} {\bibinfo {author} {\bibfnamefont {L.}~\bibnamefont
  {D.A}},\ }\bibfield  {title} {\emph {\bibinfo {title} {Review of decoherence
  free subspaces, noiseless subsystems, and dynamical decoupling},\ }}\href
  {\doibase arXiv:1208.5791} {\bibfield  {journal} {\bibinfo  {journal} {Adv.
  Chem. Phys. 154, 295}\ } (\bibinfo {year} {2012}),\
  arXiv:1208.5791}\BibitemShut {NoStop}%
\bibitem [{\citenamefont {Aharon}\ \emph {et~al.}(2019)\citenamefont {Aharon},
  \citenamefont {Spethmann}, \citenamefont {Leroux}, \citenamefont {Schmidt},\
  and\ \citenamefont {Retzker}}]{aharon_Robust_2019}%
  \BibitemOpen
  \bibfield  {author} {\bibinfo {author} {\bibfnamefont {N.}~\bibnamefont
  {Aharon}}, \bibinfo {author} {\bibfnamefont {N.}~\bibnamefont {Spethmann}},
  \bibinfo {author} {\bibfnamefont {I.~D.}\ \bibnamefont {Leroux}}, \bibinfo
  {author} {\bibfnamefont {P.~O.}\ \bibnamefont {Schmidt}}, \ and\ \bibinfo
  {author} {\bibfnamefont {A.}~\bibnamefont {Retzker}},\ }\bibfield  {title}
  {\emph {\bibinfo {title} {Robust optical clock transitions in trapped ions
  using dynamical decoupling},\ }}\href {\doibase 10.1088/1367-2630/ab3871}
  {\bibfield  {journal} {\bibinfo  {journal} {New Journal of Physics}\ }\textbf
  {\bibinfo {volume} {21}} (\bibinfo {year} {2019}),\
  10.1088/1367-2630/ab3871}\BibitemShut {NoStop}%
\bibitem [{\citenamefont {Green}\ \emph {et~al.}(2013)\citenamefont {Green},
  \citenamefont {Sastrawan}, \citenamefont {Uys},\ and\ \citenamefont
  {Biercuk}}]{green_arbitrary_2013}%
  \BibitemOpen
  \bibfield  {author} {\bibinfo {author} {\bibfnamefont {T.~J.}\ \bibnamefont
  {Green}}, \bibinfo {author} {\bibfnamefont {J.}~\bibnamefont {Sastrawan}},
  \bibinfo {author} {\bibfnamefont {H.}~\bibnamefont {Uys}}, \ and\ \bibinfo
  {author} {\bibfnamefont {M.~J.}\ \bibnamefont {Biercuk}},\ }\bibfield
  {title} {\emph {\bibinfo {title} {Arbitrary quantum control of qubits in the
  presence of universal noise},\ }}\href {\doibase
  10.1088/1367-2630/15/9/095004} {\bibfield  {journal} {\bibinfo  {journal}
  {New Journal of Physics}\ }\textbf {\bibinfo {volume} {15}},\ \bibinfo
  {pages} {095004} (\bibinfo {year} {2013})}\BibitemShut {NoStop}%
\bibitem [{\citenamefont {West}\ \emph {et~al.}(2010)\citenamefont {West},
  \citenamefont {Lidar}, \citenamefont {Fong},\ and\ \citenamefont
  {Gyure}}]{west_high_2010}%
  \BibitemOpen
  \bibfield  {author} {\bibinfo {author} {\bibfnamefont {J.~R.}\ \bibnamefont
  {West}}, \bibinfo {author} {\bibfnamefont {D.~A.}\ \bibnamefont {Lidar}},
  \bibinfo {author} {\bibfnamefont {B.~H.}\ \bibnamefont {Fong}}, \ and\
  \bibinfo {author} {\bibfnamefont {M.~F.}\ \bibnamefont {Gyure}},\ }\bibfield
  {title} {\emph {\bibinfo {title} {High {{Fidelity Quantum Gates}} via
  {{Dynamical Decoupling}}},\ }}\href {\doibase 10.1103/PhysRevLett.105.230503}
  {\bibfield  {journal} {\bibinfo  {journal} {Physical Review Letters}\
  }\textbf {\bibinfo {volume} {105}},\ \bibinfo {pages} {230503} (\bibinfo
  {year} {2010})}\BibitemShut {NoStop}%
\bibitem [{\citenamefont {Uhrig}(2007)}]{uhrig_keeping_2007}%
  \BibitemOpen
  \bibfield  {author} {\bibinfo {author} {\bibfnamefont {G.~S.}\ \bibnamefont
  {Uhrig}},\ }\bibfield  {title} {\emph {\bibinfo {title} {Keeping a quantum
  bit alive by optimized $\ensuremath{\pi}$-pulse sequences},\ }}\href
  {\doibase 10.1103/PhysRevLett.98.100504} {\bibfield  {journal} {\bibinfo
  {journal} {Physical Review Letters}\ }\textbf {\bibinfo {volume} {98}},\
  \bibinfo {pages} {100504} (\bibinfo {year} {2007})}\BibitemShut {NoStop}%
\bibitem [{\citenamefont {Haeberlen}\ and\ \citenamefont
  {Waugh}(1968)}]{haeberlen_coherent_1968}%
  \BibitemOpen
  \bibfield  {author} {\bibinfo {author} {\bibfnamefont {U.}~\bibnamefont
  {Haeberlen}}\ and\ \bibinfo {author} {\bibfnamefont {J.~S.}\ \bibnamefont
  {Waugh}},\ }\bibfield  {title} {\emph {\bibinfo {title} {Coherent {{Averaging
  Effects}} in {{Magnetic Resonance}}},\ }}\href {\doibase
  10.1103/PhysRev.175.453} {\bibfield  {journal} {\bibinfo  {journal} {Physical
  Review}\ }\textbf {\bibinfo {volume} {175}},\ \bibinfo {pages} {453}
  (\bibinfo {year} {1968})}\BibitemShut {NoStop}%
\bibitem [{\citenamefont {Biercuk}\ \emph {et~al.}(2009)\citenamefont
  {Biercuk}, \citenamefont {Uys}, \citenamefont {VanDevender}, \citenamefont
  {Shiga}, \citenamefont {Itano},\ and\ \citenamefont
  {Bollinger}}]{Biercuk2009}%
  \BibitemOpen
  \bibfield  {author} {\bibinfo {author} {\bibfnamefont {M.~J.}\ \bibnamefont
  {Biercuk}}, \bibinfo {author} {\bibfnamefont {H.}~\bibnamefont {Uys}},
  \bibinfo {author} {\bibfnamefont {A.~P.}\ \bibnamefont {VanDevender}},
  \bibinfo {author} {\bibfnamefont {N.}~\bibnamefont {Shiga}}, \bibinfo
  {author} {\bibfnamefont {W.~M.}\ \bibnamefont {Itano}}, \ and\ \bibinfo
  {author} {\bibfnamefont {J.~J.}\ \bibnamefont {Bollinger}},\ }\bibfield
  {title} {\emph {\bibinfo {title} {Optimized dynamical decoupling in a model
  quantum memory},\ }}\href {\doibase 10.1038/nature07951} {\bibfield
  {journal} {\bibinfo  {journal} {Nature}\ }\textbf {\bibinfo {volume} {458}}
  (\bibinfo {year} {2009}),\ 10.1038/nature07951}\BibitemShut {NoStop}%
\bibitem [{\citenamefont {Du}\ \emph {et~al.}(2009)\citenamefont {Du},
  \citenamefont {Rong}, \citenamefont {Zhao}, \citenamefont {Wang},
  \citenamefont {Yang},\ and\ \citenamefont {Liu}}]{Du2009}%
  \BibitemOpen
  \bibfield  {author} {\bibinfo {author} {\bibfnamefont {J.}~\bibnamefont
  {Du}}, \bibinfo {author} {\bibfnamefont {X.}~\bibnamefont {Rong}}, \bibinfo
  {author} {\bibfnamefont {N.}~\bibnamefont {Zhao}}, \bibinfo {author}
  {\bibfnamefont {Y.}~\bibnamefont {Wang}}, \bibinfo {author} {\bibfnamefont
  {J.}~\bibnamefont {Yang}}, \ and\ \bibinfo {author} {\bibfnamefont {R.~B.}\
  \bibnamefont {Liu}},\ }\bibfield  {title} {\emph {\bibinfo {title}
  {Preserving electron spin coherence in solids by optimal dynamical
  decoupling},\ }}\href {\doibase 10.1038/nature08470} {\bibfield  {journal}
  {\bibinfo  {journal} {Nature}\ }\textbf {\bibinfo {volume} {461}} (\bibinfo
  {year} {2009}),\ 10.1038/nature08470}\BibitemShut {NoStop}%
\bibitem [{\citenamefont {Damodarakurup}\ \emph {et~al.}(2009)\citenamefont
  {Damodarakurup}, \citenamefont {Lucamarini}, \citenamefont {Giuseppe},
  \citenamefont {Vitali},\ and\ \citenamefont {Tombesi}}]{Damodarakurup2009}%
  \BibitemOpen
  \bibfield  {author} {\bibinfo {author} {\bibfnamefont {S.}~\bibnamefont
  {Damodarakurup}}, \bibinfo {author} {\bibfnamefont {M.}~\bibnamefont
  {Lucamarini}}, \bibinfo {author} {\bibfnamefont {G.~D.}\ \bibnamefont
  {Giuseppe}}, \bibinfo {author} {\bibfnamefont {D.}~\bibnamefont {Vitali}}, \
  and\ \bibinfo {author} {\bibfnamefont {P.}~\bibnamefont {Tombesi}},\
  }\bibfield  {title} {\emph {\bibinfo {title} {Experimental inhibition of
  decoherence on flying qubits via ``{{Bang-Bang}}'' control},\ }}\href
  {\doibase 10.1103/physrevlett.103.040502} {\bibfield  {journal} {\bibinfo
  {journal} {Physical Review Letters}\ }\textbf {\bibinfo {volume} {103}}
  (\bibinfo {year} {2009}),\ 10.1103/physrevlett.103.040502}\BibitemShut
  {NoStop}%
\bibitem [{\citenamefont {{de Lange}}\ \emph {et~al.}(2010)\citenamefont {{de
  Lange}}, \citenamefont {Wang}, \citenamefont {Rist{\`e}}, \citenamefont
  {Dobrovitski},\ and\ \citenamefont {Hanson}}]{lange_universal_2010}%
  \BibitemOpen
  \bibfield  {author} {\bibinfo {author} {\bibfnamefont {G.}~\bibnamefont {{de
  Lange}}}, \bibinfo {author} {\bibfnamefont {Z.~H.}\ \bibnamefont {Wang}},
  \bibinfo {author} {\bibfnamefont {D.}~\bibnamefont {Rist{\`e}}}, \bibinfo
  {author} {\bibfnamefont {V.~V.}\ \bibnamefont {Dobrovitski}}, \ and\ \bibinfo
  {author} {\bibfnamefont {R.}~\bibnamefont {Hanson}},\ }\bibfield  {title}
  {\emph {\bibinfo {title} {Universal {{Dynamical Decoupling}} of a {{Single
  Solid-State Spin}} from a {{Spin Bath}}},\ }}\href {\doibase
  10.1126/science.1192739} {\bibfield  {journal} {\bibinfo  {journal} {Science
  (New York, N.Y.)}\ }\textbf {\bibinfo {volume} {330}},\ \bibinfo {pages} {60}
  (\bibinfo {year} {2010})}\BibitemShut {NoStop}%
\bibitem [{\citenamefont {Souza}\ \emph {et~al.}(2011)\citenamefont {Souza},
  \citenamefont {{\'A}lvarez},\ and\ \citenamefont {Suter}}]{Souza2011}%
  \BibitemOpen
  \bibfield  {author} {\bibinfo {author} {\bibfnamefont {A.~M.}\ \bibnamefont
  {Souza}}, \bibinfo {author} {\bibfnamefont {G.~A.}\ \bibnamefont
  {{\'A}lvarez}}, \ and\ \bibinfo {author} {\bibfnamefont {D.}~\bibnamefont
  {Suter}},\ }\bibfield  {title} {\emph {\bibinfo {title} {Robust dynamical
  decoupling for quantum computing and quantum memory},\ }}\href {\doibase
  10.1103/physrevlett.106.240501} {\bibfield  {journal} {\bibinfo  {journal}
  {Physical Review Letters}\ }\textbf {\bibinfo {volume} {106}} (\bibinfo
  {year} {2011}),\ 10.1103/physrevlett.106.240501}\BibitemShut {NoStop}%
\bibitem [{\citenamefont {Naydenov}\ \emph {et~al.}(2011)\citenamefont
  {Naydenov}, \citenamefont {Dolde}, \citenamefont {Hall}, \citenamefont
  {Shin}, \citenamefont {Fedder}, \citenamefont {Hollenberg}, \citenamefont
  {Jelezko},\ and\ \citenamefont {Wrachtrup}}]{Naydenov2011}%
  \BibitemOpen
  \bibfield  {author} {\bibinfo {author} {\bibfnamefont {B.}~\bibnamefont
  {Naydenov}}, \bibinfo {author} {\bibfnamefont {F.}~\bibnamefont {Dolde}},
  \bibinfo {author} {\bibfnamefont {L.~T.}\ \bibnamefont {Hall}}, \bibinfo
  {author} {\bibfnamefont {C.}~\bibnamefont {Shin}}, \bibinfo {author}
  {\bibfnamefont {H.}~\bibnamefont {Fedder}}, \bibinfo {author} {\bibfnamefont
  {L.~C.~L.}\ \bibnamefont {Hollenberg}}, \bibinfo {author} {\bibfnamefont
  {F.}~\bibnamefont {Jelezko}}, \ and\ \bibinfo {author} {\bibfnamefont
  {J.}~\bibnamefont {Wrachtrup}},\ }\bibfield  {title} {\emph {\bibinfo {title}
  {Dynamical decoupling of a single-electron spin at room temperature},\
  }}\href {\doibase 10.1103/physrevb.83.081201} {\bibfield  {journal} {\bibinfo
   {journal} {Physical Review B}\ }\textbf {\bibinfo {volume} {83}} (\bibinfo
  {year} {2011}),\ 10.1103/physrevb.83.081201}\BibitemShut {NoStop}%
\bibitem [{\citenamefont {{van der Sar}}\ \emph {et~al.}(2012)\citenamefont
  {{van der Sar}}, \citenamefont {Wang}, \citenamefont {Blok}, \citenamefont
  {Bernien}, \citenamefont {Taminiau}, \citenamefont {Toyli}, \citenamefont
  {Lidar}, \citenamefont {Awschalom}, \citenamefont {Hanson},\ and\
  \citenamefont {Dobrovitski}}]{Sar_2012}%
  \BibitemOpen
  \bibfield  {author} {\bibinfo {author} {\bibfnamefont {T.}~\bibnamefont {{van
  der Sar}}}, \bibinfo {author} {\bibfnamefont {Z.~H.}\ \bibnamefont {Wang}},
  \bibinfo {author} {\bibfnamefont {M.~S.}\ \bibnamefont {Blok}}, \bibinfo
  {author} {\bibfnamefont {H.}~\bibnamefont {Bernien}}, \bibinfo {author}
  {\bibfnamefont {T.~H.}\ \bibnamefont {Taminiau}}, \bibinfo {author}
  {\bibfnamefont {D.~M.}\ \bibnamefont {Toyli}}, \bibinfo {author}
  {\bibfnamefont {D.~A.}\ \bibnamefont {Lidar}}, \bibinfo {author}
  {\bibfnamefont {D.~D.}\ \bibnamefont {Awschalom}}, \bibinfo {author}
  {\bibfnamefont {R.}~\bibnamefont {Hanson}}, \ and\ \bibinfo {author}
  {\bibfnamefont {V.~V.}\ \bibnamefont {Dobrovitski}},\ }\bibfield  {title}
  {\emph {\bibinfo {title} {Decoherence-protected quantum gates for a hybrid
  solid-state spin register},\ }}\href {\doibase 10.1038/nature10900}
  {\bibfield  {journal} {\bibinfo  {journal} {Nature}\ }\textbf {\bibinfo
  {volume} {484}} (\bibinfo {year} {2012}),\ 10.1038/nature10900}\BibitemShut
  {NoStop}%
\bibitem [{\citenamefont {Shaniv}\ \emph {et~al.}(2019)\citenamefont {Shaniv},
  \citenamefont {Akerman}, \citenamefont {Manovitz}, \citenamefont {Shapira},\
  and\ \citenamefont {Ozeri}}]{shaniv_quadrupole_2019}%
  \BibitemOpen
  \bibfield  {author} {\bibinfo {author} {\bibfnamefont {R.}~\bibnamefont
  {Shaniv}}, \bibinfo {author} {\bibfnamefont {N.}~\bibnamefont {Akerman}},
  \bibinfo {author} {\bibfnamefont {T.}~\bibnamefont {Manovitz}}, \bibinfo
  {author} {\bibfnamefont {Y.}~\bibnamefont {Shapira}}, \ and\ \bibinfo
  {author} {\bibfnamefont {R.}~\bibnamefont {Ozeri}},\ }\bibfield  {title}
  {\emph {\bibinfo {title} {Quadrupole {{Shift Cancellation Using Dynamic
  Decoupling}}},\ }}\href {\doibase 10.1103/PhysRevLett.122.223204} {\bibfield
  {journal} {\bibinfo  {journal} {Physical Review Letters}\ }\textbf {\bibinfo
  {volume} {122}},\ \bibinfo {pages} {223204} (\bibinfo {year}
  {2019})}\BibitemShut {NoStop}%
\bibitem [{\citenamefont {Wang}\ \emph {et~al.}(2017)\citenamefont {Wang},
  \citenamefont {Um}, \citenamefont {Zhang}, \citenamefont {An}, \citenamefont
  {Lyu}, \citenamefont {Zhang}, \citenamefont {Duan}, \citenamefont {Yum},\
  and\ \citenamefont {Kim}}]{wang_single-qubit_2017}%
  \BibitemOpen
  \bibfield  {author} {\bibinfo {author} {\bibfnamefont {Y.}~\bibnamefont
  {Wang}}, \bibinfo {author} {\bibfnamefont {M.}~\bibnamefont {Um}}, \bibinfo
  {author} {\bibfnamefont {J.}~\bibnamefont {Zhang}}, \bibinfo {author}
  {\bibfnamefont {S.}~\bibnamefont {An}}, \bibinfo {author} {\bibfnamefont
  {M.}~\bibnamefont {Lyu}}, \bibinfo {author} {\bibfnamefont {J.-N.}\
  \bibnamefont {Zhang}}, \bibinfo {author} {\bibfnamefont {L.-M.}\ \bibnamefont
  {Duan}}, \bibinfo {author} {\bibfnamefont {D.}~\bibnamefont {Yum}}, \ and\
  \bibinfo {author} {\bibfnamefont {K.}~\bibnamefont {Kim}},\ }\bibfield
  {title} {\emph {\bibinfo {title} {Single-qubit quantum memory exceeding
  ten-minute coherence time},\ }}\href {\doibase 10.1038/s41566-017-0007-1}
  {\bibfield  {journal} {\bibinfo  {journal} {Nature Photonics}\ }\textbf
  {\bibinfo {volume} {11}},\ \bibinfo {pages} {646} (\bibinfo {year}
  {2017})}\BibitemShut {NoStop}%
\bibitem [{\citenamefont {Qiu}\ \emph {et~al.}(2021)\citenamefont {Qiu},
  \citenamefont {Zhou}, \citenamefont {Hu}, \citenamefont {Yuan}, \citenamefont
  {Zhang}, \citenamefont {Chu}, \citenamefont {Huang}, \citenamefont {Liu},
  \citenamefont {Luo}, \citenamefont {Ni}, \citenamefont {Pan}, \citenamefont
  {Yang}, \citenamefont {Zhang}, \citenamefont {Chen}, \citenamefont {Deng},
  \citenamefont {Hu}, \citenamefont {Li}, \citenamefont {Niu}, \citenamefont
  {Xu}, \citenamefont {Yan}, \citenamefont {Zhong}, \citenamefont {Liu},
  \citenamefont {Yan},\ and\ \citenamefont {Yu}}]{qiu_suppressing_2021}%
  \BibitemOpen
  \bibfield  {author} {\bibinfo {author} {\bibfnamefont {J.}~\bibnamefont
  {Qiu}}, \bibinfo {author} {\bibfnamefont {Y.}~\bibnamefont {Zhou}}, \bibinfo
  {author} {\bibfnamefont {C.-K.}\ \bibnamefont {Hu}}, \bibinfo {author}
  {\bibfnamefont {J.}~\bibnamefont {Yuan}}, \bibinfo {author} {\bibfnamefont
  {L.}~\bibnamefont {Zhang}}, \bibinfo {author} {\bibfnamefont
  {J.}~\bibnamefont {Chu}}, \bibinfo {author} {\bibfnamefont {W.}~\bibnamefont
  {Huang}}, \bibinfo {author} {\bibfnamefont {W.}~\bibnamefont {Liu}}, \bibinfo
  {author} {\bibfnamefont {K.}~\bibnamefont {Luo}}, \bibinfo {author}
  {\bibfnamefont {Z.}~\bibnamefont {Ni}}, \bibinfo {author} {\bibfnamefont
  {X.}~\bibnamefont {Pan}}, \bibinfo {author} {\bibfnamefont {Z.}~\bibnamefont
  {Yang}}, \bibinfo {author} {\bibfnamefont {Y.}~\bibnamefont {Zhang}},
  \bibinfo {author} {\bibfnamefont {Y.}~\bibnamefont {Chen}}, \bibinfo {author}
  {\bibfnamefont {X.-H.}\ \bibnamefont {Deng}}, \bibinfo {author}
  {\bibfnamefont {L.}~\bibnamefont {Hu}}, \bibinfo {author} {\bibfnamefont
  {J.}~\bibnamefont {Li}}, \bibinfo {author} {\bibfnamefont {J.}~\bibnamefont
  {Niu}}, \bibinfo {author} {\bibfnamefont {Y.}~\bibnamefont {Xu}}, \bibinfo
  {author} {\bibfnamefont {T.}~\bibnamefont {Yan}}, \bibinfo {author}
  {\bibfnamefont {Y.}~\bibnamefont {Zhong}}, \bibinfo {author} {\bibfnamefont
  {S.}~\bibnamefont {Liu}}, \bibinfo {author} {\bibfnamefont {F.}~\bibnamefont
  {Yan}}, \ and\ \bibinfo {author} {\bibfnamefont {D.}~\bibnamefont {Yu}},\
  }\bibfield  {title} {\emph {\bibinfo {title} {Suppressing {{Coherent
  Two-Qubit Errors}} via {{Dynamical Decoupling}}},\ }}\href {\doibase
  10.1103/PhysRevApplied.16.054047} {\bibfield  {journal} {\bibinfo  {journal}
  {Physical Review Applied}\ }\textbf {\bibinfo {volume} {16}},\ \bibinfo
  {pages} {054047} (\bibinfo {year} {2021})}\BibitemShut {NoStop}%
\bibitem [{\citenamefont {Zhou}\ \emph {et~al.}(2020)\citenamefont {Zhou},
  \citenamefont {Choi}, \citenamefont {Choi}, \citenamefont {Landig},
  \citenamefont {Douglas}, \citenamefont {Isoya}, \citenamefont {Jelezko},
  \citenamefont {Onoda}, \citenamefont {Sumiya}, \citenamefont {Cappellaro},
  \citenamefont {Knowles}, \citenamefont {Park},\ and\ \citenamefont
  {Lukin}}]{zhou_quantum_2020}%
  \BibitemOpen
  \bibfield  {author} {\bibinfo {author} {\bibfnamefont {H.}~\bibnamefont
  {Zhou}}, \bibinfo {author} {\bibfnamefont {J.}~\bibnamefont {Choi}}, \bibinfo
  {author} {\bibfnamefont {S.}~\bibnamefont {Choi}}, \bibinfo {author}
  {\bibfnamefont {R.}~\bibnamefont {Landig}}, \bibinfo {author} {\bibfnamefont
  {A.~M.}\ \bibnamefont {Douglas}}, \bibinfo {author} {\bibfnamefont
  {J.}~\bibnamefont {Isoya}}, \bibinfo {author} {\bibfnamefont
  {F.}~\bibnamefont {Jelezko}}, \bibinfo {author} {\bibfnamefont
  {S.}~\bibnamefont {Onoda}}, \bibinfo {author} {\bibfnamefont
  {H.}~\bibnamefont {Sumiya}}, \bibinfo {author} {\bibfnamefont
  {P.}~\bibnamefont {Cappellaro}}, \bibinfo {author} {\bibfnamefont {H.~S.}\
  \bibnamefont {Knowles}}, \bibinfo {author} {\bibfnamefont {H.}~\bibnamefont
  {Park}}, \ and\ \bibinfo {author} {\bibfnamefont {M.~D.}\ \bibnamefont
  {Lukin}},\ }\bibfield  {title} {\emph {\bibinfo {title} {Quantum
  {{Metrology}} with {{Strongly Interacting Spin Systems}}},\ }}\href {\doibase
  10.1103/PhysRevX.10.031003} {\bibfield  {journal} {\bibinfo  {journal}
  {Physical Review X}\ }\textbf {\bibinfo {volume} {10}},\ \bibinfo {pages}
  {031003} (\bibinfo {year} {2020})}\BibitemShut {NoStop}%
\bibitem [{\citenamefont {Manovitz}\ \emph {et~al.}(2017)\citenamefont
  {Manovitz}, \citenamefont {Rotem}, \citenamefont {Shaniv}, \citenamefont
  {Cohen}, \citenamefont {Shapira}, \citenamefont {Akerman}, \citenamefont
  {Retzker},\ and\ \citenamefont {Ozeri}}]{manovitz_fast_2017}%
  \BibitemOpen
  \bibfield  {author} {\bibinfo {author} {\bibfnamefont {T.}~\bibnamefont
  {Manovitz}}, \bibinfo {author} {\bibfnamefont {A.}~\bibnamefont {Rotem}},
  \bibinfo {author} {\bibfnamefont {R.}~\bibnamefont {Shaniv}}, \bibinfo
  {author} {\bibfnamefont {I.}~\bibnamefont {Cohen}}, \bibinfo {author}
  {\bibfnamefont {Y.}~\bibnamefont {Shapira}}, \bibinfo {author} {\bibfnamefont
  {N.}~\bibnamefont {Akerman}}, \bibinfo {author} {\bibfnamefont
  {A.}~\bibnamefont {Retzker}}, \ and\ \bibinfo {author} {\bibfnamefont
  {R.}~\bibnamefont {Ozeri}},\ }\bibfield  {title} {\emph {\bibinfo {title}
  {Fast dynamical decoupling of the {M}{\o}lmer-{S}{\o}rensen entangling
  gate},\ }}\href {\doibase 10.1103/PhysRevLett.119.220505} {\bibfield
  {journal} {\bibinfo  {journal} {Physical Review Letters}\ }\textbf {\bibinfo
  {volume} {119}},\ \bibinfo {pages} {220505} (\bibinfo {year}
  {2017})}\BibitemShut {NoStop}%
\bibitem [{\citenamefont {Shaniv}\ \emph {et~al.}(2016)\citenamefont {Shaniv},
  \citenamefont {Akerman},\ and\ \citenamefont {Ozeri}}]{shaniv_atomic_2016}%
  \BibitemOpen
  \bibfield  {author} {\bibinfo {author} {\bibfnamefont {R.}~\bibnamefont
  {Shaniv}}, \bibinfo {author} {\bibfnamefont {N.}~\bibnamefont {Akerman}}, \
  and\ \bibinfo {author} {\bibfnamefont {R.}~\bibnamefont {Ozeri}},\ }\bibfield
   {title} {\emph {\bibinfo {title} {Atomic {{Quadrupole Moment Measurement
  Using Dynamic Decoupling}}},\ }}\href {\doibase
  10.1103/PhysRevLett.116.140801} {\bibfield  {journal} {\bibinfo  {journal}
  {Physical Review Letters}\ }\textbf {\bibinfo {volume} {116}},\ \bibinfo
  {pages} {140801} (\bibinfo {year} {2016})}\BibitemShut {NoStop}%
\bibitem [{\citenamefont {Piltz}\ \emph {et~al.}(2013)\citenamefont {Piltz},
  \citenamefont {Scharfenberger}, \citenamefont {Khromova}, \citenamefont
  {Var{\'o}n},\ and\ \citenamefont {Wunderlich}}]{piltz_protecting_2013}%
  \BibitemOpen
  \bibfield  {author} {\bibinfo {author} {\bibfnamefont {C.}~\bibnamefont
  {Piltz}}, \bibinfo {author} {\bibfnamefont {B.}~\bibnamefont
  {Scharfenberger}}, \bibinfo {author} {\bibfnamefont {A.}~\bibnamefont
  {Khromova}}, \bibinfo {author} {\bibfnamefont {A.~F.}\ \bibnamefont
  {Var{\'o}n}}, \ and\ \bibinfo {author} {\bibfnamefont {C.}~\bibnamefont
  {Wunderlich}},\ }\bibfield  {title} {\emph {\bibinfo {title} {Protecting
  {{Conditional Quantum Gates}} by {{Robust Dynamical Decoupling}}},\ }}\href
  {\doibase 10.1103/PhysRevLett.110.200501} {\bibfield  {journal} {\bibinfo
  {journal} {Physical Review Letters}\ }\textbf {\bibinfo {volume} {110}},\
  \bibinfo {pages} {200501} (\bibinfo {year} {2013})}\BibitemShut {NoStop}%
\bibitem [{\citenamefont {Kuwahara}\ \emph {et~al.}(2016)\citenamefont
  {Kuwahara}, \citenamefont {Mori},\ and\ \citenamefont
  {Saito}}]{Kuwahara_2016}%
  \BibitemOpen
  \bibfield  {author} {\bibinfo {author} {\bibfnamefont {T.}~\bibnamefont
  {Kuwahara}}, \bibinfo {author} {\bibfnamefont {T.}~\bibnamefont {Mori}}, \
  and\ \bibinfo {author} {\bibfnamefont {K.}~\bibnamefont {Saito}},\ }\bibfield
   {title} {\emph {\bibinfo {title} {Floquet\textendash{{Magnus}} theory and
  generic transient dynamics in periodically driven many-body quantum
  systems},\ }}\href {\doibase 10.1016/j.aop.2016.01.012} {\bibfield  {journal}
  {\bibinfo  {journal} {Annals of Physics}\ }\textbf {\bibinfo {volume}
  {367}},\ \bibinfo {pages} {96} (\bibinfo {year} {2016})}\BibitemShut
  {NoStop}%
\bibitem [{\citenamefont {{Fonseca-Romero}}\ \emph {et~al.}(2005)\citenamefont
  {{Fonseca-Romero}}, \citenamefont {Kohler},\ and\ \citenamefont
  {H{\"a}nggi}}]{Fonseca_2005}%
  \BibitemOpen
  \bibfield  {author} {\bibinfo {author} {\bibfnamefont {K.~M.}\ \bibnamefont
  {{Fonseca-Romero}}}, \bibinfo {author} {\bibfnamefont {S.}~\bibnamefont
  {Kohler}}, \ and\ \bibinfo {author} {\bibfnamefont {P.}~\bibnamefont
  {H{\"a}nggi}},\ }\bibfield  {title} {\emph {\bibinfo {title} {Coherence
  stabilization of a two-qubit gate by ac fields},\ }}\href {\doibase
  10.1103/PhysRevLett.95.140502} {\bibfield  {journal} {\bibinfo  {journal}
  {Physical Review Letters}\ }\textbf {\bibinfo {volume} {95}} (\bibinfo {year}
  {2005}),\ 10.1103/PhysRevLett.95.140502}\BibitemShut {NoStop}%
\bibitem [{\citenamefont {Chen}(2006)}]{Chen_2006}%
  \BibitemOpen
  \bibfield  {author} {\bibinfo {author} {\bibfnamefont {P.}~\bibnamefont
  {Chen}},\ }\bibfield  {title} {\emph {\bibinfo {title} {Geometric continuous
  dynamical decoupling with bounded controls},\ }}\href {\doibase
  10.1103/PhysRevA.73.022343} {\bibfield  {journal} {\bibinfo  {journal}
  {Physical Review A: Atomic, Molecular, and Optical Physics}\ }\textbf
  {\bibinfo {volume} {73}} (\bibinfo {year} {2006}),\
  10.1103/PhysRevA.73.022343}\BibitemShut {NoStop}%
\bibitem [{\citenamefont {Yal{\c c}{\i}nkaya}\ \emph
  {et~al.}(2019)\citenamefont {Yal{\c c}{\i}nkaya}, \citenamefont {{\c
  C}akmak}, \citenamefont {Karpat},\ and\ \citenamefont
  {Fanchini}}]{yalcinkaya_Continuous_2019}%
  \BibitemOpen
  \bibfield  {author} {\bibinfo {author} {\bibfnamefont {{\.I}.}~\bibnamefont
  {Yal{\c c}{\i}nkaya}}, \bibinfo {author} {\bibfnamefont {B.}~\bibnamefont
  {{\c C}akmak}}, \bibinfo {author} {\bibfnamefont {G.}~\bibnamefont {Karpat}},
  \ and\ \bibinfo {author} {\bibfnamefont {F.~F.}\ \bibnamefont {Fanchini}},\
  }\bibfield  {title} {\emph {\bibinfo {title} {Continuous dynamical decoupling
  and decoherence-free subspaces for qubits with tunable interaction},\ }}\href
  {\doibase 10.1007/s11128-019-2271-0} {\bibfield  {journal} {\bibinfo
  {journal} {Quantum Information Processing}\ }\textbf {\bibinfo {volume} {18}}
  (\bibinfo {year} {2019}),\ 10.1007/s11128-019-2271-0}\BibitemShut {NoStop}%
\bibitem [{\citenamefont {Clausen}\ \emph {et~al.}(2010)\citenamefont
  {Clausen}, \citenamefont {Bensky},\ and\ \citenamefont
  {Kurizki}}]{Clausen_2010}%
  \BibitemOpen
  \bibfield  {author} {\bibinfo {author} {\bibfnamefont {J.}~\bibnamefont
  {Clausen}}, \bibinfo {author} {\bibfnamefont {G.}~\bibnamefont {Bensky}}, \
  and\ \bibinfo {author} {\bibfnamefont {G.}~\bibnamefont {Kurizki}},\
  }\bibfield  {title} {\emph {\bibinfo {title} {Bath-optimized minimal-energy
  protection of quantum operations from decoherence},\ }}\href {\doibase
  10.1103/PhysRevLett.104.040401} {\bibfield  {journal} {\bibinfo  {journal}
  {Physical Review Letters}\ }\textbf {\bibinfo {volume} {104}} (\bibinfo
  {year} {2010}),\ 10.1103/PhysRevLett.104.040401}\BibitemShut {NoStop}%
\bibitem [{\citenamefont {Xu}\ \emph {et~al.}(2012)\citenamefont {Xu},
  \citenamefont {Wang}, \citenamefont {Duan}, \citenamefont {Huang},
  \citenamefont {Wang}, \citenamefont {Wang}, \citenamefont {Xu}, \citenamefont
  {Kong}, \citenamefont {Shi}, \citenamefont {Rong},\ and\ \citenamefont
  {Du}}]{xu_coherence-protected_2012}%
  \BibitemOpen
  \bibfield  {author} {\bibinfo {author} {\bibfnamefont {X.}~\bibnamefont
  {Xu}}, \bibinfo {author} {\bibfnamefont {Z.}~\bibnamefont {Wang}}, \bibinfo
  {author} {\bibfnamefont {C.}~\bibnamefont {Duan}}, \bibinfo {author}
  {\bibfnamefont {P.}~\bibnamefont {Huang}}, \bibinfo {author} {\bibfnamefont
  {P.}~\bibnamefont {Wang}}, \bibinfo {author} {\bibfnamefont {Y.}~\bibnamefont
  {Wang}}, \bibinfo {author} {\bibfnamefont {N.}~\bibnamefont {Xu}}, \bibinfo
  {author} {\bibfnamefont {X.}~\bibnamefont {Kong}}, \bibinfo {author}
  {\bibfnamefont {F.}~\bibnamefont {Shi}}, \bibinfo {author} {\bibfnamefont
  {X.}~\bibnamefont {Rong}}, \ and\ \bibinfo {author} {\bibfnamefont
  {J.}~\bibnamefont {Du}},\ }\bibfield  {title} {\emph {\bibinfo {title}
  {Coherence-{{Protected Quantum Gate}} by {{Continuous Dynamical Decoupling}}
  in {{Diamond}}},\ }}\href {\doibase 10.1103/PhysRevLett.109.070502}
  {\bibfield  {journal} {\bibinfo  {journal} {Physical Review Letters}\
  }\textbf {\bibinfo {volume} {109}},\ \bibinfo {pages} {070502} (\bibinfo
  {year} {2012})}\BibitemShut {NoStop}%
\bibitem [{\citenamefont {Fanchini}\ \emph {et~al.}(2007)\citenamefont
  {Fanchini}, \citenamefont {Hornos},\ and\ \citenamefont
  {Napolitano}}]{Fanchini_2007}%
  \BibitemOpen
  \bibfield  {author} {\bibinfo {author} {\bibfnamefont {F.~F.}\ \bibnamefont
  {Fanchini}}, \bibinfo {author} {\bibfnamefont {J.~E.~M.}\ \bibnamefont
  {Hornos}}, \ and\ \bibinfo {author} {\bibfnamefont {R.~d.~J.}\ \bibnamefont
  {Napolitano}},\ }\bibfield  {title} {\emph {\bibinfo {title} {Continuously
  decoupling single-qubit operations from a perturbing thermal bath of scalar
  bosons},\ }}\href {\doibase 10.1103/PhysRevA.75.022329} {\bibfield  {journal}
  {\bibinfo  {journal} {Physical Review A: Atomic, Molecular, and Optical
  Physics}\ }\textbf {\bibinfo {volume} {75}} (\bibinfo {year} {2007}),\
  10.1103/PhysRevA.75.022329}\BibitemShut {NoStop}%
\bibitem [{\citenamefont {Fanchini}\ and\ \citenamefont
  {Napolitano}(2007)}]{Fanchini_prot_2007}%
  \BibitemOpen
  \bibfield  {author} {\bibinfo {author} {\bibfnamefont {F.~F.}\ \bibnamefont
  {Fanchini}}\ and\ \bibinfo {author} {\bibfnamefont {R.~d.~J.}\ \bibnamefont
  {Napolitano}},\ }\bibfield  {title} {\emph {\bibinfo {title} {Continuous
  dynamical protection of two-qubit entanglement from uncorrelated dephasing,
  bit flipping, and dissipation},\ }}\href {\doibase
  10.1103/PhysRevA.76.062306} {\bibfield  {journal} {\bibinfo  {journal}
  {Physical Review A: Atomic, Molecular, and Optical Physics}\ }\textbf
  {\bibinfo {volume} {76}} (\bibinfo {year} {2007}),\
  10.1103/PhysRevA.76.062306}\BibitemShut {NoStop}%
\bibitem [{\citenamefont {Fanchini}\ \emph {et~al.}(2015)\citenamefont
  {Fanchini}, \citenamefont {Napolitano}, \citenamefont {{\c C}{\c C}akmak},\
  and\ \citenamefont {Caldeira}}]{Fanchini_2015}%
  \BibitemOpen
  \bibfield  {author} {\bibinfo {author} {\bibfnamefont {F.~F.}\ \bibnamefont
  {Fanchini}}, \bibinfo {author} {\bibfnamefont {R.~d.~J.}\ \bibnamefont
  {Napolitano}}, \bibinfo {author} {\bibfnamefont {B.}~\bibnamefont {{\c C}{\c
  C}akmak}}, \ and\ \bibinfo {author} {\bibfnamefont {A.~O.}\ \bibnamefont
  {Caldeira}},\ }\bibfield  {title} {\emph {\bibinfo {title} {Protecting the
  {{SWAP}} operation from general and residual errors by continuous dynamical
  decoupling},\ }}\href {\doibase 10.1103/PhysRevA.91.042325} {\bibfield
  {journal} {\bibinfo  {journal} {Physical Review A: Atomic, Molecular, and
  Optical Physics}\ }\textbf {\bibinfo {volume} {91}} (\bibinfo {year}
  {2015}),\ 10.1103/PhysRevA.91.042325}\BibitemShut {NoStop}%
\bibitem [{\citenamefont {Rabl}\ \emph {et~al.}(2009)\citenamefont {Rabl},
  \citenamefont {Cappellaro}, \citenamefont {Dutt}, \citenamefont {Jiang},
  \citenamefont {Maze},\ and\ \citenamefont {Lukin}}]{Rabl_2009}%
  \BibitemOpen
  \bibfield  {author} {\bibinfo {author} {\bibfnamefont {P.}~\bibnamefont
  {Rabl}}, \bibinfo {author} {\bibfnamefont {P.}~\bibnamefont {Cappellaro}},
  \bibinfo {author} {\bibfnamefont {M.~V.~G.}\ \bibnamefont {Dutt}}, \bibinfo
  {author} {\bibfnamefont {L.}~\bibnamefont {Jiang}}, \bibinfo {author}
  {\bibfnamefont {J.~R.}\ \bibnamefont {Maze}}, \ and\ \bibinfo {author}
  {\bibfnamefont {M.~D.}\ \bibnamefont {Lukin}},\ }\bibfield  {title} {\emph
  {\bibinfo {title} {Strong magnetic coupling between an electronic spin qubit
  and a mechanical resonator},\ }}\href {\doibase 10.1103/PhysRevB.79.041302}
  {\bibfield  {journal} {\bibinfo  {journal} {Physical Review B}\ }\textbf
  {\bibinfo {volume} {79}} (\bibinfo {year} {2009}),\
  10.1103/PhysRevB.79.041302}\BibitemShut {NoStop}%
\bibitem [{\citenamefont {Chaudhry}\ and\ \citenamefont
  {Gong}(2012)}]{Chaudhry_2012}%
  \BibitemOpen
  \bibfield  {author} {\bibinfo {author} {\bibfnamefont {A.~Z.}\ \bibnamefont
  {Chaudhry}}\ and\ \bibinfo {author} {\bibfnamefont {J.}~\bibnamefont
  {Gong}},\ }\bibfield  {title} {\emph {\bibinfo {title} {Decoherence control:
  {{Universal}} protection of two-qubit states and two-qubit gates using
  continuous driving fields},\ }}\href {\doibase 10.1103/PhysRevA.85.012315}
  {\bibfield  {journal} {\bibinfo  {journal} {Physical Review A: Atomic,
  Molecular, and Optical Physics}\ }\textbf {\bibinfo {volume} {85}} (\bibinfo
  {year} {2012}),\ 10.1103/PhysRevA.85.012315}\BibitemShut {NoStop}%
\bibitem [{\citenamefont {Cai}\ \emph {et~al.}(2012)\citenamefont {Cai},
  \citenamefont {Naydenov}, \citenamefont {Pfeiffer}, \citenamefont
  {McGuinness}, \citenamefont {Jahnke}, \citenamefont {Jelezko}, \citenamefont
  {Plenio},\ and\ \citenamefont {Retzker}}]{Cai_2012}%
  \BibitemOpen
  \bibfield  {author} {\bibinfo {author} {\bibfnamefont {J.-M.}\ \bibnamefont
  {Cai}}, \bibinfo {author} {\bibfnamefont {B.}~\bibnamefont {Naydenov}},
  \bibinfo {author} {\bibfnamefont {R.}~\bibnamefont {Pfeiffer}}, \bibinfo
  {author} {\bibfnamefont {L.~P.}\ \bibnamefont {McGuinness}}, \bibinfo
  {author} {\bibfnamefont {K.~D.}\ \bibnamefont {Jahnke}}, \bibinfo {author}
  {\bibfnamefont {F.}~\bibnamefont {Jelezko}}, \bibinfo {author} {\bibfnamefont
  {M.~B.}\ \bibnamefont {Plenio}}, \ and\ \bibinfo {author} {\bibfnamefont
  {A.}~\bibnamefont {Retzker}},\ }\bibfield  {title} {\emph {\bibinfo {title}
  {Robust dynamical decoupling with concatenated continuous driving},\ }}\href
  {\doibase 10.1088/1367-2630/14/11/113023} {\bibfield  {journal} {\bibinfo
  {journal} {New Journal of Physics}\ }\textbf {\bibinfo {volume} {14}}
  (\bibinfo {year} {2012}),\ 10.1088/1367-2630/14/11/113023}\BibitemShut
  {NoStop}%
\bibitem [{\citenamefont {Laraoui}\ and\ \citenamefont
  {Meriles}(2011)}]{Laraoui_2011}%
  \BibitemOpen
  \bibfield  {author} {\bibinfo {author} {\bibfnamefont {A.}~\bibnamefont
  {Laraoui}}\ and\ \bibinfo {author} {\bibfnamefont {C.~A.}\ \bibnamefont
  {Meriles}},\ }\bibfield  {title} {\emph {\bibinfo {title} {Rotating frame
  spin dynamics of a nitrogen-vacancy center in a diamond nanocrystal},\
  }}\href {\doibase 10.1103/PhysRevB.84.161403} {\bibfield  {journal} {\bibinfo
   {journal} {Physical Review B}\ }\textbf {\bibinfo {volume} {84}} (\bibinfo
  {year} {2011}),\ 10.1103/PhysRevB.84.161403}\BibitemShut {NoStop}%
\bibitem [{\citenamefont {Bermudez}\ \emph {et~al.}(2011)\citenamefont
  {Bermudez}, \citenamefont {Jelezko}, \citenamefont {Plenio},\ and\
  \citenamefont {Retzker}}]{Bermudez_2011}%
  \BibitemOpen
  \bibfield  {author} {\bibinfo {author} {\bibfnamefont {A.}~\bibnamefont
  {Bermudez}}, \bibinfo {author} {\bibfnamefont {F.}~\bibnamefont {Jelezko}},
  \bibinfo {author} {\bibfnamefont {M.~B.}\ \bibnamefont {Plenio}}, \ and\
  \bibinfo {author} {\bibfnamefont {A.}~\bibnamefont {Retzker}},\ }\bibfield
  {title} {\emph {\bibinfo {title} {Electron-mediated nuclear-spin interactions
  between distant nitrogen-vacancy centers},\ }}\href {\doibase
  10.1103/PhysRevLett.107.150503} {\bibfield  {journal} {\bibinfo  {journal}
  {Physical Review Letters}\ }\textbf {\bibinfo {volume} {107}} (\bibinfo
  {year} {2011}),\ 10.1103/PhysRevLett.107.150503}\BibitemShut {NoStop}%
\bibitem [{\citenamefont {Bermudez}\ \emph {et~al.}(2012)\citenamefont
  {Bermudez}, \citenamefont {Schmidt}, \citenamefont {Plenio},\ and\
  \citenamefont {Retzker}}]{Bermudez_2012}%
  \BibitemOpen
  \bibfield  {author} {\bibinfo {author} {\bibfnamefont {A.}~\bibnamefont
  {Bermudez}}, \bibinfo {author} {\bibfnamefont {P.~O.}\ \bibnamefont
  {Schmidt}}, \bibinfo {author} {\bibfnamefont {M.~B.}\ \bibnamefont {Plenio}},
  \ and\ \bibinfo {author} {\bibfnamefont {A.}~\bibnamefont {Retzker}},\
  }\bibfield  {title} {\emph {\bibinfo {title} {Robust trapped-ion quantum
  logic gates by continuous dynamical decoupling},\ }}\href {\doibase
  10.1103/PhysRevA.85.040302} {\bibfield  {journal} {\bibinfo  {journal}
  {Physical Review A: Atomic, Molecular, and Optical Physics}\ }\textbf
  {\bibinfo {volume} {85}} (\bibinfo {year} {2012}),\
  10.1103/PhysRevA.85.040302}\BibitemShut {NoStop}%
\bibitem [{\citenamefont {Timoney}\ \emph {et~al.}(2011)\citenamefont
  {Timoney}, \citenamefont {Baumgart}, \citenamefont {Johanning}, \citenamefont
  {Var{\'o}n}, \citenamefont {Plenio}, \citenamefont {Retzker},\ and\
  \citenamefont {Wunderlich}}]{Timoney_2011}%
  \BibitemOpen
  \bibfield  {author} {\bibinfo {author} {\bibfnamefont {N.}~\bibnamefont
  {Timoney}}, \bibinfo {author} {\bibfnamefont {I.}~\bibnamefont {Baumgart}},
  \bibinfo {author} {\bibfnamefont {M.}~\bibnamefont {Johanning}}, \bibinfo
  {author} {\bibfnamefont {A.~F.}\ \bibnamefont {Var{\'o}n}}, \bibinfo {author}
  {\bibfnamefont {M.~B.}\ \bibnamefont {Plenio}}, \bibinfo {author}
  {\bibfnamefont {A.}~\bibnamefont {Retzker}}, \ and\ \bibinfo {author}
  {\bibfnamefont {C.}~\bibnamefont {Wunderlich}},\ }\bibfield  {title} {\emph
  {\bibinfo {title} {Quantum gates and memory using microwave-dressed states},\
  }}\href {\doibase 10.1038/nature10319} {\bibfield  {journal} {\bibinfo
  {journal} {Nature}\ }\textbf {\bibinfo {volume} {476}} (\bibinfo {year}
  {2011}),\ 10.1038/nature10319}\BibitemShut {NoStop}%
\bibitem [{\citenamefont {Doherty}\ \emph {et~al.}(2013)\citenamefont
  {Doherty}, \citenamefont {Manson}, \citenamefont {Delaney}, \citenamefont
  {Jelezko}, \citenamefont {Wrachtrup},\ and\ \citenamefont
  {Hollenberg}}]{Doherty_2013}%
  \BibitemOpen
  \bibfield  {author} {\bibinfo {author} {\bibfnamefont {M.~W.}\ \bibnamefont
  {Doherty}}, \bibinfo {author} {\bibfnamefont {N.~B.}\ \bibnamefont {Manson}},
  \bibinfo {author} {\bibfnamefont {P.}~\bibnamefont {Delaney}}, \bibinfo
  {author} {\bibfnamefont {F.}~\bibnamefont {Jelezko}}, \bibinfo {author}
  {\bibfnamefont {J.}~\bibnamefont {Wrachtrup}}, \ and\ \bibinfo {author}
  {\bibfnamefont {L.~C.}\ \bibnamefont {Hollenberg}},\ }\bibfield  {title}
  {\emph {\bibinfo {title} {The nitrogen-vacancy colour centre in diamond},\
  }}\href {\doibase 10.1016/j.physrep.2013.02.001} {\bibfield  {journal}
  {\bibinfo  {journal} {Physics Reports}\ }\textbf {\bibinfo {volume} {528}}
  (\bibinfo {year} {2013}),\ 10.1016/j.physrep.2013.02.001}\BibitemShut
  {NoStop}%
\bibitem [{\citenamefont {Albrecht}\ \emph {et~al.}(2014)\citenamefont
  {Albrecht}, \citenamefont {Koplovitz}, \citenamefont {Retzker}, \citenamefont
  {Jelezko}, \citenamefont {Yochelis}, \citenamefont {Porath}, \citenamefont
  {Nevo}, \citenamefont {Shoseyov}, \citenamefont {Paltiel},\ and\
  \citenamefont {B~Plenio}}]{Albrecht_2014}%
  \BibitemOpen
  \bibfield  {author} {\bibinfo {author} {\bibfnamefont {A.}~\bibnamefont
  {Albrecht}}, \bibinfo {author} {\bibfnamefont {G.}~\bibnamefont {Koplovitz}},
  \bibinfo {author} {\bibfnamefont {A.}~\bibnamefont {Retzker}}, \bibinfo
  {author} {\bibfnamefont {F.}~\bibnamefont {Jelezko}}, \bibinfo {author}
  {\bibfnamefont {S.}~\bibnamefont {Yochelis}}, \bibinfo {author}
  {\bibfnamefont {D.}~\bibnamefont {Porath}}, \bibinfo {author} {\bibfnamefont
  {Y.}~\bibnamefont {Nevo}}, \bibinfo {author} {\bibfnamefont {O.}~\bibnamefont
  {Shoseyov}}, \bibinfo {author} {\bibfnamefont {Y.}~\bibnamefont {Paltiel}}, \
  and\ \bibinfo {author} {\bibfnamefont {M.}~\bibnamefont {B~Plenio}},\
  }\bibfield  {title} {\emph {\bibinfo {title} {Self-assembling hybrid
  diamond\textendash biological quantum devices},\ }}\href {\doibase
  10.1088/1367-2630/16/9/093002} {\bibfield  {journal} {\bibinfo  {journal}
  {New Journal of Physics}\ }\textbf {\bibinfo {volume} {16}} (\bibinfo {year}
  {2014}),\ 10.1088/1367-2630/16/9/093002}\BibitemShut {NoStop}%
\bibitem [{\citenamefont {Golter}\ \emph {et~al.}(2014)\citenamefont {Golter},
  \citenamefont {Baldwin},\ and\ \citenamefont {Wang}}]{Golter_2014}%
  \BibitemOpen
  \bibfield  {author} {\bibinfo {author} {\bibfnamefont {D.~A.}\ \bibnamefont
  {Golter}}, \bibinfo {author} {\bibfnamefont {T.~K.}\ \bibnamefont {Baldwin}},
  \ and\ \bibinfo {author} {\bibfnamefont {H.}~\bibnamefont {Wang}},\
  }\bibfield  {title} {\emph {\bibinfo {title} {Protecting a solid-state spin
  from decoherence using dressed spin states},\ }}\href {\doibase
  10.1103/PhysRevLett.113.237601} {\bibfield  {journal} {\bibinfo  {journal}
  {Physical Review Letters}\ }\textbf {\bibinfo {volume} {113}} (\bibinfo
  {year} {2014}),\ 10.1103/PhysRevLett.113.237601}\BibitemShut {NoStop}%
\bibitem [{\citenamefont {Finkelstein}\ \emph {et~al.}(2021)\citenamefont
  {Finkelstein}, \citenamefont {Lahad}, \citenamefont {Cohen}, \citenamefont
  {Davidson}, \citenamefont {Kiriati}, \citenamefont {Poem},\ and\
  \citenamefont {Firstenberg}}]{finkelstein_continuous_2021}%
  \BibitemOpen
  \bibfield  {author} {\bibinfo {author} {\bibfnamefont {R.}~\bibnamefont
  {Finkelstein}}, \bibinfo {author} {\bibfnamefont {O.}~\bibnamefont {Lahad}},
  \bibinfo {author} {\bibfnamefont {I.}~\bibnamefont {Cohen}}, \bibinfo
  {author} {\bibfnamefont {O.}~\bibnamefont {Davidson}}, \bibinfo {author}
  {\bibfnamefont {S.}~\bibnamefont {Kiriati}}, \bibinfo {author} {\bibfnamefont
  {E.}~\bibnamefont {Poem}}, \ and\ \bibinfo {author} {\bibfnamefont
  {O.}~\bibnamefont {Firstenberg}},\ }\bibfield  {title} {\emph {\bibinfo
  {title} {Continuous {{Protection}} of a {{Collective State}} from
  {{Inhomogeneous Dephasing}}},\ }}\href {\doibase 10.1103/PhysRevX.11.011008}
  {\bibfield  {journal} {\bibinfo  {journal} {Physical Review X}\ }\textbf
  {\bibinfo {volume} {11}},\ \bibinfo {pages} {011008} (\bibinfo {year}
  {2021})}\BibitemShut {NoStop}%
\bibitem [{\citenamefont {Trypogeorgos}\ \emph {et~al.}(2018)\citenamefont
  {Trypogeorgos}, \citenamefont {{Vald{\'e}s-Curiel}}, \citenamefont
  {Lundblad},\ and\ \citenamefont {Spielman}}]{trypogeorgos_synthetic_2018}%
  \BibitemOpen
  \bibfield  {author} {\bibinfo {author} {\bibfnamefont {D.}~\bibnamefont
  {Trypogeorgos}}, \bibinfo {author} {\bibfnamefont {A.}~\bibnamefont
  {{Vald{\'e}s-Curiel}}}, \bibinfo {author} {\bibfnamefont {N.}~\bibnamefont
  {Lundblad}}, \ and\ \bibinfo {author} {\bibfnamefont {I.~B.}\ \bibnamefont
  {Spielman}},\ }\bibfield  {title} {\emph {\bibinfo {title} {Synthetic clock
  transitions via continuous dynamical decoupling},\ }}\href {\doibase
  10.1103/PhysRevA.97.013407} {\bibfield  {journal} {\bibinfo  {journal}
  {Physical Review A: Atomic, Molecular, and Optical Physics}\ }\textbf
  {\bibinfo {volume} {97}},\ \bibinfo {pages} {013407} (\bibinfo {year}
  {2018})}\BibitemShut {NoStop}%
\bibitem [{\citenamefont {Anderson}\ \emph {et~al.}(2018)\citenamefont
  {Anderson}, \citenamefont {Kewming},\ and\ \citenamefont
  {Turner}}]{anderson_continuously_2018}%
  \BibitemOpen
  \bibfield  {author} {\bibinfo {author} {\bibfnamefont {R.~P.}\ \bibnamefont
  {Anderson}}, \bibinfo {author} {\bibfnamefont {M.~J.}\ \bibnamefont
  {Kewming}}, \ and\ \bibinfo {author} {\bibfnamefont {L.~D.}\ \bibnamefont
  {Turner}},\ }\bibfield  {title} {\emph {\bibinfo {title} {Continuously
  observing a dynamically decoupled spin-1 quantum gas},\ }}\href {\doibase
  10.1103/PhysRevA.97.013408} {\bibfield  {journal} {\bibinfo  {journal}
  {Physical Review A: Atomic, Molecular, and Optical Physics}\ }\textbf
  {\bibinfo {volume} {97}},\ \bibinfo {pages} {013408} (\bibinfo {year}
  {2018})}\BibitemShut {NoStop}%
\bibitem [{\citenamefont {Laucht}\ \emph {et~al.}(2017)\citenamefont {Laucht},
  \citenamefont {Kalra}, \citenamefont {Simmons}, \citenamefont {Dehollain},
  \citenamefont {Muhonen}, \citenamefont {Mohiyaddin}, \citenamefont {Freer},
  \citenamefont {Hudson}, \citenamefont {Itoh}, \citenamefont {Jamieson},
  \citenamefont {McCallum}, \citenamefont {Dzurak},\ and\ \citenamefont
  {Morello}}]{laucht_dressed_2017}%
  \BibitemOpen
  \bibfield  {author} {\bibinfo {author} {\bibfnamefont {A.}~\bibnamefont
  {Laucht}}, \bibinfo {author} {\bibfnamefont {R.}~\bibnamefont {Kalra}},
  \bibinfo {author} {\bibfnamefont {S.}~\bibnamefont {Simmons}}, \bibinfo
  {author} {\bibfnamefont {J.~P.}\ \bibnamefont {Dehollain}}, \bibinfo {author}
  {\bibfnamefont {J.~T.}\ \bibnamefont {Muhonen}}, \bibinfo {author}
  {\bibfnamefont {F.~A.}\ \bibnamefont {Mohiyaddin}}, \bibinfo {author}
  {\bibfnamefont {S.}~\bibnamefont {Freer}}, \bibinfo {author} {\bibfnamefont
  {F.~E.}\ \bibnamefont {Hudson}}, \bibinfo {author} {\bibfnamefont {K.~M.}\
  \bibnamefont {Itoh}}, \bibinfo {author} {\bibfnamefont {D.~N.}\ \bibnamefont
  {Jamieson}}, \bibinfo {author} {\bibfnamefont {J.~C.}\ \bibnamefont
  {McCallum}}, \bibinfo {author} {\bibfnamefont {A.~S.}\ \bibnamefont
  {Dzurak}}, \ and\ \bibinfo {author} {\bibfnamefont {A.}~\bibnamefont
  {Morello}},\ }\bibfield  {title} {\emph {\bibinfo {title} {A dressed spin
  qubit in silicon},\ }}\href {\doibase 10.1038/nnano.2016.178} {\bibfield
  {journal} {\bibinfo  {journal} {Nature Nanotechnology}\ }\textbf {\bibinfo
  {volume} {12}},\ \bibinfo {pages} {61} (\bibinfo {year} {2017})}\BibitemShut
  {NoStop}%
\bibitem [{\citenamefont {S{\'a}rk{\'a}ny}\ \emph {et~al.}(2014)\citenamefont
  {S{\'a}rk{\'a}ny}, \citenamefont {Weiss}, \citenamefont {Hattermann},\ and\
  \citenamefont {Fort{\'a}gh}}]{sarkany_controlling_2014}%
  \BibitemOpen
  \bibfield  {author} {\bibinfo {author} {\bibfnamefont {L.}~\bibnamefont
  {S{\'a}rk{\'a}ny}}, \bibinfo {author} {\bibfnamefont {P.}~\bibnamefont
  {Weiss}}, \bibinfo {author} {\bibfnamefont {H.}~\bibnamefont {Hattermann}}, \
  and\ \bibinfo {author} {\bibfnamefont {J.}~\bibnamefont {Fort{\'a}gh}},\
  }\bibfield  {title} {\emph {\bibinfo {title} {Controlling the magnetic-field
  sensitivity of atomic-clock states by microwave dressing},\ }}\href {\doibase
  10.1103/PhysRevA.90.053416} {\bibfield  {journal} {\bibinfo  {journal}
  {Physical Review A: Atomic, Molecular, and Optical Physics}\ }\textbf
  {\bibinfo {volume} {90}},\ \bibinfo {pages} {053416} (\bibinfo {year}
  {2014})}\BibitemShut {NoStop}%
\bibitem [{\citenamefont {Webster}\ \emph {et~al.}(2013)\citenamefont
  {Webster}, \citenamefont {Weidt}, \citenamefont {Lake}, \citenamefont
  {McLoughlin},\ and\ \citenamefont {Hensinger}}]{webster_simple_2013}%
  \BibitemOpen
  \bibfield  {author} {\bibinfo {author} {\bibfnamefont {S.~C.}\ \bibnamefont
  {Webster}}, \bibinfo {author} {\bibfnamefont {S.}~\bibnamefont {Weidt}},
  \bibinfo {author} {\bibfnamefont {K.}~\bibnamefont {Lake}}, \bibinfo {author}
  {\bibfnamefont {J.~J.}\ \bibnamefont {McLoughlin}}, \ and\ \bibinfo {author}
  {\bibfnamefont {W.~K.}\ \bibnamefont {Hensinger}},\ }\bibfield  {title}
  {\emph {\bibinfo {title} {Simple {{Manipulation}} of a {{Microwave
  Dressed-State Ion Qubit}}},\ }}\href {\doibase
  10.1103/PhysRevLett.111.140501} {\bibfield  {journal} {\bibinfo  {journal}
  {Physical Review Letters}\ }\textbf {\bibinfo {volume} {111}},\ \bibinfo
  {pages} {140501} (\bibinfo {year} {2013})}\BibitemShut {NoStop}%
\bibitem [{\citenamefont {Tan}\ \emph {et~al.}(2013)\citenamefont {Tan},
  \citenamefont {Gaebler}, \citenamefont {Bowler}, \citenamefont {Lin},
  \citenamefont {Jost}, \citenamefont {Leibfried},\ and\ \citenamefont
  {Wineland}}]{tan_demonstration_2013}%
  \BibitemOpen
  \bibfield  {author} {\bibinfo {author} {\bibfnamefont {T.~R.}\ \bibnamefont
  {Tan}}, \bibinfo {author} {\bibfnamefont {J.~P.}\ \bibnamefont {Gaebler}},
  \bibinfo {author} {\bibfnamefont {R.}~\bibnamefont {Bowler}}, \bibinfo
  {author} {\bibfnamefont {Y.}~\bibnamefont {Lin}}, \bibinfo {author}
  {\bibfnamefont {J.~D.}\ \bibnamefont {Jost}}, \bibinfo {author}
  {\bibfnamefont {D.}~\bibnamefont {Leibfried}}, \ and\ \bibinfo {author}
  {\bibfnamefont {D.~J.}\ \bibnamefont {Wineland}},\ }\bibfield  {title} {\emph
  {\bibinfo {title} {Demonstration of a {{Dressed-State Phase Gate}} for
  {{Trapped Ions}}},\ }}\href {\doibase 10.1103/PhysRevLett.110.263002}
  {\bibfield  {journal} {\bibinfo  {journal} {Physical Review Letters}\
  }\textbf {\bibinfo {volume} {110}},\ \bibinfo {pages} {263002} (\bibinfo
  {year} {2013})}\BibitemShut {NoStop}%
\bibitem [{\citenamefont {Aharon}\ \emph {et~al.}(2013)\citenamefont {Aharon},
  \citenamefont {Drewsen},\ and\ \citenamefont
  {Retzker}}]{aharon_general_2013}%
  \BibitemOpen
  \bibfield  {author} {\bibinfo {author} {\bibfnamefont {N.}~\bibnamefont
  {Aharon}}, \bibinfo {author} {\bibfnamefont {M.}~\bibnamefont {Drewsen}}, \
  and\ \bibinfo {author} {\bibfnamefont {A.}~\bibnamefont {Retzker}},\
  }\bibfield  {title} {\emph {\bibinfo {title} {General {{Scheme}} for the
  {{Construction}} of a {{Protected Qubit Subspace}}},\ }}\href {\doibase
  10.1103/PhysRevLett.111.230507} {\bibfield  {journal} {\bibinfo  {journal}
  {Physical Review Letters}\ }\textbf {\bibinfo {volume} {111}},\ \bibinfo
  {pages} {230507} (\bibinfo {year} {2013})}\BibitemShut {NoStop}%
\bibitem [{\citenamefont {{Zanon-Willette}}\ \emph {et~al.}(2012)\citenamefont
  {{Zanon-Willette}}, \citenamefont {{de Clercq}},\ and\ \citenamefont
  {Arimondo}}]{zanon-willette_magic_2012}%
  \BibitemOpen
  \bibfield  {author} {\bibinfo {author} {\bibfnamefont {T.}~\bibnamefont
  {{Zanon-Willette}}}, \bibinfo {author} {\bibfnamefont {E.}~\bibnamefont {{de
  Clercq}}}, \ and\ \bibinfo {author} {\bibfnamefont {E.}~\bibnamefont
  {Arimondo}},\ }\bibfield  {title} {\emph {\bibinfo {title} {Magic
  {{Radio-Frequency Dressing}} of {{Nuclear Spins}} in {{High-Accuracy Optical
  Clocks}}},\ }}\href {\doibase 10.1103/PhysRevLett.109.223003} {\bibfield
  {journal} {\bibinfo  {journal} {Physical Review Letters}\ }\textbf {\bibinfo
  {volume} {109}},\ \bibinfo {pages} {223003} (\bibinfo {year}
  {2012})}\BibitemShut {NoStop}%
\bibitem [{\citenamefont {Kessler}\ \emph {et~al.}(2014)\citenamefont
  {Kessler}, \citenamefont {K{\'o}m{\'a}r}, \citenamefont {Bishof},
  \citenamefont {Jiang}, \citenamefont {S{\o}rensen}, \citenamefont {Ye},\ and\
  \citenamefont {Lukin}}]{kessler_heisenberg-limited_2014}%
  \BibitemOpen
  \bibfield  {author} {\bibinfo {author} {\bibfnamefont {E.~M.}\ \bibnamefont
  {Kessler}}, \bibinfo {author} {\bibfnamefont {P.}~\bibnamefont
  {K{\'o}m{\'a}r}}, \bibinfo {author} {\bibfnamefont {M.}~\bibnamefont
  {Bishof}}, \bibinfo {author} {\bibfnamefont {L.}~\bibnamefont {Jiang}},
  \bibinfo {author} {\bibfnamefont {A.~S.}\ \bibnamefont {S{\o}rensen}},
  \bibinfo {author} {\bibfnamefont {J.}~\bibnamefont {Ye}}, \ and\ \bibinfo
  {author} {\bibfnamefont {M.~D.}\ \bibnamefont {Lukin}},\ }\bibfield  {title}
  {\emph {\bibinfo {title} {Heisenberg-{{Limited Atom Clocks Based}} on
  {{Entangled Qubits}}},\ }}\href {\doibase 10.1103/PhysRevLett.112.190403}
  {\bibfield  {journal} {\bibinfo  {journal} {Physical Review Letters}\
  }\textbf {\bibinfo {volume} {112}},\ \bibinfo {pages} {190403} (\bibinfo
  {year} {2014})}\BibitemShut {NoStop}%
\bibitem [{\citenamefont {Peik}\ \emph {et~al.}(2005)\citenamefont {Peik},
  \citenamefont {Schneider},\ and\ \citenamefont {Tamm}}]{Peik_2005}%
  \BibitemOpen
  \bibfield  {author} {\bibinfo {author} {\bibfnamefont {E.}~\bibnamefont
  {Peik}}, \bibinfo {author} {\bibfnamefont {T.}~\bibnamefont {Schneider}}, \
  and\ \bibinfo {author} {\bibfnamefont {C.}~\bibnamefont {Tamm}},\ }\bibfield
  {title} {\emph {\bibinfo {title} {Laser frequency stabilization to a single
  ion},\ }}\href {\doibase 10.1088/0953-4075/39/1/012} {\bibfield  {journal}
  {\bibinfo  {journal} {Journal of Physics B: Atomic, Molecular and Optical
  Physics}\ }\textbf {\bibinfo {volume} {39}} (\bibinfo {year} {2005}),\
  10.1088/0953-4075/39/1/012}\BibitemShut {NoStop}%
\bibitem [{\citenamefont {Leroux}\ \emph {et~al.}(2017)\citenamefont {Leroux},
  \citenamefont {Scharnhorst}, \citenamefont {Hannig}, \citenamefont {Kramer},
  \citenamefont {Pelzer}, \citenamefont {Stepanova},\ and\ \citenamefont
  {Schmidt}}]{Leroux_2017}%
  \BibitemOpen
  \bibfield  {author} {\bibinfo {author} {\bibfnamefont {I.~D.}\ \bibnamefont
  {Leroux}}, \bibinfo {author} {\bibfnamefont {N.}~\bibnamefont {Scharnhorst}},
  \bibinfo {author} {\bibfnamefont {S.}~\bibnamefont {Hannig}}, \bibinfo
  {author} {\bibfnamefont {J.}~\bibnamefont {Kramer}}, \bibinfo {author}
  {\bibfnamefont {L.}~\bibnamefont {Pelzer}}, \bibinfo {author} {\bibfnamefont
  {M.}~\bibnamefont {Stepanova}}, \ and\ \bibinfo {author} {\bibfnamefont
  {P.~O.}\ \bibnamefont {Schmidt}},\ }\bibfield  {title} {\emph {\bibinfo
  {title} {On-line estimation of local oscillator noise and optimisation of
  servo parameters in atomic clocks},\ }}\href {\doibase
  10.1088/1681-7575/aa66e9} {\bibfield  {journal} {\bibinfo  {journal}
  {Metrologia}\ }\textbf {\bibinfo {volume} {54}} (\bibinfo {year} {2017}),\
  10.1088/1681-7575/aa66e9}\BibitemShut {NoStop}%
\bibitem [{\citenamefont {Keller}\ \emph {et~al.}(2019)\citenamefont {Keller},
  \citenamefont {Kalincev}, \citenamefont {Burgermeister}, \citenamefont
  {Kulosa}, \citenamefont {Didier}, \citenamefont {Nordmann}, \citenamefont
  {Kiethe},\ and\ \citenamefont {Mehlst{\"a}ubler}}]{Keller_ProbingTime_2019}%
  \BibitemOpen
  \bibfield  {author} {\bibinfo {author} {\bibfnamefont {J.}~\bibnamefont
  {Keller}}, \bibinfo {author} {\bibfnamefont {D.}~\bibnamefont {Kalincev}},
  \bibinfo {author} {\bibfnamefont {T.}~\bibnamefont {Burgermeister}}, \bibinfo
  {author} {\bibfnamefont {A.~P.}\ \bibnamefont {Kulosa}}, \bibinfo {author}
  {\bibfnamefont {A.}~\bibnamefont {Didier}}, \bibinfo {author} {\bibfnamefont
  {T.}~\bibnamefont {Nordmann}}, \bibinfo {author} {\bibfnamefont
  {J.}~\bibnamefont {Kiethe}}, \ and\ \bibinfo {author} {\bibfnamefont
  {T.}~\bibnamefont {Mehlst{\"a}ubler}},\ }\bibfield  {title} {\emph {\bibinfo
  {title} {Probing time dilation in coulomb crystals in a high-precision ion
  trap},\ }}\href {\doibase 10.1103/PhysRevApplied.11.011002} {\bibfield
  {journal} {\bibinfo  {journal} {Physical Review Applied}\ }\textbf {\bibinfo
  {volume} {11}} (\bibinfo {year} {2019}),\
  10.1103/PhysRevApplied.11.011002}\BibitemShut {NoStop}%
\bibitem [{\citenamefont {Arnold}\ \emph {et~al.}(2015)\citenamefont {Arnold},
  \citenamefont {Hajiyev}, \citenamefont {Paez}, \citenamefont {Lee},
  \citenamefont {Barrett},\ and\ \citenamefont
  {Bollinger}}]{arnold_prospects_2015}%
  \BibitemOpen
  \bibfield  {author} {\bibinfo {author} {\bibfnamefont {K.}~\bibnamefont
  {Arnold}}, \bibinfo {author} {\bibfnamefont {E.}~\bibnamefont {Hajiyev}},
  \bibinfo {author} {\bibfnamefont {E.}~\bibnamefont {Paez}}, \bibinfo {author}
  {\bibfnamefont {C.~H.}\ \bibnamefont {Lee}}, \bibinfo {author} {\bibfnamefont
  {M.~D.}\ \bibnamefont {Barrett}}, \ and\ \bibinfo {author} {\bibfnamefont
  {J.}~\bibnamefont {Bollinger}},\ }\bibfield  {title} {\emph {\bibinfo {title}
  {Prospects for atomic clocks based on large ion crystals},\ }}\href {\doibase
  10.1103/PhysRevA.92.032108} {\bibfield  {journal} {\bibinfo  {journal}
  {Physical Review A: Atomic, Molecular, and Optical Physics}\ }\textbf
  {\bibinfo {volume} {92}},\ \bibinfo {pages} {032108} (\bibinfo {year}
  {2015})}\BibitemShut {NoStop}%
\bibitem [{\citenamefont {Herschbach}\ \emph {et~al.}(2012)\citenamefont
  {Herschbach}, \citenamefont {Pyka}, \citenamefont {Keller},\ and\
  \citenamefont {Mehlst{\"a}ubler}}]{herschbach_Linear_2012}%
  \BibitemOpen
  \bibfield  {author} {\bibinfo {author} {\bibfnamefont {N.}~\bibnamefont
  {Herschbach}}, \bibinfo {author} {\bibfnamefont {K.}~\bibnamefont {Pyka}},
  \bibinfo {author} {\bibfnamefont {J.}~\bibnamefont {Keller}}, \ and\ \bibinfo
  {author} {\bibfnamefont {T.~E.}\ \bibnamefont {Mehlst{\"a}ubler}},\
  }\bibfield  {title} {\emph {\bibinfo {title} {Linear {{Paul}} trap design for
  an optical clock with {{Coulomb}} crystals},\ }}\href {\doibase
  10.1007/s00340-011-4790-y} {\bibfield  {journal} {\bibinfo  {journal}
  {Applied Physics B}\ }\textbf {\bibinfo {volume} {107}} (\bibinfo {year}
  {2012}),\ 10.1007/s00340-011-4790-y}\BibitemShut {NoStop}%
\bibitem [{\citenamefont {Champenois}\ \emph {et~al.}(2010)\citenamefont
  {Champenois}, \citenamefont {Marciante}, \citenamefont
  {{Pedregosa-Gutierrez}}, \citenamefont {Houssin}, \citenamefont {Knoop},\
  and\ \citenamefont {Kajita}}]{Champenois_2010}%
  \BibitemOpen
  \bibfield  {author} {\bibinfo {author} {\bibfnamefont {C.}~\bibnamefont
  {Champenois}}, \bibinfo {author} {\bibfnamefont {M.}~\bibnamefont
  {Marciante}}, \bibinfo {author} {\bibfnamefont {J.}~\bibnamefont
  {{Pedregosa-Gutierrez}}}, \bibinfo {author} {\bibfnamefont {M.}~\bibnamefont
  {Houssin}}, \bibinfo {author} {\bibfnamefont {M.}~\bibnamefont {Knoop}}, \
  and\ \bibinfo {author} {\bibfnamefont {M.}~\bibnamefont {Kajita}},\
  }\bibfield  {title} {\emph {\bibinfo {title} {Ion ring in a linear multipole
  trap for optical frequency metrology},\ }}\href {\doibase
  10.1103/PhysRevA.81.043410} {\bibfield  {journal} {\bibinfo  {journal}
  {Physical Review A: Atomic, Molecular, and Optical Physics}\ }\textbf
  {\bibinfo {volume} {81}} (\bibinfo {year} {2010}),\
  10.1103/PhysRevA.81.043410}\BibitemShut {NoStop}%
\bibitem [{\citenamefont {Itano}(2000)}]{Itano2000}%
  \BibitemOpen
  \bibfield  {author} {\bibinfo {author} {\bibfnamefont {W.}~\bibnamefont
  {Itano}},\ }\bibfield  {title} {\emph {\bibinfo {title} {External-field
  shifts of the {$^{199}$}{{Hg}}{$^+$} optical frequency standard},\ }}\href
  {\doibase 10.6028/jres.105.065} {\bibfield  {journal} {\bibinfo  {journal}
  {Journal of Research of the National Institute of Standards and Technology}\
  }\textbf {\bibinfo {volume} {105}} (\bibinfo {year} {2000}),\
  10.6028/jres.105.065}\BibitemShut {NoStop}%
\bibitem [{\citenamefont {Berkeland}\ \emph {et~al.}(1998)\citenamefont
  {Berkeland}, \citenamefont {Miller}, \citenamefont {Bergquist}, \citenamefont
  {Itano},\ and\ \citenamefont {Wineland}}]{berkeland_Minimization_1998}%
  \BibitemOpen
  \bibfield  {author} {\bibinfo {author} {\bibfnamefont {D.~J.}\ \bibnamefont
  {Berkeland}}, \bibinfo {author} {\bibfnamefont {J.~D.}\ \bibnamefont
  {Miller}}, \bibinfo {author} {\bibfnamefont {J.~C.}\ \bibnamefont
  {Bergquist}}, \bibinfo {author} {\bibfnamefont {W.~M.}\ \bibnamefont
  {Itano}}, \ and\ \bibinfo {author} {\bibfnamefont {D.~J.}\ \bibnamefont
  {Wineland}},\ }\bibfield  {title} {\emph {\bibinfo {title} {Minimization of
  ion micromotion in a {{Paul}} trap},\ }}\href {\doibase 10.1063/1.367318}
  {\bibfield  {journal} {\bibinfo  {journal} {Journal of Applied Physics}\
  }\textbf {\bibinfo {volume} {83}} (\bibinfo {year} {1998}),\
  10.1063/1.367318}\BibitemShut {NoStop}%
\bibitem [{\citenamefont {Schneider}\ \emph {et~al.}(2005)\citenamefont
  {Schneider}, \citenamefont {Peik},\ and\ \citenamefont
  {Tamm}}]{Schneider2005}%
  \BibitemOpen
  \bibfield  {author} {\bibinfo {author} {\bibfnamefont {T.}~\bibnamefont
  {Schneider}}, \bibinfo {author} {\bibfnamefont {E.}~\bibnamefont {Peik}}, \
  and\ \bibinfo {author} {\bibfnamefont {C.}~\bibnamefont {Tamm}},\ }\bibfield
  {title} {\emph {\bibinfo {title} {Sub-hertz optical frequency comparisons
  between two trapped {$^{171}$}{{Yb}}{$^+$} ions},\ }}\href {\doibase
  10.1103/physrevlett.94.230801} {\bibfield  {journal} {\bibinfo  {journal}
  {Physical Review Letters}\ }\textbf {\bibinfo {volume} {94}} (\bibinfo {year}
  {2005}),\ 10.1103/physrevlett.94.230801}\BibitemShut {NoStop}%
\bibitem [{\citenamefont {Dub{\'e}}\ \emph {et~al.}(2005)\citenamefont
  {Dub{\'e}}, \citenamefont {Madej}, \citenamefont {Bernard}, \citenamefont
  {Marmet}, \citenamefont {Boulanger},\ and\ \citenamefont
  {Cundy}}]{dube_electric_2005}%
  \BibitemOpen
  \bibfield  {author} {\bibinfo {author} {\bibfnamefont {P.}~\bibnamefont
  {Dub{\'e}}}, \bibinfo {author} {\bibfnamefont {A.}~\bibnamefont {Madej}},
  \bibinfo {author} {\bibfnamefont {J.}~\bibnamefont {Bernard}}, \bibinfo
  {author} {\bibfnamefont {L.}~\bibnamefont {Marmet}}, \bibinfo {author}
  {\bibfnamefont {J.-S.}\ \bibnamefont {Boulanger}}, \ and\ \bibinfo {author}
  {\bibfnamefont {S.}~\bibnamefont {Cundy}},\ }\bibfield  {title} {\emph
  {\bibinfo {title} {Electric {{Quadrupole Shift Cancellation}} in {{Single-Ion
  Optical Frequency Standards}}},\ }}\href {\doibase
  10.1103/PhysRevLett.95.033001} {\bibfield  {journal} {\bibinfo  {journal}
  {Physical Review Letters}\ }\textbf {\bibinfo {volume} {95}},\ \bibinfo
  {pages} {033001} (\bibinfo {year} {2005})}\BibitemShut {NoStop}%
\bibitem [{\citenamefont {Tan}\ \emph {et~al.}(2019)\citenamefont {Tan},
  \citenamefont {Kaewuam}, \citenamefont {Arnold}, \citenamefont {Chanu},
  \citenamefont {Zhang}, \citenamefont {Safronova},\ and\ \citenamefont
  {Barrett}}]{tan_suppressing_2019}%
  \BibitemOpen
  \bibfield  {author} {\bibinfo {author} {\bibfnamefont {T.~R.}\ \bibnamefont
  {Tan}}, \bibinfo {author} {\bibfnamefont {R.}~\bibnamefont {Kaewuam}},
  \bibinfo {author} {\bibfnamefont {K.~J.}\ \bibnamefont {Arnold}}, \bibinfo
  {author} {\bibfnamefont {S.~R.}\ \bibnamefont {Chanu}}, \bibinfo {author}
  {\bibfnamefont {Z.}~\bibnamefont {Zhang}}, \bibinfo {author} {\bibfnamefont
  {M.~S.}\ \bibnamefont {Safronova}}, \ and\ \bibinfo {author} {\bibfnamefont
  {M.~D.}\ \bibnamefont {Barrett}},\ }\bibfield  {title} {\emph {\bibinfo
  {title} {Suppressing {{Inhomogeneous Broadening}} in a {{Lutetium Multi-ion
  Optical Clock}}},\ }}\href {\doibase 10.1103/PhysRevLett.123.063201}
  {\bibfield  {journal} {\bibinfo  {journal} {Physical Review Letters}\
  }\textbf {\bibinfo {volume} {123}},\ \bibinfo {pages} {063201} (\bibinfo
  {year} {2019})}\BibitemShut {NoStop}%
\bibitem [{\citenamefont {Lange}\ \emph {et~al.}(2020)\citenamefont {Lange},
  \citenamefont {Huntemann}, \citenamefont {Sanner}, \citenamefont {Shao},
  \citenamefont {Lipphardt}, \citenamefont {Tamm},\ and\ \citenamefont
  {Peik}}]{lange_Coherent_2020}%
  \BibitemOpen
  \bibfield  {author} {\bibinfo {author} {\bibfnamefont {R.}~\bibnamefont
  {Lange}}, \bibinfo {author} {\bibfnamefont {N.}~\bibnamefont {Huntemann}},
  \bibinfo {author} {\bibfnamefont {C.}~\bibnamefont {Sanner}}, \bibinfo
  {author} {\bibfnamefont {H.}~\bibnamefont {Shao}}, \bibinfo {author}
  {\bibfnamefont {B.}~\bibnamefont {Lipphardt}}, \bibinfo {author}
  {\bibfnamefont {C.}~\bibnamefont {Tamm}}, \ and\ \bibinfo {author}
  {\bibfnamefont {E.}~\bibnamefont {Peik}},\ }\bibfield  {title} {\emph
  {\bibinfo {title} {Coherent {{Suppression}} of {{Tensor Frequency Shifts}}
  through {{Magnetic Field Rotation}}},\ }}\href {\doibase
  10.1103/PhysRevLett.125.143201} {\bibfield  {journal} {\bibinfo  {journal}
  {Physical Review Letters}\ }\textbf {\bibinfo {volume} {125}} (\bibinfo
  {year} {2020}),\ 10.1103/PhysRevLett.125.143201}\BibitemShut {NoStop}%
\bibitem [{\citenamefont {Andrew}\ \emph {et~al.}(1958)\citenamefont {Andrew},
  \citenamefont {Bradbury},\ and\ \citenamefont {Eades}}]{andrew_Nuclear_1958}%
  \BibitemOpen
  \bibfield  {author} {\bibinfo {author} {\bibfnamefont {E.~R.}\ \bibnamefont
  {Andrew}}, \bibinfo {author} {\bibfnamefont {A.}~\bibnamefont {Bradbury}}, \
  and\ \bibinfo {author} {\bibfnamefont {R.~G.}\ \bibnamefont {Eades}},\
  }\bibfield  {title} {\emph {\bibinfo {title} {Nuclear {{Magnetic Resonance
  Spectra}} from a {{Crystal}} rotated at {{High Speed}}},\ }}\href {\doibase
  10.1038/1821659a0} {\bibfield  {journal} {\bibinfo  {journal} {Nature}\
  }\textbf {\bibinfo {volume} {182}} (\bibinfo {year} {1958}),\
  10.1038/1821659a0}\BibitemShut {NoStop}%
\bibitem [{\citenamefont {Kaewuam}\ \emph {et~al.}(2020)\citenamefont
  {Kaewuam}, \citenamefont {Tan}, \citenamefont {Arnold}, \citenamefont
  {Chanu}, \citenamefont {Zhang},\ and\ \citenamefont
  {Barrett}}]{kaewuam_hyperfine_2020}%
  \BibitemOpen
  \bibfield  {author} {\bibinfo {author} {\bibfnamefont {R.}~\bibnamefont
  {Kaewuam}}, \bibinfo {author} {\bibfnamefont {T.~R.}\ \bibnamefont {Tan}},
  \bibinfo {author} {\bibfnamefont {K.~J.}\ \bibnamefont {Arnold}}, \bibinfo
  {author} {\bibfnamefont {S.~R.}\ \bibnamefont {Chanu}}, \bibinfo {author}
  {\bibfnamefont {Z.}~\bibnamefont {Zhang}}, \ and\ \bibinfo {author}
  {\bibfnamefont {M.~D.}\ \bibnamefont {Barrett}},\ }\bibfield  {title} {\emph
  {\bibinfo {title} {Hyperfine {{Averaging}} by {{Dynamic Decoupling}} in a
  {{Multi-Ion Lutetium Clock}}},\ }}\href {\doibase
  10.1103/PhysRevLett.124.083202} {\bibfield  {journal} {\bibinfo  {journal}
  {Physical Review Letters}\ }\textbf {\bibinfo {volume} {124}},\ \bibinfo
  {pages} {083202} (\bibinfo {year} {2020})}\BibitemShut {NoStop}%
\bibitem [{\citenamefont {Mart{\'i}nez}(2022)}]{Martinez_thesis}%
  \BibitemOpen
  \bibfield  {author} {\bibinfo {author} {\bibfnamefont {V.}~\bibnamefont
  {Mart{\'i}nez}},\ }\emph {\bibinfo {title} {Relativistic Corrections and
  Dynamic Decoupling in Trapped Ion Optical Atomic Clocks}},\ \href@noop {}
  {Ph.D. thesis},\ \bibinfo  {school} {Leibniz University Hannover} (\bibinfo
  {year} {2022})\BibitemShut {NoStop}%
\bibitem [{\citenamefont {James}(1998)}]{James1998}%
  \BibitemOpen
  \bibfield  {author} {\bibinfo {author} {\bibfnamefont {D.}~\bibnamefont
  {James}},\ }\bibfield  {title} {\emph {\bibinfo {title} {Quantum dynamics of
  cold trapped ions with application to quantum computation},\ }}\href
  {\doibase 10.1007/s003400050373} {\bibfield  {journal} {\bibinfo  {journal}
  {Applied Physics B: Lasers and Optics}\ }\textbf {\bibinfo {volume} {66}}
  (\bibinfo {year} {1998}),\ 10.1007/s003400050373}\BibitemShut {NoStop}%
\bibitem [{\citenamefont {Dalibard}\ and\ \citenamefont
  {{Cohen-Tannoudji}}(1985)}]{dalibard_dressed-atom_1985}%
  \BibitemOpen
  \bibfield  {author} {\bibinfo {author} {\bibfnamefont {J.}~\bibnamefont
  {Dalibard}}\ and\ \bibinfo {author} {\bibfnamefont {C.}~\bibnamefont
  {{Cohen-Tannoudji}}},\ }\bibfield  {title} {\emph {\bibinfo {title}
  {Dressed-atom approach to atomic motion in laser light: {{The}} dipole force
  revisited},\ }}\href@noop {} {\bibfield  {journal} {\bibinfo  {journal} {JOSA
  B}\ }\textbf {\bibinfo {volume} {2}},\ \bibinfo {pages} {1707} (\bibinfo
  {year} {1985})}\BibitemShut {NoStop}%
\bibitem [{\citenamefont {Tommaseo}\ \emph {et~al.}(2003)\citenamefont
  {Tommaseo}, \citenamefont {Pfeil}, \citenamefont {Revalde}, \citenamefont
  {Werth}, \citenamefont {Indelicato},\ and\ \citenamefont
  {Desclaux}}]{tommaseo_factor_2003}%
  \BibitemOpen
  \bibfield  {author} {\bibinfo {author} {\bibfnamefont {G.}~\bibnamefont
  {Tommaseo}}, \bibinfo {author} {\bibfnamefont {T.}~\bibnamefont {Pfeil}},
  \bibinfo {author} {\bibfnamefont {G.}~\bibnamefont {Revalde}}, \bibinfo
  {author} {\bibfnamefont {G.}~\bibnamefont {Werth}}, \bibinfo {author}
  {\bibfnamefont {P.}~\bibnamefont {Indelicato}}, \ and\ \bibinfo {author}
  {\bibfnamefont {J.~P.}\ \bibnamefont {Desclaux}},\ }\bibfield  {title} {\emph
  {\bibinfo {title} {The g{{{\textsubscript{J}}}}-factor in the ground state of
  {{Ca}}{$^{+}$}},\ }}\href {\doibase 10.1140/epjd/e2003-00096-6} {\bibfield
  {journal} {\bibinfo  {journal} {The European Physical Journal D - Atomic,
  Molecular, Optical and Plasma Physics}\ }\textbf {\bibinfo {volume} {25}}
  (\bibinfo {year} {2003}),\ 10.1140/epjd/e2003-00096-6}\BibitemShut {NoStop}%
\bibitem [{\citenamefont {Chwalla}\ \emph {et~al.}(2009)\citenamefont
  {Chwalla}, \citenamefont {Benhelm}, \citenamefont {Kim}, \citenamefont
  {Kirchmair}, \citenamefont {Monz}, \citenamefont {Riebe}, \citenamefont
  {Schindler}, \citenamefont {Villar}, \citenamefont {H{\"a}nsel},
  \citenamefont {Roos}, \citenamefont {Blatt}, \citenamefont {Abgrall},
  \citenamefont {Santarelli}, \citenamefont {Rovera},\ and\ \citenamefont
  {Laurent}}]{chwalla_absolute_2009}%
  \BibitemOpen
  \bibfield  {author} {\bibinfo {author} {\bibfnamefont {M.}~\bibnamefont
  {Chwalla}}, \bibinfo {author} {\bibfnamefont {J.}~\bibnamefont {Benhelm}},
  \bibinfo {author} {\bibfnamefont {K.}~\bibnamefont {Kim}}, \bibinfo {author}
  {\bibfnamefont {G.}~\bibnamefont {Kirchmair}}, \bibinfo {author}
  {\bibfnamefont {T.}~\bibnamefont {Monz}}, \bibinfo {author} {\bibfnamefont
  {M.}~\bibnamefont {Riebe}}, \bibinfo {author} {\bibfnamefont
  {P.}~\bibnamefont {Schindler}}, \bibinfo {author} {\bibfnamefont
  {A.}~\bibnamefont {Villar}}, \bibinfo {author} {\bibfnamefont
  {W.}~\bibnamefont {H{\"a}nsel}}, \bibinfo {author} {\bibfnamefont
  {C.}~\bibnamefont {Roos}}, \bibinfo {author} {\bibfnamefont {R.}~\bibnamefont
  {Blatt}}, \bibinfo {author} {\bibfnamefont {M.}~\bibnamefont {Abgrall}},
  \bibinfo {author} {\bibfnamefont {G.}~\bibnamefont {Santarelli}}, \bibinfo
  {author} {\bibfnamefont {G.}~\bibnamefont {Rovera}}, \ and\ \bibinfo {author}
  {\bibfnamefont {P.}~\bibnamefont {Laurent}},\ }\bibfield  {title} {\emph
  {\bibinfo {title} {Absolute frequency measurement of the $^{40}${Ca}$^{+}$
  $4s ^{2}{S}_{1/2}-3d^{2}{D}_{5/2}$ clock transition},\ }}\href {\doibase
  10.1103/PhysRevLett.102.023002} {\bibfield  {journal} {\bibinfo  {journal}
  {Physical Review Letters}\ }\textbf {\bibinfo {volume} {102}},\ \bibinfo
  {pages} {023002} (\bibinfo {year} {2009})}\BibitemShut {NoStop}%
\bibitem [{\citenamefont {Monz}\ \emph {et~al.}(2011)\citenamefont {Monz},
  \citenamefont {Schindler}, \citenamefont {Barreiro}, \citenamefont {Chwalla},
  \citenamefont {Nigg}, \citenamefont {Coish}, \citenamefont {Harlander},
  \citenamefont {H{\"a}nsel}, \citenamefont {Hennrich},\ and\ \citenamefont
  {Blatt}}]{monz_14qubit_2011}%
  \BibitemOpen
  \bibfield  {author} {\bibinfo {author} {\bibfnamefont {T.}~\bibnamefont
  {Monz}}, \bibinfo {author} {\bibfnamefont {P.}~\bibnamefont {Schindler}},
  \bibinfo {author} {\bibfnamefont {J.~T.}\ \bibnamefont {Barreiro}}, \bibinfo
  {author} {\bibfnamefont {M.}~\bibnamefont {Chwalla}}, \bibinfo {author}
  {\bibfnamefont {D.}~\bibnamefont {Nigg}}, \bibinfo {author} {\bibfnamefont
  {W.~A.}\ \bibnamefont {Coish}}, \bibinfo {author} {\bibfnamefont
  {M.}~\bibnamefont {Harlander}}, \bibinfo {author} {\bibfnamefont
  {W.}~\bibnamefont {H{\"a}nsel}}, \bibinfo {author} {\bibfnamefont
  {M.}~\bibnamefont {Hennrich}}, \ and\ \bibinfo {author} {\bibfnamefont
  {R.}~\bibnamefont {Blatt}},\ }\bibfield  {title} {\emph {\bibinfo {title}
  {14-{{Qubit Entanglement}}: {{Creation}} and {{Coherence}}},\ }}\href
  {\doibase 10.1103/PhysRevLett.106.130506} {\bibfield  {journal} {\bibinfo
  {journal} {Physical Review Letters}\ }\textbf {\bibinfo {volume} {106}},\
  \bibinfo {pages} {130506} (\bibinfo {year} {2011})}\BibitemShut {NoStop}%
\bibitem [{\citenamefont {Kaushal}\ \emph {et~al.}(2020)\citenamefont
  {Kaushal}, \citenamefont {Lekitsch}, \citenamefont {Stahl}, \citenamefont
  {Hilder}, \citenamefont {Pijn}, \citenamefont {Schmiegelow}, \citenamefont
  {Bermudez}, \citenamefont {M{\"u}ller}, \citenamefont {{Schmidt-Kaler}},\
  and\ \citenamefont {Poschinger}}]{kaushal_shuttlingbased_2020}%
  \BibitemOpen
  \bibfield  {author} {\bibinfo {author} {\bibfnamefont {V.}~\bibnamefont
  {Kaushal}}, \bibinfo {author} {\bibfnamefont {B.}~\bibnamefont {Lekitsch}},
  \bibinfo {author} {\bibfnamefont {A.}~\bibnamefont {Stahl}}, \bibinfo
  {author} {\bibfnamefont {J.}~\bibnamefont {Hilder}}, \bibinfo {author}
  {\bibfnamefont {D.}~\bibnamefont {Pijn}}, \bibinfo {author} {\bibfnamefont
  {C.}~\bibnamefont {Schmiegelow}}, \bibinfo {author} {\bibfnamefont
  {A.}~\bibnamefont {Bermudez}}, \bibinfo {author} {\bibfnamefont
  {M.}~\bibnamefont {M{\"u}ller}}, \bibinfo {author} {\bibfnamefont
  {F.}~\bibnamefont {{Schmidt-Kaler}}}, \ and\ \bibinfo {author} {\bibfnamefont
  {U.}~\bibnamefont {Poschinger}},\ }\bibfield  {title} {\emph {\bibinfo
  {title} {Shuttling-based trapped-ion quantum information processing},\
  }}\href {\doibase 10.1116/1.5126186} {\bibfield  {journal} {\bibinfo
  {journal} {AVS Quantum Science}\ }\textbf {\bibinfo {volume} {2}},\ \bibinfo
  {pages} {014101} (\bibinfo {year} {2020})}\BibitemShut {NoStop}%
\bibitem [{\citenamefont {Ringbauer}\ \emph {et~al.}(2022)\citenamefont
  {Ringbauer}, \citenamefont {Meth}, \citenamefont {Postler}, \citenamefont
  {Stricker}, \citenamefont {Blatt}, \citenamefont {Schindler},\ and\
  \citenamefont {Monz}}]{ringbauer_universal_2022}%
  \BibitemOpen
  \bibfield  {author} {\bibinfo {author} {\bibfnamefont {M.}~\bibnamefont
  {Ringbauer}}, \bibinfo {author} {\bibfnamefont {M.}~\bibnamefont {Meth}},
  \bibinfo {author} {\bibfnamefont {L.}~\bibnamefont {Postler}}, \bibinfo
  {author} {\bibfnamefont {R.}~\bibnamefont {Stricker}}, \bibinfo {author}
  {\bibfnamefont {R.}~\bibnamefont {Blatt}}, \bibinfo {author} {\bibfnamefont
  {P.}~\bibnamefont {Schindler}}, \ and\ \bibinfo {author} {\bibfnamefont
  {T.}~\bibnamefont {Monz}},\ }\bibfield  {title} {\emph {\bibinfo {title} {A
  universal qudit quantum processor with trapped ions},\ }}\href {\doibase
  10.1038/s41567-022-01658-0} {\bibfield  {journal} {\bibinfo  {journal}
  {Nature Physics}\ }\textbf {\bibinfo {volume} {18}},\ \bibinfo {pages} {1053}
  (\bibinfo {year} {2022})}\BibitemShut {NoStop}%
\bibitem [{\citenamefont {Pogorelov}\ \emph {et~al.}(2021)\citenamefont
  {Pogorelov}, \citenamefont {Feldker}, \citenamefont {Marciniak},
  \citenamefont {Postler}, \citenamefont {Jacob}, \citenamefont
  {Krieglsteiner}, \citenamefont {Podlesnic}, \citenamefont {Meth},
  \citenamefont {Negnevitsky}, \citenamefont {Stadler}, \citenamefont
  {H{\"o}fer}, \citenamefont {W{\"a}chter}, \citenamefont {Lakhmanskiy},
  \citenamefont {Blatt}, \citenamefont {Schindler},\ and\ \citenamefont
  {Monz}}]{pogorelov_compact_2021}%
  \BibitemOpen
  \bibfield  {author} {\bibinfo {author} {\bibfnamefont {I.}~\bibnamefont
  {Pogorelov}}, \bibinfo {author} {\bibfnamefont {T.}~\bibnamefont {Feldker}},
  \bibinfo {author} {\bibfnamefont {C.~D.}\ \bibnamefont {Marciniak}}, \bibinfo
  {author} {\bibfnamefont {L.}~\bibnamefont {Postler}}, \bibinfo {author}
  {\bibfnamefont {G.}~\bibnamefont {Jacob}}, \bibinfo {author} {\bibfnamefont
  {O.}~\bibnamefont {Krieglsteiner}}, \bibinfo {author} {\bibfnamefont
  {V.}~\bibnamefont {Podlesnic}}, \bibinfo {author} {\bibfnamefont
  {M.}~\bibnamefont {Meth}}, \bibinfo {author} {\bibfnamefont {V.}~\bibnamefont
  {Negnevitsky}}, \bibinfo {author} {\bibfnamefont {M.}~\bibnamefont
  {Stadler}}, \bibinfo {author} {\bibfnamefont {B.}~\bibnamefont {H{\"o}fer}},
  \bibinfo {author} {\bibfnamefont {C.}~\bibnamefont {W{\"a}chter}}, \bibinfo
  {author} {\bibfnamefont {K.}~\bibnamefont {Lakhmanskiy}}, \bibinfo {author}
  {\bibfnamefont {R.}~\bibnamefont {Blatt}}, \bibinfo {author} {\bibfnamefont
  {P.}~\bibnamefont {Schindler}}, \ and\ \bibinfo {author} {\bibfnamefont
  {T.}~\bibnamefont {Monz}},\ }\bibfield  {title} {\emph {\bibinfo {title}
  {Compact {{Ion-Trap Quantum Computing Demonstrator}}},\ }}\href {\doibase
  10.1103/PRXQuantum.2.020343} {\bibfield  {journal} {\bibinfo  {journal} {PRX
  Quantum}\ }\textbf {\bibinfo {volume} {2}},\ \bibinfo {pages} {020343}
  (\bibinfo {year} {2021})}\BibitemShut {NoStop}%
\bibitem [{\citenamefont {Hilder}\ \emph {et~al.}(2022)\citenamefont {Hilder},
  \citenamefont {Pijn}, \citenamefont {Onishchenko}, \citenamefont {Stahl},
  \citenamefont {Orth}, \citenamefont {Lekitsch}, \citenamefont
  {{Rodriguez-Blanco}}, \citenamefont {M{\"u}ller}, \citenamefont
  {{Schmidt-Kaler}},\ and\ \citenamefont
  {Poschinger}}]{hilder_fault-tolerant_2022}%
  \BibitemOpen
  \bibfield  {author} {\bibinfo {author} {\bibfnamefont {J.}~\bibnamefont
  {Hilder}}, \bibinfo {author} {\bibfnamefont {D.}~\bibnamefont {Pijn}},
  \bibinfo {author} {\bibfnamefont {O.}~\bibnamefont {Onishchenko}}, \bibinfo
  {author} {\bibfnamefont {A.}~\bibnamefont {Stahl}}, \bibinfo {author}
  {\bibfnamefont {M.}~\bibnamefont {Orth}}, \bibinfo {author} {\bibfnamefont
  {B.}~\bibnamefont {Lekitsch}}, \bibinfo {author} {\bibfnamefont
  {A.}~\bibnamefont {{Rodriguez-Blanco}}}, \bibinfo {author} {\bibfnamefont
  {M.}~\bibnamefont {M{\"u}ller}}, \bibinfo {author} {\bibfnamefont
  {F.}~\bibnamefont {{Schmidt-Kaler}}}, \ and\ \bibinfo {author} {\bibfnamefont
  {U.~G.}\ \bibnamefont {Poschinger}},\ }\bibfield  {title} {\emph {\bibinfo
  {title} {Fault-{{Tolerant Parity Readout}} on a {{Shuttling-Based Trapped-Ion
  Quantum Computer}}},\ }}\href {\doibase 10.1103/PhysRevX.12.011032}
  {\bibfield  {journal} {\bibinfo  {journal} {Physical Review X}\ }\textbf
  {\bibinfo {volume} {12}},\ \bibinfo {pages} {011032} (\bibinfo {year}
  {2022})}\BibitemShut {NoStop}%
\bibitem [{\citenamefont {Joshi}\ \emph {et~al.}(2022)\citenamefont {Joshi},
  \citenamefont {Kranzl}, \citenamefont {Schuckert}, \citenamefont {Lovas},
  \citenamefont {Maier}, \citenamefont {Blatt}, \citenamefont {Knap},\ and\
  \citenamefont {Roos}}]{joshi_observing_2022}%
  \BibitemOpen
  \bibfield  {author} {\bibinfo {author} {\bibfnamefont {M.~K.}\ \bibnamefont
  {Joshi}}, \bibinfo {author} {\bibfnamefont {F.}~\bibnamefont {Kranzl}},
  \bibinfo {author} {\bibfnamefont {A.}~\bibnamefont {Schuckert}}, \bibinfo
  {author} {\bibfnamefont {I.}~\bibnamefont {Lovas}}, \bibinfo {author}
  {\bibfnamefont {C.}~\bibnamefont {Maier}}, \bibinfo {author} {\bibfnamefont
  {R.}~\bibnamefont {Blatt}}, \bibinfo {author} {\bibfnamefont
  {M.}~\bibnamefont {Knap}}, \ and\ \bibinfo {author} {\bibfnamefont {C.~F.}\
  \bibnamefont {Roos}},\ }\bibfield  {title} {\emph {\bibinfo {title}
  {Observing emergent hydrodynamics in a long-range quantum magnet},\ }}\href
  {\doibase 10.1126/science.abk2400} {\bibfield  {journal} {\bibinfo  {journal}
  {Science (New York, N.Y.)}\ }\textbf {\bibinfo {volume} {376}},\ \bibinfo
  {pages} {720} (\bibinfo {year} {2022})}\BibitemShut {NoStop}%
\bibitem [{\citenamefont {Kokail}\ \emph {et~al.}(2019)\citenamefont {Kokail},
  \citenamefont {Maier}, \citenamefont {{van Bijnen}}, \citenamefont {Brydges},
  \citenamefont {Joshi}, \citenamefont {Jurcevic}, \citenamefont {Muschik},
  \citenamefont {Silvi}, \citenamefont {Blatt}, \citenamefont {Roos},\ and\
  \citenamefont {Zoller}}]{kokail_self-verifying_2019}%
  \BibitemOpen
  \bibfield  {author} {\bibinfo {author} {\bibfnamefont {C.}~\bibnamefont
  {Kokail}}, \bibinfo {author} {\bibfnamefont {C.}~\bibnamefont {Maier}},
  \bibinfo {author} {\bibfnamefont {R.}~\bibnamefont {{van Bijnen}}}, \bibinfo
  {author} {\bibfnamefont {T.}~\bibnamefont {Brydges}}, \bibinfo {author}
  {\bibfnamefont {M.~K.}\ \bibnamefont {Joshi}}, \bibinfo {author}
  {\bibfnamefont {P.}~\bibnamefont {Jurcevic}}, \bibinfo {author}
  {\bibfnamefont {C.~A.}\ \bibnamefont {Muschik}}, \bibinfo {author}
  {\bibfnamefont {P.}~\bibnamefont {Silvi}}, \bibinfo {author} {\bibfnamefont
  {R.}~\bibnamefont {Blatt}}, \bibinfo {author} {\bibfnamefont {C.~F.}\
  \bibnamefont {Roos}}, \ and\ \bibinfo {author} {\bibfnamefont
  {P.}~\bibnamefont {Zoller}},\ }\bibfield  {title} {\emph {\bibinfo {title}
  {Self-verifying variational quantum simulation of lattice models},\ }}\href
  {\doibase 10.1038/s41586-019-1177-4} {\bibfield  {journal} {\bibinfo
  {journal} {Nature}\ }\textbf {\bibinfo {volume} {569}},\ \bibinfo {pages}
  {355} (\bibinfo {year} {2019})}\BibitemShut {NoStop}%
\bibitem [{\citenamefont {Hempel}\ \emph {et~al.}(2018)\citenamefont {Hempel},
  \citenamefont {Maier}, \citenamefont {Romero}, \citenamefont {McClean},
  \citenamefont {Monz}, \citenamefont {Shen}, \citenamefont {Jurcevic},
  \citenamefont {Lanyon}, \citenamefont {Love}, \citenamefont {Babbush},
  \citenamefont {{Aspuru-Guzik}}, \citenamefont {Blatt},\ and\ \citenamefont
  {Roos}}]{hempel_quantum_2018}%
  \BibitemOpen
  \bibfield  {author} {\bibinfo {author} {\bibfnamefont {C.}~\bibnamefont
  {Hempel}}, \bibinfo {author} {\bibfnamefont {C.}~\bibnamefont {Maier}},
  \bibinfo {author} {\bibfnamefont {J.}~\bibnamefont {Romero}}, \bibinfo
  {author} {\bibfnamefont {J.}~\bibnamefont {McClean}}, \bibinfo {author}
  {\bibfnamefont {T.}~\bibnamefont {Monz}}, \bibinfo {author} {\bibfnamefont
  {H.}~\bibnamefont {Shen}}, \bibinfo {author} {\bibfnamefont {P.}~\bibnamefont
  {Jurcevic}}, \bibinfo {author} {\bibfnamefont {B.~P.}\ \bibnamefont
  {Lanyon}}, \bibinfo {author} {\bibfnamefont {P.}~\bibnamefont {Love}},
  \bibinfo {author} {\bibfnamefont {R.}~\bibnamefont {Babbush}}, \bibinfo
  {author} {\bibfnamefont {A.}~\bibnamefont {{Aspuru-Guzik}}}, \bibinfo
  {author} {\bibfnamefont {R.}~\bibnamefont {Blatt}}, \ and\ \bibinfo {author}
  {\bibfnamefont {C.~F.}\ \bibnamefont {Roos}},\ }\bibfield  {title} {\emph
  {\bibinfo {title} {Quantum {{Chemistry Calculations}} on a {{Trapped-Ion
  Quantum Simulator}}},\ }}\href {\doibase 10.1103/PhysRevX.8.031022}
  {\bibfield  {journal} {\bibinfo  {journal} {Physical Review X}\ }\textbf
  {\bibinfo {volume} {8}},\ \bibinfo {pages} {031022} (\bibinfo {year}
  {2018})}\BibitemShut {NoStop}%
\bibitem [{\citenamefont {Matsubara}\ \emph {et~al.}(2012)\citenamefont
  {Matsubara}, \citenamefont {Hachisu}, \citenamefont {Li}, \citenamefont
  {Nagano}, \citenamefont {Locke}, \citenamefont {Nogami}, \citenamefont
  {Kajita}, \citenamefont {Hayasaka}, \citenamefont {Ido},\ and\ \citenamefont
  {Hosokawa}}]{matsubara_direct_2012}%
  \BibitemOpen
  \bibfield  {author} {\bibinfo {author} {\bibfnamefont {K.}~\bibnamefont
  {Matsubara}}, \bibinfo {author} {\bibfnamefont {H.}~\bibnamefont {Hachisu}},
  \bibinfo {author} {\bibfnamefont {Y.}~\bibnamefont {Li}}, \bibinfo {author}
  {\bibfnamefont {S.}~\bibnamefont {Nagano}}, \bibinfo {author} {\bibfnamefont
  {C.}~\bibnamefont {Locke}}, \bibinfo {author} {\bibfnamefont
  {A.}~\bibnamefont {Nogami}}, \bibinfo {author} {\bibfnamefont
  {M.}~\bibnamefont {Kajita}}, \bibinfo {author} {\bibfnamefont
  {K.}~\bibnamefont {Hayasaka}}, \bibinfo {author} {\bibfnamefont
  {T.}~\bibnamefont {Ido}}, \ and\ \bibinfo {author} {\bibfnamefont
  {M.}~\bibnamefont {Hosokawa}},\ }\bibfield  {title} {\emph {\bibinfo {title}
  {Direct comparison of a {{Ca}}{$^+$} single-ion clock against a {{Sr}}
  lattice clock to verify the absolute frequency measurement},\ }}\href
  {\doibase 10.1364/OE.20.022034} {\bibfield  {journal} {\bibinfo  {journal}
  {Optics Express}\ }\textbf {\bibinfo {volume} {20}},\ \bibinfo {pages}
  {22034} (\bibinfo {year} {2012})}\BibitemShut {NoStop}%
\bibitem [{\citenamefont {Huang}\ \emph {et~al.}(2019)\citenamefont {Huang},
  \citenamefont {Guan}, \citenamefont {Zeng}, \citenamefont {Tang},\ and\
  \citenamefont {Gao}}]{huang_ca_2019}%
  \BibitemOpen
  \bibfield  {author} {\bibinfo {author} {\bibfnamefont {Y.}~\bibnamefont
  {Huang}}, \bibinfo {author} {\bibfnamefont {H.}~\bibnamefont {Guan}},
  \bibinfo {author} {\bibfnamefont {M.}~\bibnamefont {Zeng}}, \bibinfo {author}
  {\bibfnamefont {L.}~\bibnamefont {Tang}}, \ and\ \bibinfo {author}
  {\bibfnamefont {K.}~\bibnamefont {Gao}},\ }\bibfield  {title} {\emph
  {\bibinfo {title} {$^{40}${Ca}{$^{+}$} ion optical clock with
  micromotion-induced shifts below $1 \times {10}^{\ensuremath{-}18}$},\
  }}\href {\doibase 10.1103/PhysRevA.99.011401} {\bibfield  {journal} {\bibinfo
   {journal} {Physical Review A}\ }\textbf {\bibinfo {volume} {99}},\ \bibinfo
  {pages} {011401} (\bibinfo {year} {2019})}\BibitemShut {NoStop}%
\bibitem [{\citenamefont {Huang}\ \emph {et~al.}(2021)\citenamefont {Huang},
  \citenamefont {Zhang}, \citenamefont {Zeng}, \citenamefont {Hao},
  \citenamefont {Zhang}, \citenamefont {Guan}, \citenamefont {Chen},
  \citenamefont {Wang},\ and\ \citenamefont {Gao}}]{huang_liquid_2021}%
  \BibitemOpen
  \bibfield  {author} {\bibinfo {author} {\bibfnamefont {Y.}~\bibnamefont
  {Huang}}, \bibinfo {author} {\bibfnamefont {B.}~\bibnamefont {Zhang}},
  \bibinfo {author} {\bibfnamefont {M.}~\bibnamefont {Zeng}}, \bibinfo {author}
  {\bibfnamefont {Y.}~\bibnamefont {Hao}}, \bibinfo {author} {\bibfnamefont
  {H.}~\bibnamefont {Zhang}}, \bibinfo {author} {\bibfnamefont
  {H.}~\bibnamefont {Guan}}, \bibinfo {author} {\bibfnamefont {Z.}~\bibnamefont
  {Chen}}, \bibinfo {author} {\bibfnamefont {M.}~\bibnamefont {Wang}}, \ and\
  \bibinfo {author} {\bibfnamefont {K.}~\bibnamefont {Gao}},\ }\bibfield
  {title} {\emph {\bibinfo {title} {A liquid nitrogen-cooled {{Ca}}{$^+$}
  optical clock with systematic uncertainty of {$3*10^{-18}$}},\ }}\href@noop
  {} {\bibfield  {journal} {\bibinfo  {journal} {ArXiv210308913 Phys.}\ }
  (\bibinfo {year} {2021})},\ \Eprint {http://arxiv.org/abs/2103.08913}
  {arXiv:2103.08913 [physics]} \BibitemShut {NoStop}%
\bibitem [{\citenamefont {Cao}\ \emph {et~al.}(2017)\citenamefont {Cao},
  \citenamefont {Zhang}, \citenamefont {Shang}, \citenamefont {Cui},
  \citenamefont {Yuan}, \citenamefont {Chao}, \citenamefont {Wang},
  \citenamefont {Shu},\ and\ \citenamefont {Huang}}]{cao_compact_2017}%
  \BibitemOpen
  \bibfield  {author} {\bibinfo {author} {\bibfnamefont {J.}~\bibnamefont
  {Cao}}, \bibinfo {author} {\bibfnamefont {P.}~\bibnamefont {Zhang}}, \bibinfo
  {author} {\bibfnamefont {J.}~\bibnamefont {Shang}}, \bibinfo {author}
  {\bibfnamefont {K.}~\bibnamefont {Cui}}, \bibinfo {author} {\bibfnamefont
  {J.}~\bibnamefont {Yuan}}, \bibinfo {author} {\bibfnamefont {S.}~\bibnamefont
  {Chao}}, \bibinfo {author} {\bibfnamefont {S.}~\bibnamefont {Wang}}, \bibinfo
  {author} {\bibfnamefont {H.}~\bibnamefont {Shu}}, \ and\ \bibinfo {author}
  {\bibfnamefont {X.}~\bibnamefont {Huang}},\ }\bibfield  {title} {\emph
  {\bibinfo {title} {A compact, transportable single-ion optical clock with
  7.8\texttimes 10{$^{-17}$} systematic uncertainty},\ }}\href {\doibase
  10.1007/s00340-017-6671-5} {\bibfield  {journal} {\bibinfo  {journal}
  {Applied Physics B: Photophysics and Laser Chemistry}\ }\textbf {\bibinfo
  {volume} {123}},\ \bibinfo {pages} {112} (\bibinfo {year}
  {2017})}\BibitemShut {NoStop}%
\bibitem [{\citenamefont {Li}\ \emph {et~al.}(2022)\citenamefont {Li},
  \citenamefont {Wolf}, \citenamefont {Klein}, \citenamefont {Budker},
  \citenamefont {D{\"u}llmann},\ and\ \citenamefont
  {{Schmidt-Kaler}}}]{li_robust_2022}%
  \BibitemOpen
  \bibfield  {author} {\bibinfo {author} {\bibfnamefont {W.}~\bibnamefont
  {Li}}, \bibinfo {author} {\bibfnamefont {S.}~\bibnamefont {Wolf}}, \bibinfo
  {author} {\bibfnamefont {L.}~\bibnamefont {Klein}}, \bibinfo {author}
  {\bibfnamefont {D.}~\bibnamefont {Budker}}, \bibinfo {author} {\bibfnamefont
  {C.~E.}\ \bibnamefont {D{\"u}llmann}}, \ and\ \bibinfo {author}
  {\bibfnamefont {F.}~\bibnamefont {{Schmidt-Kaler}}},\ }\bibfield  {title}
  {\emph {\bibinfo {title} {Robust polarization gradient cooling of trapped
  ions},\ }}\href {\doibase 10.1088/1367-2630/ac6233} {\bibfield  {journal}
  {\bibinfo  {journal} {New Journal of Physics}\ }\textbf {\bibinfo {volume}
  {24}},\ \bibinfo {pages} {043028} (\bibinfo {year} {2022})}\BibitemShut
  {NoStop}%
\bibitem [{\citenamefont {Morigi}\ \emph {et~al.}(2000)\citenamefont {Morigi},
  \citenamefont {Eschner},\ and\ \citenamefont {Keitel}}]{morigi_Ground_2000}%
  \BibitemOpen
  \bibfield  {author} {\bibinfo {author} {\bibfnamefont {G.}~\bibnamefont
  {Morigi}}, \bibinfo {author} {\bibfnamefont {J.}~\bibnamefont {Eschner}}, \
  and\ \bibinfo {author} {\bibfnamefont {C.~H.}\ \bibnamefont {Keitel}},\
  }\bibfield  {title} {\emph {\bibinfo {title} {Ground {{State Laser Cooling
  Using Electromagnetically Induced Transparency}}},\ }}\href {\doibase
  10.1103/PhysRevLett.85.4458} {\bibfield  {journal} {\bibinfo  {journal}
  {Physical Review Letters}\ }\textbf {\bibinfo {volume} {85}} (\bibinfo {year}
  {2000}),\ 10.1103/PhysRevLett.85.4458}\BibitemShut {NoStop}%
\bibitem [{\citenamefont {Scharnhorst}\ \emph {et~al.}(2018)\citenamefont
  {Scharnhorst}, \citenamefont {Cerrillo}, \citenamefont {Kramer},
  \citenamefont {Leroux}, \citenamefont {W{\"u}bbena}, \citenamefont
  {Retzker},\ and\ \citenamefont {Schmidt}}]{scharnhorst_Experimental_2018}%
  \BibitemOpen
  \bibfield  {author} {\bibinfo {author} {\bibfnamefont {N.}~\bibnamefont
  {Scharnhorst}}, \bibinfo {author} {\bibfnamefont {J.}~\bibnamefont
  {Cerrillo}}, \bibinfo {author} {\bibfnamefont {J.}~\bibnamefont {Kramer}},
  \bibinfo {author} {\bibfnamefont {I.~D.}\ \bibnamefont {Leroux}}, \bibinfo
  {author} {\bibfnamefont {J.~B.}\ \bibnamefont {W{\"u}bbena}}, \bibinfo
  {author} {\bibfnamefont {A.}~\bibnamefont {Retzker}}, \ and\ \bibinfo
  {author} {\bibfnamefont {P.~O.}\ \bibnamefont {Schmidt}},\ }\bibfield
  {title} {\emph {\bibinfo {title} {Experimental and theoretical investigation
  of a multimode cooling scheme using multiple
  electromagnetically-induced-transparency resonances},\ }}\href {\doibase
  10.1103/PhysRevA.98.023424} {\bibfield  {journal} {\bibinfo  {journal}
  {Physical Review A}\ }\textbf {\bibinfo {volume} {98}} (\bibinfo {year}
  {2018}),\ 10.1103/PhysRevA.98.023424}\BibitemShut {NoStop}%
\bibitem [{\citenamefont {Lechner}\ \emph {et~al.}(2016)\citenamefont
  {Lechner}, \citenamefont {Maier}, \citenamefont {Hempel}, \citenamefont
  {Jurcevic}, \citenamefont {Lanyon}, \citenamefont {Monz}, \citenamefont
  {Brownnutt}, \citenamefont {Blatt},\ and\ \citenamefont
  {Roos}}]{lechner_electromagnetically-induced-transparency_2016}%
  \BibitemOpen
  \bibfield  {author} {\bibinfo {author} {\bibfnamefont {R.}~\bibnamefont
  {Lechner}}, \bibinfo {author} {\bibfnamefont {C.}~\bibnamefont {Maier}},
  \bibinfo {author} {\bibfnamefont {C.}~\bibnamefont {Hempel}}, \bibinfo
  {author} {\bibfnamefont {P.}~\bibnamefont {Jurcevic}}, \bibinfo {author}
  {\bibfnamefont {B.~P.}\ \bibnamefont {Lanyon}}, \bibinfo {author}
  {\bibfnamefont {T.}~\bibnamefont {Monz}}, \bibinfo {author} {\bibfnamefont
  {M.}~\bibnamefont {Brownnutt}}, \bibinfo {author} {\bibfnamefont
  {R.}~\bibnamefont {Blatt}}, \ and\ \bibinfo {author} {\bibfnamefont {C.~F.}\
  \bibnamefont {Roos}},\ }\bibfield  {title} {\emph {\bibinfo {title}
  {Electromagnetically-induced-transparency ground-state cooling of long ion
  strings},\ }}\href {\doibase 10.1103/PhysRevA.93.053401} {\bibfield
  {journal} {\bibinfo  {journal} {Physical Review A: Atomic, Molecular, and
  Optical Physics}\ }\textbf {\bibinfo {volume} {93}} (\bibinfo {year}
  {2016}),\ 10.1103/PhysRevA.93.053401}\BibitemShut {NoStop}%
\bibitem [{\citenamefont {Hannig}\ \emph {et~al.}(2019)\citenamefont {Hannig},
  \citenamefont {Pelzer}, \citenamefont {Scharnhorst}, \citenamefont {Kramer},
  \citenamefont {Stepanova}, \citenamefont {Xu}, \citenamefont {Spethmann},
  \citenamefont {Leroux}, \citenamefont {Mehlst{\"a}ubler},\ and\ \citenamefont
  {Schmidt}}]{hannig_transportable_2019}%
  \BibitemOpen
  \bibfield  {author} {\bibinfo {author} {\bibfnamefont {S.}~\bibnamefont
  {Hannig}}, \bibinfo {author} {\bibfnamefont {L.}~\bibnamefont {Pelzer}},
  \bibinfo {author} {\bibfnamefont {N.}~\bibnamefont {Scharnhorst}}, \bibinfo
  {author} {\bibfnamefont {J.}~\bibnamefont {Kramer}}, \bibinfo {author}
  {\bibfnamefont {M.}~\bibnamefont {Stepanova}}, \bibinfo {author}
  {\bibfnamefont {Z.~T.}\ \bibnamefont {Xu}}, \bibinfo {author} {\bibfnamefont
  {N.}~\bibnamefont {Spethmann}}, \bibinfo {author} {\bibfnamefont {I.~D.}\
  \bibnamefont {Leroux}}, \bibinfo {author} {\bibfnamefont {T.~E.}\
  \bibnamefont {Mehlst{\"a}ubler}}, \ and\ \bibinfo {author} {\bibfnamefont
  {P.~O.}\ \bibnamefont {Schmidt}},\ }\bibfield  {title} {\emph {\bibinfo
  {title} {Towards a transportable aluminium ion quantum logic optical clock},\
  }}\href {\doibase 10.1063/1.5090583} {\bibfield  {journal} {\bibinfo
  {journal} {Review of Scientific Instruments}\ }\textbf {\bibinfo {volume}
  {90}} (\bibinfo {year} {2019}),\ 10.1063/1.5090583}\BibitemShut {NoStop}%
\bibitem [{Note1()}]{Note1}%
  \BibitemOpen
  \bibinfo {note} {High Finesse U10}\BibitemShut {NoStop}%
\bibitem [{Note2()}]{Note2}%
  \BibitemOpen
  \bibinfo {note} {TA pro, Toptica}\BibitemShut {NoStop}%
\bibitem [{\citenamefont {Drever}\ \emph {et~al.}(1983)\citenamefont {Drever},
  \citenamefont {Hall}, \citenamefont {Kowalski}, \citenamefont {Hough},
  \citenamefont {Ford}, \citenamefont {Munley},\ and\ \citenamefont
  {Ward}}]{drever_Laser_1983}%
  \BibitemOpen
  \bibfield  {author} {\bibinfo {author} {\bibfnamefont {R.~W.~P.}\
  \bibnamefont {Drever}}, \bibinfo {author} {\bibfnamefont {J.~L.}\
  \bibnamefont {Hall}}, \bibinfo {author} {\bibfnamefont {F.~V.}\ \bibnamefont
  {Kowalski}}, \bibinfo {author} {\bibfnamefont {J.}~\bibnamefont {Hough}},
  \bibinfo {author} {\bibfnamefont {G.~M.}\ \bibnamefont {Ford}}, \bibinfo
  {author} {\bibfnamefont {A.~J.}\ \bibnamefont {Munley}}, \ and\ \bibinfo
  {author} {\bibfnamefont {H.}~\bibnamefont {Ward}},\ }\bibfield  {title}
  {\emph {\bibinfo {title} {Laser phase and frequency stabilization using an
  optical resonator},\ }}\href {\doibase 10.1007/BF00702605} {\bibfield
  {journal} {\bibinfo  {journal} {Applied Physics B}\ }\textbf {\bibinfo
  {volume} {31}} (\bibinfo {year} {1983}),\ 10.1007/BF00702605}\BibitemShut
  {NoStop}%
\bibitem [{\citenamefont {Scharnhorst}\ \emph {et~al.}(2015)\citenamefont
  {Scharnhorst}, \citenamefont {W{\"u}bbena}, \citenamefont {Hannig},
  \citenamefont {Jakobsen}, \citenamefont {Kramer}, \citenamefont {Leroux},\
  and\ \citenamefont {Schmidt}}]{scharnhorst_Highbandwidth_2015}%
  \BibitemOpen
  \bibfield  {author} {\bibinfo {author} {\bibfnamefont {N.}~\bibnamefont
  {Scharnhorst}}, \bibinfo {author} {\bibfnamefont {J.~B.}\ \bibnamefont
  {W{\"u}bbena}}, \bibinfo {author} {\bibfnamefont {S.}~\bibnamefont {Hannig}},
  \bibinfo {author} {\bibfnamefont {K.}~\bibnamefont {Jakobsen}}, \bibinfo
  {author} {\bibfnamefont {J.}~\bibnamefont {Kramer}}, \bibinfo {author}
  {\bibfnamefont {I.~D.}\ \bibnamefont {Leroux}}, \ and\ \bibinfo {author}
  {\bibfnamefont {P.~O.}\ \bibnamefont {Schmidt}},\ }\bibfield  {title} {\emph
  {\bibinfo {title} {High-bandwidth transfer of phase stability through a fiber
  frequency comb},\ }}\href {\doibase 10.1364/OE.23.019771} {\bibfield
  {journal} {\bibinfo  {journal} {Optics Express}\ }\textbf {\bibinfo {volume}
  {23}} (\bibinfo {year} {2015}),\ 10.1364/OE.23.019771}\BibitemShut {NoStop}%
\bibitem [{\citenamefont {Matei}\ \emph {et~al.}(2017)\citenamefont {Matei},
  \citenamefont {Legero}, \citenamefont {H{\"a}fner}, \citenamefont {Grebing},
  \citenamefont {Weyrich}, \citenamefont {Zhang}, \citenamefont {Sonderhouse},
  \citenamefont {Robinson}, \citenamefont {Ye}, \citenamefont {Riehle},\ and\
  \citenamefont {Sterr}}]{matei_Lasers_2017}%
  \BibitemOpen
  \bibfield  {author} {\bibinfo {author} {\bibfnamefont {D.~G.}\ \bibnamefont
  {Matei}}, \bibinfo {author} {\bibfnamefont {T.}~\bibnamefont {Legero}},
  \bibinfo {author} {\bibfnamefont {S.}~\bibnamefont {H{\"a}fner}}, \bibinfo
  {author} {\bibfnamefont {C.}~\bibnamefont {Grebing}}, \bibinfo {author}
  {\bibfnamefont {R.}~\bibnamefont {Weyrich}}, \bibinfo {author} {\bibfnamefont
  {W.}~\bibnamefont {Zhang}}, \bibinfo {author} {\bibfnamefont
  {L.}~\bibnamefont {Sonderhouse}}, \bibinfo {author} {\bibfnamefont {J.~M.}\
  \bibnamefont {Robinson}}, \bibinfo {author} {\bibfnamefont {J.}~\bibnamefont
  {Ye}}, \bibinfo {author} {\bibfnamefont {F.}~\bibnamefont {Riehle}}, \ and\
  \bibinfo {author} {\bibfnamefont {U.}~\bibnamefont {Sterr}},\ }\bibfield
  {title} {\emph {\bibinfo {title} {1.5{{$\mu$m}} {{Lasers}} with {{Sub-10 mHz
  Linewidth}}},\ }}\href {\doibase 10.1103/PhysRevLett.118.263202} {\bibfield
  {journal} {\bibinfo  {journal} {Physical Review Letters}\ }\textbf {\bibinfo
  {volume} {118}} (\bibinfo {year} {2017}),\
  10.1103/PhysRevLett.118.263202}\BibitemShut {NoStop}%
\bibitem [{\citenamefont {Benkler}\ \emph {et~al.}(2019)\citenamefont
  {Benkler}, \citenamefont {Lipphardt}, \citenamefont {Puppe}, \citenamefont
  {Wilk}, \citenamefont {Rohde},\ and\ \citenamefont
  {Sterr}}]{benkler_Endtoend_2019}%
  \BibitemOpen
  \bibfield  {author} {\bibinfo {author} {\bibfnamefont {E.}~\bibnamefont
  {Benkler}}, \bibinfo {author} {\bibfnamefont {B.}~\bibnamefont {Lipphardt}},
  \bibinfo {author} {\bibfnamefont {T.}~\bibnamefont {Puppe}}, \bibinfo
  {author} {\bibfnamefont {R.}~\bibnamefont {Wilk}}, \bibinfo {author}
  {\bibfnamefont {F.}~\bibnamefont {Rohde}}, \ and\ \bibinfo {author}
  {\bibfnamefont {U.}~\bibnamefont {Sterr}},\ }\bibfield  {title} {\emph
  {\bibinfo {title} {End-to-end topology for fiber comb based optical frequency
  transfer at the 10{$^{-21}$} level},\ }}\href {\doibase 10.1364/OE.27.036886}
  {\bibfield  {journal} {\bibinfo  {journal} {Optics Express}\ }\textbf
  {\bibinfo {volume} {27}} (\bibinfo {year} {2019}),\
  10.1364/OE.27.036886}\BibitemShut {NoStop}%
\bibitem [{\citenamefont {Schindler}(2008)}]{schindler_Frequency_2008}%
  \BibitemOpen
  \bibfield  {author} {\bibinfo {author} {\bibfnamefont {P.}~\bibnamefont
  {Schindler}},\ }\emph {\bibinfo {title} {Frequency Synthesis and Pulse
  Shaping for Quantum Information Processing with Trapped Ions}},\ \href@noop
  {} {\bibinfo {type} {Diploma {{Thesis}}}},\ \bibinfo  {school} {University of
  Innsbruck}, \bibinfo {address} {{Innsbruck, Austria}} (\bibinfo {year}
  {2008})\BibitemShut {NoStop}%
\bibitem [{\citenamefont {Pham}(2005)}]{pham_generalpurpose_2005}%
  \BibitemOpen
  \bibfield  {author} {\bibinfo {author} {\bibfnamefont {P.~T.~T.}\
  \bibnamefont {Pham}},\ }\emph {\bibinfo {title} {A General-Purpose Pulse
  Sequencer for Quantum Computing}},\ \href@noop {} {Ph.D. thesis},\ \bibinfo
  {school} {Massachusetts Institute of Technology}, \bibinfo {address}
  {{Cambridge, Massachusetts, USA}} (\bibinfo {year} {2005})\BibitemShut
  {NoStop}%
\bibitem [{Note3()}]{Note3}%
  \BibitemOpen
  \bibinfo {note} {Keysight 33622A}\BibitemShut {NoStop}%
\bibitem [{\citenamefont {Roos}\ \emph {et~al.}(2000)\citenamefont {Roos},
  \citenamefont {Leibfried}, \citenamefont {Mundt}, \citenamefont
  {{Schmidt-Kaler}}, \citenamefont {Eschner},\ and\ \citenamefont
  {Blatt}}]{roos_experimental_2000}%
  \BibitemOpen
  \bibfield  {author} {\bibinfo {author} {\bibfnamefont {C.~F.}\ \bibnamefont
  {Roos}}, \bibinfo {author} {\bibfnamefont {D.}~\bibnamefont {Leibfried}},
  \bibinfo {author} {\bibfnamefont {A.}~\bibnamefont {Mundt}}, \bibinfo
  {author} {\bibfnamefont {F.}~\bibnamefont {{Schmidt-Kaler}}}, \bibinfo
  {author} {\bibfnamefont {J.}~\bibnamefont {Eschner}}, \ and\ \bibinfo
  {author} {\bibfnamefont {R.}~\bibnamefont {Blatt}},\ }\bibfield  {title}
  {\emph {\bibinfo {title} {Experimental demonstration of ground state laser
  cooling with electromagnetically induced transparency},\ }}\href@noop {}
  {\bibfield  {journal} {\bibinfo  {journal} {Physical Review Letters}\
  }\textbf {\bibinfo {volume} {85}},\ \bibinfo {pages} {5547} (\bibinfo {year}
  {2000})}\BibitemShut {NoStop}%
\bibitem [{\citenamefont {Wunderlich}\ \emph {et~al.}(2007)\citenamefont
  {Wunderlich}, \citenamefont {Hannemann}, \citenamefont {K{\"o}rber},
  \citenamefont {H{\"a}ffner}, \citenamefont {Roos}, \citenamefont
  {H{\"a}nsel}, \citenamefont {Blatt},\ and\ \citenamefont
  {{Schmidt-Kaler}}}]{wunderlich_Robust_2007}%
  \BibitemOpen
  \bibfield  {author} {\bibinfo {author} {\bibfnamefont {C.}~\bibnamefont
  {Wunderlich}}, \bibinfo {author} {\bibfnamefont {T.}~\bibnamefont
  {Hannemann}}, \bibinfo {author} {\bibfnamefont {T.}~\bibnamefont
  {K{\"o}rber}}, \bibinfo {author} {\bibfnamefont {H.}~\bibnamefont
  {H{\"a}ffner}}, \bibinfo {author} {\bibfnamefont {C.}~\bibnamefont {Roos}},
  \bibinfo {author} {\bibfnamefont {W.}~\bibnamefont {H{\"a}nsel}}, \bibinfo
  {author} {\bibfnamefont {R.}~\bibnamefont {Blatt}}, \ and\ \bibinfo {author}
  {\bibfnamefont {F.}~\bibnamefont {{Schmidt-Kaler}}},\ }\bibfield  {title}
  {\emph {\bibinfo {title} {Robust state preparation of a single trapped ion by
  adiabatic passage},\ }}\href {\doibase 10.1080/09500340600741082} {\bibfield
  {journal} {\bibinfo  {journal} {Journal of Modern Optics}\ }\textbf {\bibinfo
  {volume} {54}} (\bibinfo {year} {2007}),\
  10.1080/09500340600741082}\BibitemShut {NoStop}%
\bibitem [{\citenamefont {Leibfried}\ \emph {et~al.}(2004)\citenamefont
  {Leibfried}, \citenamefont {Barrett}, \citenamefont {Schaetz}, \citenamefont
  {Britton}, \citenamefont {Chiaverini}, \citenamefont {Itano}, \citenamefont
  {Jost}, \citenamefont {Langer},\ and\ \citenamefont
  {Wineland}}]{leibfried_toward_2004}%
  \BibitemOpen
  \bibfield  {author} {\bibinfo {author} {\bibfnamefont {D.}~\bibnamefont
  {Leibfried}}, \bibinfo {author} {\bibfnamefont {M.~D.}\ \bibnamefont
  {Barrett}}, \bibinfo {author} {\bibfnamefont {T.}~\bibnamefont {Schaetz}},
  \bibinfo {author} {\bibfnamefont {J.}~\bibnamefont {Britton}}, \bibinfo
  {author} {\bibfnamefont {J.}~\bibnamefont {Chiaverini}}, \bibinfo {author}
  {\bibfnamefont {W.~M.}\ \bibnamefont {Itano}}, \bibinfo {author}
  {\bibfnamefont {J.~D.}\ \bibnamefont {Jost}}, \bibinfo {author}
  {\bibfnamefont {C.}~\bibnamefont {Langer}}, \ and\ \bibinfo {author}
  {\bibfnamefont {D.~J.}\ \bibnamefont {Wineland}},\ }\bibfield  {title} {\emph
  {\bibinfo {title} {Toward {{Heisenberg-Limited Spectroscopy}} with
  {{Multiparticle Entangled States}}},\ }}\href {\doibase
  10.1126/science.1097576} {\bibfield  {journal} {\bibinfo  {journal} {Science
  (New York, N.Y.)}\ }\textbf {\bibinfo {volume} {304}},\ \bibinfo {pages}
  {1476} (\bibinfo {year} {2004})}\BibitemShut {NoStop}%
\bibitem [{\citenamefont {Nichol}\ \emph {et~al.}(2022)\citenamefont {Nichol},
  \citenamefont {Srinivas}, \citenamefont {Nadlinger}, \citenamefont {Drmota},
  \citenamefont {Main}, \citenamefont {Araneda}, \citenamefont {Ballance},\
  and\ \citenamefont {Lucas}}]{nichol_elementary_2022}%
  \BibitemOpen
  \bibfield  {author} {\bibinfo {author} {\bibfnamefont {B.~C.}\ \bibnamefont
  {Nichol}}, \bibinfo {author} {\bibfnamefont {R.}~\bibnamefont {Srinivas}},
  \bibinfo {author} {\bibfnamefont {D.~P.}\ \bibnamefont {Nadlinger}}, \bibinfo
  {author} {\bibfnamefont {P.}~\bibnamefont {Drmota}}, \bibinfo {author}
  {\bibfnamefont {D.}~\bibnamefont {Main}}, \bibinfo {author} {\bibfnamefont
  {G.}~\bibnamefont {Araneda}}, \bibinfo {author} {\bibfnamefont {C.~J.}\
  \bibnamefont {Ballance}}, \ and\ \bibinfo {author} {\bibfnamefont {D.~M.}\
  \bibnamefont {Lucas}},\ }\bibfield  {title} {\emph {\bibinfo {title} {An
  elementary quantum network of entangled optical atomic clocks},\ }}\href
  {\doibase 10.1038/s41586-022-05088-z} {\bibfield  {journal} {\bibinfo
  {journal} {Nature}\ ,\ \bibinfo {pages} {1}} (\bibinfo {year}
  {2022})}\BibitemShut {NoStop}%
\bibitem [{\citenamefont {Schulte}\ \emph {et~al.}(2020)\citenamefont
  {Schulte}, \citenamefont {Lisdat}, \citenamefont {Schmidt}, \citenamefont
  {Sterr},\ and\ \citenamefont {Hammerer}}]{schulte_prospects_2020}%
  \BibitemOpen
  \bibfield  {author} {\bibinfo {author} {\bibfnamefont {M.}~\bibnamefont
  {Schulte}}, \bibinfo {author} {\bibfnamefont {C.}~\bibnamefont {Lisdat}},
  \bibinfo {author} {\bibfnamefont {P.~O.}\ \bibnamefont {Schmidt}}, \bibinfo
  {author} {\bibfnamefont {U.}~\bibnamefont {Sterr}}, \ and\ \bibinfo {author}
  {\bibfnamefont {K.}~\bibnamefont {Hammerer}},\ }\bibfield  {title} {\emph
  {\bibinfo {title} {Prospects and challenges for squeezing-enhanced optical
  atomic clocks},\ }}\href {\doibase 10.1038/s41467-020-19403-7} {\bibfield
  {journal} {\bibinfo  {journal} {Nature Communications}\ }\textbf {\bibinfo
  {volume} {11}},\ \bibinfo {pages} {5955} (\bibinfo {year}
  {2020})}\BibitemShut {NoStop}%
\bibitem [{\citenamefont {Häffner}\ \emph {et~al.}(2008)\citenamefont
  {Häffner}, \citenamefont {Roos},\ and\ \citenamefont
  {Blatt}}]{HAFFNER_2008}%
  \BibitemOpen
  \bibfield  {author} {\bibinfo {author} {\bibfnamefont {H.}~\bibnamefont
  {Häffner}}, \bibinfo {author} {\bibfnamefont {C.}~\bibnamefont {Roos}}, \
  and\ \bibinfo {author} {\bibfnamefont {R.}~\bibnamefont {Blatt}},\ }\bibfield
   {title} {\emph {\bibinfo {title} {Quantum computing with trapped ions},\
  }}\href {\doibase 10.1016/j.physrep.2008.09.003} {\bibfield  {journal}
  {\bibinfo  {journal} {Physics Reports}\ }\textbf {\bibinfo {volume} {469}},\
  \bibinfo {pages} {155} (\bibinfo {year} {2008})}\BibitemShut {NoStop}%
\bibitem [{\citenamefont {Magnus}(1954)}]{MagnusExpansion}%
  \BibitemOpen
  \bibfield  {author} {\bibinfo {author} {\bibfnamefont {W.}~\bibnamefont
  {Magnus}},\ }\bibfield  {title} {\emph {\bibinfo {title} {On the exponential
  solution of differential equations for a linear operator},\ }}\href {\doibase
  10.1002/cpa.3160070404} {\bibfield  {journal} {\bibinfo  {journal}
  {Communications on Pure and Applied Mathematics}\ }\textbf {\bibinfo {volume}
  {7}} (\bibinfo {year} {1954}),\ 10.1002/cpa.3160070404}\BibitemShut {NoStop}%
\bibitem [{\citenamefont {Galindo}\ and\ \citenamefont
  {Pascual}(2012)}]{Quantummechanics1A.GalindoP.Pascual}%
  \BibitemOpen
  \bibfield  {author} {\bibinfo {author} {\bibfnamefont {A.}~\bibnamefont
  {Galindo}}\ and\ \bibinfo {author} {\bibfnamefont {P.}~\bibnamefont
  {Pascual}},\ }\href {\doibase https://doi.org/10.1007/978-3-642-83854-5}
  {\emph {\bibinfo {title} {Quantum Mechanics {{I}}}}}\ (\bibinfo  {publisher}
  {Springer Berlin, Heidelberg},\ \bibinfo {year} {2012})\BibitemShut {NoStop}%
\end{thebibliography}
\end{document}